



\def\rarrow{\rightarrow}

\def\Dt{\spose{\raise 1.5ex\hbox{\hskip3pt$\mathchar"201$}}}    
\def\dt{\spose{\raise 1.0ex\hbox{\hskip2pt$\mathchar"201$}}}    

\def\lta{\mathrel{\spose{\lower 3pt\hbox{$\mathchar"218$}}
 \raise 2.0pt\hbox{$\mathchar"13C$}}}
\def\gta{\mathrel{\spose{\lower 3pt\hbox{$\mathchar"218$}}
 \raise 2.0pt\hbox{$\mathchar"13E$}}}
\def\recip#1{^{-#1}} 



\def\new{{\rm\chaphead\the\eqnumber}\global\advance\eqnumber by 1}
\def\ref#1{\advance\eqnumber by -#1\chaphead\the\eqnumber\advance\eqnumber
by #1 }
\def\last{\advance\eqnumber by -1 {\rm\chaphead\the\eqnumber}\advance
 \eqnumber by 1}
\def\eqnam#1{\xdef#1{\chaphead\the\eqnumber}}


\def\nfig{\chaphead\the\fignumber\global\advance\fignumber by 1} 
\def\nfiga#1{\chaphead\the\fignumber{#1}\global\advance\fignumber by 1}
\def\rfig#1{\advance\fignumber by -#1 \chaphead\the\fignumber
 \advance\fignumber by #1}
\def\fignam#1{\xdef#1{\chaphead\the\fignumber}}


\def\i#1{{\it #1\/}} 

\def\eg{{\it e.g.\ }}
\def\ie{{\it i.e.,\ }}

\def\etc{{\it etc.\ }}

\def\spose#1{\hbox to 0pt{#1\hss}}
\def\s{\ifmmode \widetilde \else \~\fi} 
\def\={\overline}
\def\-{\underline}


\def\dw{\Delta\omega}
\def\imax{\dot I_{\rm max}}
\def\maxi{I_{\rm max}}

\def\emin{\epsilon_{\rm min}}
\def\cost{energy cost per bit}
\def\eofp{\varepsilon (p)}
\def\ekt{\varepsilon (p)/kT}
\def\pa{p_a}
\def\nj{n_j}

\def\speed{v(\varepsilon)}
\def\ei{\varepsilon_{i\alpha}}
\def\ej{\varepsilon_{j\beta}}

\def\ket#1{|#1\negthinspace\negthinspace >}
\def\bra#1{<\negthinspace\negthinspace#1|}

\def\energy{\ei+\ej+\cdots}
\def\sh{self--heralding}
\def\te{time--energy uncertainty relation}

\def\po{{2\pi P_\omega\over\hbar\omega}}
\def\ha{\hbar\omega/kT}


\def\q#1{$^{#1}$}
\def\r#1{\par\noindent\hangindent=23pt\hangafter=1 #1.\ }
\def\pr#1#2#3#4{#1, {\nineit Phys. Rev.\/}{\ninebf  #2}  (19#4) #3}
\def\pra#1#2#3#4{#1, {\nineit Phys. Rev. A\/} {\ninebf  #2}  (19#4) #3}
\def\prd#1#2#3#4{#1, {\nineit Phys. Rev. D\/} {\ninebf  #2}  (19#4) #3}
\def\prl#1#2#3#4{#1, {\nineit Phys. Rev. Letters\/} {\ninebf  #2}  (19#4) #3}
\def\cmp#1#2#3#4{#1, {\nineit Commun. Math. Phys.\/} {\ninebf  #2}  (19#4) #3}
\def\rmp#1#2#3#4{#1, {\nineit Revs. Mod. Phys.\/} {\ninebf  #2}  (19#4)}
\def\ijtp#1#2#3#4{#1, {\nineit Int. J. Theor. Phys.\/} {\ninebf  #2}  (19#4) #3}
\def\fop#1#2#3#4{#1, {\nineit Found. Phys.\/} {\ninebf  #2}  (19#4) #3}
\def\jpa#1#2#3#4{#1, {\nineit J. Phys.\/} {\ninebf  A#2}  (19#4) #3}

\def\pla#1#2#3#4{#1, {\nineit Phys. Lett.\/}\enspace {\ninebf  #2A}  (19#4) #3}
\def\nc#1#2#3#4{#1, {\nineit Nuov. Cim.\/} {\ninebf #2}  (19#4) #3}
\def\ncl#1#2#3#4{#1, {\nineit Lett. Nuov. Cim.\/} {\ninebf  #2}  (19#4) #3}
\def\obscure#1#2#3#4#5{{#1}, {\nineit#2\/} {\ninebf #3}\thinspace(19#5) #4}
\def\book#1#2#3#4{#1, {\nineit#2,} (#3, 19#4)}
\def\inbooked#1#2#3#4#5{#1, in {\nineit#2\/}, ed. #3, (#4, 19#5)}
\def\inbookeds#1#2#3#4#5{#1, in {\nineit#2\/}, eds. #3, (#4, 19#5)}
\def\anp#1#2#3#4{#1, {\nineit Ann. Phys.\/} {\ninebf  #2}  (19#4) #3}
\def\zp#1#2#3#4{#1, {\nineit Z. Phys.\/} {\ninebf  #2}  (19#4) #3}
\def\grg#1#2#3#4{#1, {\nineit General Relativity and Gravitation\/} {\ninebf 
#2}  (19#4) #3} \def\aa#1#2#3#4{#1, {\nineit Astron. Astrophys.\/} {\ninebf  #2} 
(19#4) #3}


\font\tenrm=cmr10
\font\tenit=cmti10
\font\ninebf=cmbx9
\font\ninerm=cmr9
\font\nineit=cmti9

\font\eightrm=cmr8
\font\eightit=cmti8

\hsize=6.5truein
\vsize=9.9truein
\parskip=0.0pt
\parindent=15pt
\baselineskip=10pt
\def\qed{\hbox{${\vcenter{\vbox{\hrule height 0.4pt\hbox{\vrule width 0.4pt
height 6pt\kern5pt\vrule width 0.4pt}\hrule height 0.4pt}}}$}}
\line{\eightrm International Journal of Modern Physics C Vol. 1, No. 4
355-422 (1990)\hfil}
\line{\eightrm $\copyright\,$ World Scientific Publishing Company \hfil}
\vglue 5pc
\baselineskip=13pt
\centerline{\tenbf QUANTUM LIMITATIONS ON THE}
\centerline{\tenbf STORAGE AND TRANSMISSION OF INFORMATION}
\baselineskip=10pt
\vglue 24pt
\centerline{\eightrm JACOB D. BEKENSTEIN}
\baselineskip=12pt
\centerline{\eightit Physics Department, Ben Gurion University, Beersheva, 84105
ISRAEL}
\baselineskip=10pt
\centerline{\eightit and}
\centerline{\eightit Racah Institute of Physics, The Hebrew University of Jerusalem}
\centerline{\eightit Givat Ram, Jerusalem, 91904 ISRAEL
\footnote{$^\ast$}{\eightrm\baselineskip=10pt Present address.}} \vglue 10pt
\centerline{\rm and} \vglue 10pt
\centerline{\eightrm MARCELO SCHIFFER}
\baselineskip=12pt
\centerline{\eightit Insituto de F\'isica Te\'orica, Rua Pamplona 145, S\~ao
Paulo, S.P. 01405 BRAZIL}
\vglue 20pt
\centerline{\eightrm Received  7 November 1990}
\vglue 16pt
\centerline{\eightrm ABSTRACT}
{\rightskip=1.5pc
\leftskip=1.5pc
\eightrm
\baselineskip=10pt
\parindent=1pc
\noindent
Information must take up space, must weigh, and its flux must be limited. Quantum limits on
communication and information storage leading to these conclusions are here described. 
Quantum channel capacity theory is reviewed for both steady state and burst communication.  An
analytic approximation is given for the maximum signal information possible with occupation number
signal states as a function of mean signal energy. A theorem guaranteeing that
these states are optimal for communication is proved.  A heuristic ``proof'' of the linear
bound on communication is given, followed by rigorous proofs for signals with specified mean energy,
and for signals with given energy budget.  And systems of many parallel quantum channels are shown
to obey the linear bound for a natural channel architecture.   The time--energy
uncertainty principle is reformulated in information language by means of the linear bound.  The
quantum bound on information storage capacity of quantum mechanical and quantum field
devices is reviewed.  A simplified version of the analytic proof for the bound is given for the
latter case.  Solitons as information caches are discussed, as is information storage in
one dimensional systems.  The influence of signal self--gravitation on communication is considerd.
Finally, it is shown that acceleration of a receiver acts to block information transfer.

\vglue 5pt
\noindent
{\eightit Keywords:\/Information,  entropy, coding, communication, quantum channel capacity.} \vglue
12pt}

\vglue 12pt
\baselineskip=13pt
\line{\tenbf 1. Introduction  \hfil}
\vglue 5pt

{\rm Information, its storage, and its transfer from
system to system are all crucial issues in science and technology.  They are at the crux of
computation.  They are connected with the foundations of thermodynamics.  Their influence is
felt far from the physical sciences.  Thus a fundamental aspect in the
evolution of life is the ability to store and transmit genetic information
at the level of the species.  The same is true at the level of
society. Human survival rests on the ability of society to acquire and
store large quantities of information and transmit it rapidly.  Given the
importance of the subject, a natural question is whether there are intrisic 
limitations dictated by the laws of nature on information storage and communication.\par

Does information take up space?  Does it weigh?  Can its flux
be made arbitrarily large?  These related questions must be very old.  They
evidently have immediate technological bearing.  More important, they go
right to the heart of the nature of information: is information impalpable,
or must it always be associated with material entities?  We take it as
axiomatic here that there is no such thing as disembodied information,
information in the abstract. Information, of whatever kind, must be associated
with matter, radiation or fields of some sort.  Granted this, the questions raised
above can be faced quantitatively.\par

Some aspects of the query are easy to answer.  For example, we know that
physical structures cannot travel with speed faster than that of light in
vacuum. We infer that information cannot be conveyed from point to point with
speed faster than that of light in vacuum.  In fact, if one wishes to avoid
paradoxes in relativity theory, such as that arising from the fact that
phase velocities can sometimes exceed the speed of light, one is forced to state
the appropriate principle of relativity in the form "{\it information\/}
cannot be be propagated at speeds higher than that of light in vacuum".\par

Of the trio of questions mentioned, that concerning the limitations on the
flux of information was the first to be taken up in the wake of developments in
communication technology.  The primitive answer was the 1930 Hartley--Nyquist
law in communication theory\q{1} which states that, for a single communication
channel, the peak rate of information flow in bits s$^{-1}$ equals to the bandwidth of the channel
in Hz. In Shannon's 1948 information theory\q{2} the Hartley--Nyquist law is replaced by the
famous channel capacity formula, Eq.(5) below, in which (classical) noise in the channel
limits the rate at which information can flow through it without incurring errors.  This
formula has had incalculable influence on communication technology.\par

Not until the early sixties were quantum generalizations of Shannon's formula proposed. 
Early expressions of an approximate nature for quantum capacity were proposed by Stern,\q{3}
Gordon,\q{4,5} and Marko.\q{6} In 1963 Lebedev and Levitin\q{7}, starting from thermodynamic
considerations, obtained a precise and powerful formula for channel capacity including
the effects of both quantum and thermal noise.  In the classical limit this formula
reduces to Shannon's, while in the noiseless limit it leads back to the estimate of Stein and
Gordon [see Eq.(45) below].  Much later Pendry\q{8} independently derived the noiseless limit
of the Lebedev--Levitin formula from elegant pure thermodynamic considerations.
Gordon's approach\q{5} to the quantum channel capacity by combining Shannon's capacity
formula with the time--energy uncertainty relation recurs in papers by Bremermann's\q{9-11}
which, however, led him to a an entirely new result, a linear limit on channel capacity
[see Eq.(84) below].  Subjected to heavy criticism,\q{12-14} Bremermann's work has been vindicated
to some extent, at least isofar as the linear bound can be justified by other means (see Sec.4) 
The real significance of Bremermann's limit emerges when attention shifts from steady state
communication to burst communication.\q{15}\par

Interest in fundamental quantum limits on information storage is a later occurrence.  It first
grew out of developments in black hole thermodynamics.\q{16-18} In order
to preclude contradictions with the second law of thermodynamics in systems involving black holes
and ordinary matter, it turns out necessary to assume that the entropy of an ordinary system is
limited in terms of its mass and size.  This led one of us\q{16,19} to conjecture a quantum
upper bound on specific entropy which depends only on the maximum radius of the system in question.
Because of the connection between entropy and information, such bound is equivalent to one on the
information storage capacity of a system.  And combined with causality considerations, this last
bound leads\q{12} to a limit of the Bremermann type.  Much progress has been made in establishing
the bound on information storage capacity on the basis of statistical and quantum ideas
independent of gravitational physics.\q{13,19,20-23}\par

This paper is partly a review of the mentioned developments, and partly a report on a number of
related new results obtained lately by us in relation to quantum channel capacity and bound on
information storage.  It does not give full coverage to all questions related to these
subjects.  We have not tried to go into details of devices that might implement information
storage or communication in such a way as to approach the various bounds mentioned.  There are
now several good reviews in these areas.\q{24-26}  We have also steered clear of the subject of
bounds on information processing or computation; this large area has also been thoroughly
reviewed.\q{14,27}  In our opinion the relation of the computation process at the elementary level
to communication and information storage is not sufficiently understood to allow a review of it to
be made at the same level as is possible for the separate areas.  For these reasons our list of
references on fundamental limits on information is far from complete.

In Sec.2. we review steady state classical and quantum channel capacity theory. New is a detailed
development of Lebedev and Levitin's idea of an information theoretic derivation of the quantum
capacity for a noisy narrowband channel.  In Sec.3 we describe quantum channel capacity theory for
burst signals.  New here is an accurate analytic approximation to the maximum information possible
with occupation number signal states as a function of mean signal energy. We show explicitly that
coherent signal states do worse than occupation number states, and prove the theorem that
guarantees that occupation number states are optimal in this regard.   Sec.4 reviews the linear
bound on communication.  We give a heuristic ``proof'' of the bound, followed by rigorous proofs,
one applicable to signals with specified mean energy, and the other to signals with given energy
budget.  New here is a reformulation of the time--energy uncertainty principle in terms of the
linear bound, and an discussion of channel capacity for many parallel channels.  Sec.5 reviews the
evidence for the quantum bound on information storage capacity of quantum mechanical and quantum
field systems.  We give a new, simplified, version of the analytic proof for the bound in the
latter case.  Also new here is a discussion of solitons as information caches, and of information
storage in one dimensional systems, including storage with the help of fluctuations; the bound is
shown to be obeyed in both cases.  Some aspects of the spacetime view of information are described
in Sec.6.  New here are discussions of the influence of signal self--gravitation on communication,
and of the role of acceleration as a jammer of information transfer.          

\vglue 12pt
\line{\tenbf 2. Limits on Steady State Communication \hfil}
\vglue 5pt

The simplest situation regarding communication or information transfer is when
steady state obtains.  Physically the problem is somewhat analogous to
equilibrium thermodynamics, and indeed thermodynamics has played an important
role in the development of steady state communication theory.  In reviewing it
we shall first introduce Shannon's famous information theory, mention his
classical channel capacity formula, and pass on to the subject of steady state
quantum channel capacity.

\vglue 12pt
\line{\it 2.1. Shannon's Information Theory \hfil}
\vglue 5pt

How to quantify information?  Shannon\q{2} imagined a system capable of storing information
by virtue of its possessing many distinguishable states.  Although the
system's actual state $a$ is not known {\it a priori\/}, the probabilty for it
to occur, $p_a$, is assumed known.  Shannon sought a measure of the uncertainty
about the actual state before the system is examined.  He demanded that the
uncertainty measure, called {\it entropy\/} and represented as
$H(p_1,p_2,\thinspace\dots)$, satisfy the following requirements:

\vglue 10pt
$\bullet$ $H$ shall be a continuous function of the $p_a$}.

\vglue 3pt
$\bullet$ When there are $n$ equally probable states, $H$ should monotonically
increase with $n$.  This is reasonable since more states means more
uncertainty.

\vglue 3pt 
$\bullet$ Whenever a state $a$ in the original list is found to involve several
substates $a_1, a_2,\thinspace\dots$, the original entropy must be augmented by
$p_a\, H(p_{a_1}, p_{a_2}, \thinspace\dots)$.  This means that the multiple
state is to be treated as a system all by itself, but its entropy is weighed by
that state's {\it a priori\/} probability.\par

\vglue 3pt

Shannon found that the only function satisfying the requirements is
$$
H = - K\enspace\sum_a \pa\ln\pa,  \eqno(1)
$$
where $K$ is an arbitrary positive constant, corresponding to different
choices of the unit of entropy.  The free parameter $K$ can be
traded for a free choice of logarithm base, and, therefore, can be set to
unity.  With logarithms to base 2 the entropy is said to be expressed in
``bits'' (binary information units).  With the natural logarithm it is
expressed in ``nits'' (natural information units), \etc  The reader is referred
to Shannon's original work\q{2} for proof of Eq.(1).  In what follows, in line with the
``physicist's notation'' prevalent in this review, we shall usually compute with the natural
logarithm; however, we shall reexpress important results in bits.\par

When one of the probabilities $p_a$ is unity, the entropy vanishes:  when the
state is known precisely there is no uncertainty. For $\Omega$ possible states,
the entropy is maximal when they are all equally likely, and
$$
H_{\rm max} = \log_2 \Omega\enspace {\rm bits}.      \eqno(2)
$$
 
The basic idea of information theory is that when the system in question is
examined, and the state it is in is fully determined, the amount of information
so acquired equals $H$.  For example, in old style telegraphy each signal can
be either dot or dash with equal probability, so that by  Eq.(1) there is one
bit of entropy per symbol.  When the signal is received and fully identified,
one bit of information per symbol is made available.  It also follows that if
the state is not fully identified, the information obtained is less than the
full value of the entropy.  Technically the measurement is regarded as imposing
constraints which are expressed in terms of conditional probabilities for the
various states given the results of the examination. The mean negative
logarithm of the conditional probabilities, the {\it conditional entropy},
can be proved to be less than the original entropy,\q{2} and represents the
part of the potential information that was not revealed by examination of the
system.  We shall express all this in equations in Sec.2.6\par

Shannon's entropy formula (1) parallels Boltzmann's definition of entropy as
used in the proof of the H theorem.  (This explains the use of the name
``entropy'' and the symbol $H$ in information theory, both suggested to Shannon
by von Neumann.\q{2})  Shannon's entropy corresponds to thermodynamic entropy if
one equates $K$ with Boltzmann's constant, and uses natural logarithms. 
Boltzmann's work on statistical mechanics already contained germs of the
connection between thermodynamic entropy and information.  These were amplified
later by Szilard in his discussion of the Maxwell demon.\q{28}  After Shannon's
work, Brillouin\q{29} worked out in great detail the interrelation of
information concepts with thermodynamics.  The kinship between Shannon's
entropy and the entropy of statistical physics made it possible for Jaynes to
use information theory as an axiomatic basis for statistical mechanics.\q{30}
For a pedagogical treatment the reader is referred to the monograph by
Katz.\q{31}\par

Shannon's view of information gives up any attempt to ascribe {\it value}
to information.  One bit can stand for rather unimportant information, like
specifying the tenth digit of a binary number, or it can represent crucial life
saving information, like the bit that triggers an alarm indicating that a
patient's heart has stopped. There are alternatives to Shannon's definition
which do ascribe a semblance of value to information, for example, algorithmic
entropy.\q{32}  But since most treatments of communication and information
storage have had in mind Shannon's definition, we confine our remarks to this context.\par

\vglue 12pt
\line{\it 2.2. Shannon's Classical Channel Capacity \hfil}
\vglue 5pt

Suppose one wishes to transmit a message through some channel, \eg the
telephone.  Typically the message is first converted into a continuous function
of time, \eg the electric current $J(t)$ in the telephone line. If the
current lasts time $\tau$,  $J(t)$ is completely specified by the Fourier
coefficients
$$
b_k = {1\over \tau} \int_0 ^\tau J(t) e^{ik\omega t} dt,          \eqno(3) 
$$
where $\omega = 2\pi\tau^{-1}$.  In practice the signal is restricted to a certain
band, an angular frequency range $\dw$ which may or may not extend down to zero
frequency.  In the telephone example $\dw\sim 4\,$KHz.  The specification
of the physical nature of the signal and of the bandwidth constitute a
definition of the communication channel.  When $\dw$ is small compared to the
typical frequency in the channel, one calls it a narrowband channel.  When
$\dw$ is broader one speaks of a broadband channel.  This includes the often
discussed case of infinite $\dw$.  Evidently for a narrowband channel all the
coefficients with $k\omega > \dw$ vanish, so that all the information is
completely specified by $\tau\dw/2\pi$ complex numbers $b_k$ or
$n\equiv\tau\dw/\pi$ real numbers $a_k$.   Ascribing a probability $p(a_1,
a_2,...,a_n)$ to the set of Fourier components $\{a_1, a_2, ..., a_n\}$, one is
in position to calculate the entropy {\it flow rate\/} associated with the
signal
$$
\Dt H = - {1\over \tau}\int p(a_1,a_2,...,a_n)\,\ln
p(a_1,a_2,...,a_n) da_1da_2...da_n. \eqno(4)
$$
We have glossed here over ambiguities relating to the calculation of entropy from probability {\it
densities\/} rather than probabilities for discrete variables.\q{2}\par

The above describes an ideal situation.  In practice the transmitting channel is
affected by noise, \eg electronic shot noise.  Noise has the effect that the
received signal differs in a stochastic way from the transmitted one.  This
limits the receiver's ability to recover the encoded information.  In effect
the received signal is associated with larger entropy than the transmitted one
because the noise has introduced a further measure of uncertainty.  It is
still possible, in principle, to code the information at the transmitter in such
a way that it can all be recovered at the receiver.  One may here recall the
error correcting codes in effect in intercomputer communication links.  However,
as proved by Shannon,\q{2} elimination of errors upon reception is guaranteed
only if the information is not transmitted too fast.  In effect every physical
channel is ascribed a {\it capacity\/} which represents the maximum rate in
bits $s^{-1}$ at which information can be transmitted through it with negligible
probability of error.  We shall here denote the capacity by $\imax$
(standard communication texts usually denote it by $C$).  Whenever the
actual communication rate $\Dt I$ exceeds $\imax$, the difference $\Dt I-\imax$
represents information that will be degraded by errors due to noise.\par

In Shannon's communication theory the extra uncertainty introduced into the
received signal by processes in the channel is quantified by the conditional
entropy calculated from the conditional probabilities for various received
signal states given that a specific signal state was sent.  Evidently the
uncertainty here is fully a result of stochastic events in the channel.  This
conditional entropy is to be substracted from the total entropy of the
received signal to obtain the useful information obtainable at the receiver. 
The channel capacity is the maximum of this last quantity.\par

Whenever the noise is independent of the transmitted signal, \eg thermal noise,
the conditional entropy is just the entropy constructed from the
{\it a priori} probabilities for the diverse noise states -- the noise entropy.  Supposing the
noise to be Gaussian, and to have power $N$ uniformly distributed over the
bandwidth $\dw$ of the channel (white noise), Shannon obtained from Eq.(4)
that the noise entropy $\propto \dw\log N$.  The transmitted signal carries
power $P$, so that the received power is $P+N$.  The entropy at the
receiver is maximized when the total {\it received\/} signal is itself
Gaussian.  Again from Eq.(4) it is found that the maximum received entropy is
$\propto\dw\log (P+N)$.  Subtracting the noise entropy, Shannon obtained the
famous classical capacity formula 
$$
\imax = \left(\dw\over 2\pi\right)\,\log_2\left(P + N\over N\right)\ {\rm bits\
s}^{-1}.        \eqno(5) 
$$
It may be seen that signal-to-noise ratio $P/N$ is the parameter controlling
the classical channel capacity.  Shannon's theory applies to all signals which
may be represented  by frequency limited continuous functions of time, \ie the
theory is classical.  Shannon's capacity formula successfully describes
myriad systems (telephone, fiber optics links, space telemetry, $\thinspace\dots
$).  We shall recover it as a limit of the quantum capacity formula for noisy 
narrowband channels (Sec.2.6)\par

How may we understand Eq.(5) heuristically?  The expression $\dw/2\pi$ is the
number of phase space cells passing by a given point per unit time.  How much
information can be packed in one cell?  That depends on the noise which makes
it difficult to distinguish one signal from another very close to it in
intensity.  We can argue that in the presence of noise energy $N\tau$, one can
distinguish one level of total signal (signal plus noise) from another only if
there are no more than ${(P+N)\tau\over N\tau}$ allowed levels.  We interpret
this as the number of states available.  Maximum information is attained when
the states are equally probable, and is given by Eq.(2).  We thus get back to
Shannon's capacity formula Eq.(5).  If the noise is not
white, but still Gaussian at each frequency, one can partition the channel into
many narrow bands, use Shannon's capacity for each, and convert Eq.(5) into an
integral over $\log_2[1+P(\omega)/N(\omega)]$.\q{2}\par

Shannon's capacity formula predicts that the capacity diverges logarithmically
as the noise is reduced to zero \eg by cooling the  channel for purely thermal
noise.  This divergence will be seen to disappear in the quantum theory (see
Sec.2.5).  The Shannon \cost, $P/\imax$, can be written as
$$
\emin=N\,{2^{2\pi\imax/\dw}-1\over\imax}. \eqno(6) 
$$
For given communication rate, $\emin$ can be reduced arbitrarily by suppress
the  noise.  For thermal noise and low $\imax$, $\emin\approx kT\ln 2$  
[see Eq.(7) below], where $T$ is the absolute temperature of the channel. This
reproduces Brillouin's principle\q{29} that energy $kT\ln 2$ must be dissipated
when a bit of information is acquired in an environment at temperature $T$.\par

\vglue 12pt
\line{\it 2.3. Heuristic View of Quantum Channel Capacity \hfil}
\vglue 5pt

It is seldom realized that Shannon's classical capacity formula already
suggests the {\it form} of the quantum capacity formula !  To see this
assume the noise is thermal. Then, in the classical regime, the noise is
white and its power is given by Nyquist's formula\q{29, 33}
$$
N=kT(\dw /2\pi ). \eqno(7)
$$
This formula, which merely states that classically each phase space cell
carries mean energy $kT$, is accurate for $kT\gg\hbar\omega_0$ where  $\omega_0$
is a typical frequency of the channel. Evidently $\omega_0 > \dw /2$  (the
inequality is saturated for a bandwidth extending from zero up to some cutoff if
we take $\omega_0$ as half the cutoff frequency).  The classical regime 
obtains for  $\hbar \omega_0\lta kT$.  Putting the two inequalities in  Eq.(7)
we get 
$$ 
\dw \lta (4\pi N /\hbar)^{1/2}. \eqno(8)
$$\par

This inequality is not a physical restriction on $N$ but merely a guarantee that
the classical regime  obtains with given $T$, $\dw$ and $\omega_0$.
If the inferred $\dw$ is not necessarily small, the  calculation may nevertheless
be justified provided the signal power $P$ is  frequency independent also. Then
Shannon's capacity formula is  valid for a wide band. Substituting Eq.(8) into
Eq.(5) we get
$$
\imax\lta (P/\pi\hbar)^{1/2}\ f(P/N)\ \ {\rm bits\ s}^{-1}, \eqno(9)
$$
where $f(x)\equiv x^{-1/2}\log_2(1+x)$.  Now $f(x)$ has a maximum of 
$\approx 1.16$ at $x\approx 3.92$; therefore, we find
$$
\imax \lta 0.65 (P/\hbar)^{1/2}\ \ {\rm bits\ s}^{-1}. \eqno(10)
$$
\par 

What inequality (10) claims is that on the borderline between the classical
and quantum regimes, the channel capacity scales as $(P/\hbar)^{1/2}$.  A
complete quantum treatment is necessary to see if this behavior persists
deep in the quantum regime.  At this point we should mention the common fallacy
of substituting the quantum version of Nyquist's noise power Eq.(7) into
Shannon's formula in order to derive the quantum channel capacity.\q{5} This is
incorrect since Shannon's theory describes the signal classically, so
that it is not consistent to combine it with a quantum formula for noise power.

\vglue 12pt
\line{\it 2.4.  Quantum Capacity for a Broadband Noiseless Channel\hfil} 
\vglue 5pt

In the early 1960's Gordon,\q{5} gave two early derivations of the quantum channel capacity for a
noiseless channel. One was based on the \te, a very popular though flawed approach which confuses
the time entering into the principle with the duration of the signal.  The
second approach, already criticized in Sec.2.3., combined the classical Shannon
capacity formula with Nyquist's quantum noise formula.  Neither gave the
correct coefficient in the quantum channel capacity, Eq.(16) below, but both gave
the correct dependence $\imax\propto (P/\hbar)^{1/2}$.  Stern\q{3} and
Marko\q{6} had a similar measure of success by other approaches.  Before
describing the full thermodynamic derivation of the quantum capacity due to
Lebedev and Levitin,\q{7} which includes the effects of thermal noise, we shall
review the more recent thermodynamic derivation of Pendry\q{8} which deals
specifically with a noiseless channel. It illustrates well two important issues:
the difference between boson and fermion channels, and the insensitivity of the
channel capacity to dispersion.\par

Pendry's focuses on the channel and the carrying field, rather than on the
process of detection.  His description of signals, unlike Shannon's, is a quantum
one: each possible signal is represented by a particular quantum state of the
field, \eg a particular set of occupation numbers for the various  propagating
modes in the channel.  Pendry assumes uniformity of the channel in the
direction of propagation, which allows him to label  modes by momentum $p$.  He
allows dispersion so that a quantum of momentum $p$  has some energy $\eofp$.
Then the propagation velocity of the quanta is the group velocity
$\speed=d\eofp /dp$. 

The basic assumption is that $\imax$ can be identified (apart from units) with
the unidirectional thermodynamic entropy current that the channel carries in
a thermal state. This hails back to the idea that in a thermal state the entropy
in each mode is maximal. Of course in the thermal state there is no net flow of
entropy, but all modes moving in a definite sense along the channel do carry an
entropy current. It is assumed to be maximal.

Now the entropy $s(p) $ of a boson mode of momentum $p$ in thermal equilibrium at
temperature $T$ is\q{34} 
$$
s(p)={\ekt\over e^{\ekt }-1}-\ln \left( 1-e^{-\ekt}\right).
\eqno(11)
$$
The entropy current in one direction is thus
$$
\dot H=\int^{\infty}_0 s(p)\thinspace\speed\thinspace dp/2\pi \hbar ,\eqno(12) 
$$
where $dp/2 \pi \hbar $ is the number of modes per unit length in the interval
$dp$ which go by in one direction.  This factor, when multiplied by the group
velocity, gives the unidirectional current of  modes.\par

After an integration by parts on the second term coming from (11), we can cast
the last result into the form
$$
\dot H= {2\over kT}\int^{\infty}_0 {\eofp\over e^{\ekt} -1}\thinspace {d\eofp
\over dp} \thinspace {dp\over 2\pi\hbar }. \eqno(13)
$$
The first factor in the integrand is the mean energy per mode, so that the
integral represents the  unidirectional power $P$ in the channel: 
$$
\dot H=2P/kT . \eqno(14)
$$
The integral in Eq.(13) is evaluated by cancelling the two differentials $dp$ and
assuming the energy spectrum is single valued  and extends from 0 to $\infty$. 
Then the form of the dispersion relation  $\eofp$ does not enter, and Pendry's
result is 
$$
P = \pi (kT)^2/12 \hbar. \eqno(15)
$$
\par

The last and crucial step is to eliminate $kT$ between the expressions for
$\dot H$ and $P$.  Multiplying by $\log_2 e$ to convert thermodynamic units to
bits one has
$$
\imax=(\pi P/3\hbar)^{1/2}\,\log_2 e\ \ \ {\rm bits\ s}^{-1}, \eqno(16)
$$
which is the noiseless quantum channel capacity.  (The analogous calculation for
Fermi statistics gives a capacity smaller by a factor $\surd 2$.  Pendry\q{8}
actually quotes the same capacity as for bosons, but this is because he
considers the contributions of  both particles and holes in a solid state
communication channel).  Henceforth we refer to Eq.(16) simply as Pendry's formula; it must be
borne in mind, however, that this result appeared in the earlier work of Lebedev and Levitin\q{7},
and in approximate form in Refs.3, 5 and 6.\par

Instead of Eq.(6) of Shannon's theory we have here the \cost\ 
$$
\emin =3\hbar\pi^{-1}(\ln 2)^2\,\imax. \eqno(17)
$$
(For a fermionic channel the \cost\ is a factor $\surd 2$ larger.)   Whereas the
\cost\ in classical theory rises exponentially with $\imax$, the quantum \cost\
grows only linearly.\par

It is somewhat surprising that the channel capacity is independent, not only of
the form of the mode velocity $\speed$, but also of its scale.  Phonon channel capacity
is as large as photon channel capacity despite the difference in speeds. Why? Although
phonons convey information at lower speed, the energy of a phonon is
proportionately smaller than that of a photon in the equivalent mode.  When
the capacity is expressed in terms of the energy flux, it thus turns out to
involve the same constants.  We may also offer the trivial comment that the
capacity for massive bosons must be lower than Eq.(16) since part of the
energy is locked in rest mass, and thus the range of modes available for
information carrying is smaller than in the massless case.

\vglue 12pt
\line{\it 2.5. Broadband Channel Subject to Thermal Noise
\hfil} \vglue 5pt

Lebedev and Levitin's derivation of the quantum channel capacity,\q{6} like
Pendry's much latter one, was a thermodynamic derivation.  Unlike Pendry's
approach, this one focuses on the process of detection.  Although Lebedev and
Levitin were thinking of electromagnetic transmission, their results apply to
any single channel carrying a Bose field (one polarization and fixed wave vector
electromagnetic, fixed wave vector acoustic, \thinspace\dots), and they can
easily be extended to channels carrying fermion fields.  For mathematical
convenience the signal is regarded as periodic with very long period $\tau$.
Therefore, the angular frequencies present are $\omega_j=2\pi
j\tau^{-1}$. Again each possible signal state is regarded as represented by a
specific set of occupation numbers of the various modes $\omega_j$.  The whole
communication system is regarded as subject to thermal noise characterized by a
temperature $T_1$.\par

The detector is idealized as a collection of harmonic oscillators, one
for each $\omega_j$.  The thermal energy of the oscillators before any
signal is received follows from Planck's formula
$$
E_1=E(T_1)\equiv\sum_{j=1}^{\infty} {\hbar\omega_j\over e^{\hbar\omega_j/
kT_1} -1}.\eqno(18) 
$$
The thermodynamic entropy in the oscillators is
$$
H_1 = H(E_1)\equiv\int_0^{E_1} {dE'\over T'} =\int_0^{T_1} {1\over
T'}{dE(T')\over dT'}dT'. \eqno(19)
$$\par

If the signal carries power $P$, the energy of the oscillators is changed to
$E_1+P\tau$ upon reception of a full period of signal.  The signal arrives in a
particular (pure) quantum state, and thus brings no entropy with it, so that
the detector entropy is still $H_1$.  It is clear that this is below the
maximum entropy possible with the new energy $E_2\equiv E_1+P\tau$.  
According to Brillouin's principle,\q{29} the deficit is a measure of the
maximum information $\maxi$ that can now be contained in the detector.  This
principle is, of course, merely a variant of Shannon's information principle
stated in Sec.2.1.  Accordingly one can write
$$
\maxi=k^{-1}\,[H(E_2)-H(E_1)]\log_2 e\ \ {\rm bits}, \eqno(20)
$$
where Boltzmann's constant $k$ transforms from thermodynamic units to nits,
and $\log_2 e$ from nits to bits.  The capacity $\imax$ follows by dividing
$\maxi$ by $\tau$.\par

By now it must be clear that the maximum information transmitted corresponds
to the situation when $H(E_2)$ is maximal, \ie for a thermal state
characterized by the formal temperature $T_2$ defined by $E(T_2)=E_2$.  Keeping
in mind that the noise is also thermal, it follows from Eqs.(19)--(20) that
$$
\imax=\int_{T_1}^{T_2}{1\over kT'}{dW(T')\over
dT'} dT'\ \log_2 e \ \ {\rm bits\ s}^ {-1}. \eqno(21)
$$
Here $W(T)$ is just $E(T)/\tau$, the thermal power issuing from the channel
when it is at temperature $T$.  Substituting  $\omega_j=2\pi j\tau^{-1}$ into
Eq.(18) and passing to the continuum limit by means of the rule
$\tau^{-1}\sum_{}^{} \rightarrow \int_{}^{}{2\pi}^{-1}d\omega$ one gets
$$
W(T) = {\pi (kT)^2\over 12\hbar},           \eqno(22)
$$
which is equivalent to Pendry's result Eq.(15).\par

This result is useful in two ways.  From $E_2=E_1+P\tau$ it is evident that
$W(T_2)=W(T_1)+P$.  From Eq.(22) it now follows that
$$
T_2 = T_1\left[1 + {12\hbar P\over \pi (kT_1)^2}\right]^{1/2}.   \eqno(23)
$$
In addition it follows from substituting (22) in (21) that
$$
\imax = {\pi k\over 6\hbar}(T_2 - T_1) \log_2 e\ \ {\rm bits\ s}^{-1}.  \eqno(24) 
$$ 
Elimination of $T_2$ between (23) and (24) finally gives the Lebedev--Levitin
capacity for a noisy channel at temperature $T_1$
$$
\imax ={\pi kT_1\over 6\hbar}\left\{\left[1 + {12\hbar P\over \pi
(kT_1)^2}\right]^{1/2}-1\right\}\log_2 e\ \ {\rm bits\ s}^{-1}. \eqno(25)
$$\par

In the classical (or low signal power) limit $P\hbar/(kT_1)^2 \ll 1$, this
formula reduces to
$$
\imax = (P/kT_1)\log_2 e\ \ {\rm bits\ s}^{-1},  \eqno(26)
$$
which coincides with the low signal--to--noise limit of Shannon's capacity
formula (5) when the noise power $N$ is given in terms of $T_1$ by Nyquist's
formula (7). [Strictly speaking one has to assume a white noise spectrum in
order to compare the Shannon formula with the wideband result, Eq.(26).]  In
the quantum (or high signal power) limit, Lebedev and Levitin's formula goes
over to Pendry's Eq.(16).\par

The \cost\ $P/\imax$ computed from Eq.(25) can be cast, after some algebra, into
the convenient form
$$
\emin = (kT_1 + {3\hbar\over \pi}\imax)\ln 2.  \eqno(27)
$$  
In this formula the classical and quantum contributions are neatly additive.
The first term is Brillouin's classical \cost; the second, clearly the \cost\
arising from quantum fluctuations (some say ``quantum noise''), coincides with
Eq.(17) for Pendry's noiseless quantum channel.\par

The importance of the channel capacity formula, Eq.(25), should not be overstated.  It is an upper
bound on the channel capacity {\it only if the noise is thermal\/}.  This is because the thermal
distribution maximizes entropy rate for given power.  Thus for nonthermal, \eg Poisson, noise we
would substract a smaller number in Eq.(20), and would get a larger capacity than inferred from
Eq.(25) with $T_1$ replaced by noise power $N$ according to Eq.(22). But since
it is impossible to exceed the noiseless channel capacity Eq.(16), if we wish
to be noncommittal about the nature of the noise, we should write
$$
\Bigl({\pi P\over 3\hbar}\Bigr)^{1/2}\enskip{[1 + {P/N}]^{1/2}-1\over
(P/N)^{1/2}}\log_2 e\ \ {\rm bits\ s}^{-1}\leq \imax \leq \Bigl({\pi P\over 3\hbar}
\Bigr)^{1/2}\log_2 e\ \ {\rm bits\ s}^{-1}.   \eqno(28)
$$

\vglue 12pt
\line{\it 2.6. Narrowband Channel Subject to Noise\hfil}
\vglue 5pt

Notwithstanding the conceptual simplicity of the foregoing discussion, in 
practice communication channels are narrowband channels. In attempting to deal
with the latter, it is most instructive to treat the flow of information through
a single mode of the channel.  Because usually the separate modes are decoupled,
the result for a narrowband channel will follow from summation over modes. 
Rather than follow the thermodynamic approach of Lebedev and Levitin, we
emphasize here the information-theoretic approach that may be used to deal with
noise (this method was also alluded in Lebedev and Levitin's paper).\par

Let the {\it input\/} signal contain a mean number of quanta $\=m$.  We
associate with it a probability distribution for the number of quanta
$p_{i}(m)$.  Having negotiated the channel, the signal enters the receiver
which is modeled as an harmonic oscillator of frequency $\omega$.  Due to noise
the oscillator is initially in a mixed state characterized by the mean
occupation number $\=\ell$.  Let us parametrize the noise by the parameter
$\alpha$ defined by 
$$
\=\ell = {1\over e^\alpha -1}.         \eqno(29)
$$ 
In case the noise is thermal Planck's law gives $\alpha=\hbar\omega/kT$
where $T$ is the temperature. The oscillator's entropy may be calculated by
looking for that probability distribution $r(\ell)$ which maximizes the Shannon
entropy $H=-\sum_\ell r(\ell)\ln r(\ell)$ subject to the constraint that the
mean number of quanta be $\=\ell$.  This happens to be the exponential (thermal)
distribution
$$ 
r(\ell)=(1-e^{-\alpha}) e^{-\alpha\ell}.           \eqno(30) 
$$
The corresponding noise entropy is 
$$ 
H_{n}={\alpha\over e^{\alpha-1}}-\ln (1-e^{-\alpha}).       \eqno(31) 
$$
\par

Upon reception of the signal the mean number of quanta in the oscillator goes
up to [see Eq.(29)]
$$
\=\ell = {1\over e^\alpha - 1} + \=n_{i}.               \eqno(32)
$$
How much information is now contained in the receiver?  Since the number of
quanta $n$ in it is partly a result of noise, we cannot identify the
quantity of information with the entropy $H_{o}$ of the {\it output\/} signal
as calculated from its probability distribution $p_{o}(n)$.  Neither is the
entropy of the initial signal $H_{i}$ the correct quantity; it {\it did\/}
quantify the information that could be borne by the signal, but this information
has since been adulterated by noise.\par

The procedure for dealing with this situation was
outlined by Shannon.\q{2}  There is a {\it joint\/} probability distribution
$p_{o,i}(n,m)$ for input and output numbers of quanta which supplies a complete
statistical description of the noisy system.  From it we can compute the two 
{\it marginal\/} probability distributions, one, $p_{i}(m)$, by summing out $n$,
and a second one, $p_{o}(n)$, by summing out $m$, as well as two {\it
conditional\/} distributions.  One
$$
p_{o|i}(n|m)\equiv p_{o,i}(n,m)/p_{i}(m),            \eqno(33)
$$
stands for the probability of $n$ quanta in the detector given that $m$ were
sent.  The second,
$$ 
p_{i|o}(m|n)\equiv p_{o,i}(n,m)/p_{o}(n),            \eqno(34)
$$
gives the probability that $m$ quanta were sent given that the detector
contains $n$.\par

There is an entropy for each of these distributions.  The
generic definition is 
$$
H_a \equiv \sum_{n,m} p_{o,i}(n,m)\, \log p_a({\rm indexes\ relevant\
to\ } a),           \eqno(35)
$$
where $a$ can stand for $i$, $o$, $(o,i)$, $(i|o)$ or $(o|i)$.
The following identities\q{2} are easily verified:
$$
H_{o,i} = H_{i} + H_{o|i} = H_{o} + H_{i|o}.        \eqno(36)
$$
Shannon noted that $H_{i|o}$, the conditional entropy of the input {\it when
the output is known\/}, must represent the extra uncertainty introduced by the
noise which hinders reconstruction of the initial signal even when the
output is known.  He thus interpreted  $H_{i} - H_{i|o}$ to be
the useful information $I$ that can be recovered from the output signal (by
means of appropriate coding and decoding) in the face of noise.  Another way
to understand this is to rewrite this definition with help of Eq.(36) as
$$
I = H_{o}-H_{o|i}.                  \eqno(37)
$$
We can think of $H_{o|i}$, the uncertainty in the output {\it for given
input\/}, as the effect of the noise.  Therefore, it is to be subtracted from
the full entropy of the output $H_{o}$ to get the uncertainty asociated with the
signal itself.\par

Now in the case being considered, the noise is independent of the signal and
described by distribution (30).  Therefore, $p_{o,i}(n,m)=p_{i}(m)\,r(n-m)$. 
It follows from (33) that $p_{o|i}(n|m)=r(n-m)$ so that (37) gives
$$
I = H_{o} - \sum_{n,m}\,p_{i}(m)\,r(n-m) \log r(n-m).   \eqno(38)
$$
The sum over $n\geq m$ for fixed $m$ gives just $H_{n}$, the {\it noise\/}
entropy.  For thermal noise it is given by Eq.(31).  Summation over $m$ just
multiplies by the normalization factor 1.  Thus
$$
I = H_{o} - H_{n}.                   \eqno(39)
$$
It should be clear from the foregoing that Brillouin's principle is only valid
in the case that signal and noise are statistically independent.  For
example, if the "noise" were due to stimulated emission which is influenced by
the incoming signal, Eq.(39) would not apply. \par

We must still maximize $I$ over the distribution $p_{o}(n)$ subject to the mean
number of quanta $\=\ell$ given by Eq.(32).  In analogy with the
discussion leading to Eq.(30) we find that $H_{o}$ is maximized for an
exponential distribution like (30) but with a parameter $\beta$ determined by
$$
{1\over e^\beta -1} = {1\over e^\alpha - 1} + \=n_{i}.        \eqno(40)
$$
The maximal entropy is the analog of (31); therefore,
$$
\maxi = {\beta\over e^\beta-1}-\ln (1-e^{-\beta}) - H_{n}.      \eqno(41)
$$
We know that for thermal noise $H_{n}$ takes on its maximal
value, Eq.(31), for given mean number of noise quanta $\=\ell$. This means that
$\maxi$ is actually smaller than than for any other kind of noise with the same
$\=\ell$.   Thus
$${\beta\over e^\beta-1}-\ln (1-e^{-\beta}) -{\alpha\over e^\alpha-1}+
\ln (1-e^{-\alpha})\leq\maxi\leq{\beta\over e^\beta-1}-\ln (1-e^{-\beta}),  
\eqno(42)
$$
which is the one--mode analog of (28).\par

Recall that this is the information per mode.  Now if the channel in question has
bandwidth $\dw$, a total of $\dw/2\pi$ modes reach the receiver per unit time. 
Also, we may define the differential power $P_\omega$ as the energy per unit
time per unit circular frequency.  Clearly since each quantum carries energy
$\hbar\omega$, $\=n_{i} = 2\pi P_\omega/\hbar\omega$.  Making these
substitutions in (31), (40)--(41) we get for a narrowband channel with
thermal noise 
$$
\imax={\dw\over 2\pi}
\Biggl\{\ln\Bigl[1+\po\bigl(1-e^{-\ha}\bigr)\Bigr]
+\Bigl(\po+{1\over e^{\ha} -1}\Bigr)\times
$$
$$
\ln\Bigl[1+{e^{\ha} -1\over \po\bigl(e^{\ha} -1\bigr)
+1}\Bigr]-{\ha\over e^{\ha} -1}\Biggr\} \log_2 e\ \ {\rm bits\ s}^{-1}.   
\eqno(43)
$$
A formula of this form was first given by Gordon\q{4}, and was later rederived
by Lebedev and Levitin\q{7} by the thermodynamic method reviewed in Sec.2.4.\par

The classical limit ($\hbar\omega\ll kT$) of Eq.(43) is
$$
\imax\approx{\dw\over 2\pi}\log_2\Bigl(1+{2\pi P_\omega\over kT}\Bigr)\ {\rm
bits\ s}^{-1},   \eqno(44)
$$
which coincides with Shannon's capacity formula Eq.(5) when one uses the
Nyquist formula for thermal noise Eq.(7).  However, the Shannon formula for
arbitrary noise cannot be gotten from (41) in any simple way.\par

In the extreme quantum limit ($kT\ll\hbar\omega$) (or in the noiseless case) we
get 
$$
\imax\approx{\dw\over2\pi}\Big[\log_2\big(1+\po\big)+\po\log_2\big(1+{\hbar\omega\over
2\pi P_\omega}\big)\Big]\ \ {\rm bits\ s}^{-1},           \eqno(45)
$$
a formula previously given by Stern\q{3}, Gordon\q{4}, Lebedev and
Levitin\q{7}, Yamamoto and Haus,\q{24} and Takahashi\q{25} among
others.  The two terms in brackets have interesting interpretations.\q{24} The
first dominates at large power or for occupation number large compared to
unity.  It tells us that the information delivered per mode is the logarithm of
the mean occupation number plus one.  We may call this the wave contribution
because it dominates whenever the signal can be treated as a wave.  The second
term dominates when the occupation number is small compared to unity.  Since
$\dw P_\omega/\hbar\omega$ is just the rate at which {\it quanta\/} arrive,
it attributes to each photon information equal to the logarithm of one plus the
number of {\it modes per photon\/}.  Plainly the corpuscular aspect of
the signal is manifested here.\par

Yamamoto and Haus\q{24} have discussed the information per quantum and the
\cost\ for the narrowband channel in various limits.  For general
$\hbar\omega/kT$ in the low signal power case, \hbox{$2\pi
P_\omega\ll\hbar\omega$}, one has \hbox{$\emin\approx kT \ln 2$}, which
coincides with Brillouin's term in Eq.(27) for the broadband channel.  In the
noiseless channel case the \cost\ diverges as $P_\omega\rarrow 0$ like
\hbox{$-\log(P_\omega)$}.  We shall have more to say about this in Sec.3.4.

We still have to settle one question.  In order to reach the peak
communication rate, how should one code the signal?  The mathematical question is what should be
the adopted probability distribution $p_{i}(m)$ for the signal?  It can be found with help of
the following theorem.\par

{\tenbf Theorem 1.} When an integer--valued variable $\ell$ with
exponential distribution $p(\ell)=(1-e^{-\alpha})e^{-\alpha \ell}$ is added to an
independent integer--valued variable $m$ with distribution $Q(m)$, and there
results a variable with exponential distribution with parameter $\beta$,
the distribution $Q(m)$ must be a {\it modified\/} exponential one:
$$
Q(m) = {1-e^{-\beta}\over 1-e^{-\alpha}}e^{-\beta m}\times\cases{1\ \ &
if $m=0$;\cr 1-e^{\beta -\alpha}\ \  &if $m\geq 1$.\cr}       \eqno(46)
$$
\par  

{\tenbf Proof.} The proof is given in Appendix A.\par

We now identify the noise in the receiver with the exponentially distributed
variable $\ell$, and $p_{o}(n)$ with the exponential distribution with parameter
$\beta$.  The theorem tells us that $p_{i}(m)$ must be identified with $Q(m)$ of
Eq.(46): the signal distribution must be chosen as modified exponential with the
parameters $\alpha$ and $\beta$ defined by Eqs.(29) and (40).

Thus far our discussion has been based on occupation number states as signal
states.  But, of course, there are other choices, \eg coherent states, in-phase
squeezed states, photon number squeezed states,\thinspace\dots\quad  As shown in Sec.3.5.
the maximum communication rate is lower when coherent states are used. 
Yamamoto and Haus\q{24} and Saleh and Teich\q{26} have analyzed the implementation of quite a
variety of quantum states in communication by means of quantum optical techniques. They
find that the maximal communication rate does depend on the type of
state as well as the type of measurement performed by the receiver, but
conclude that the capacity (45) cannot be exceeded. This feeling can be formalized; a general
theorem to this effect is proved in Sec.3.6.\par

\vglue 12pt
\line{\tenbf 3. Limits on Burst Communication Through Noiseless Channels \hfil}
\vglue 5pt

{\it A priori\/} there is no guarantee that the previous results Eqs.(16),
(25) and (43) apply to signals of finite duration.  This is because all of them
can be obtained by thermodynamic arguments, and thermodynamics is usually
applicable only in equilibrium.  This suggests that the mentioned capacities
are, strictly speaking, valid only for steady state communication, namely
communication using signals of very long duration where the information and
energy flow can be construed as in steady state.  So we may ask, what is the
capacity for burst communication?\par

Already Shannon\q{2} worried about departures from the simple capacity formula
(5) when the power is not steady, and worked out some bounds on capacity
expressed in terms of mean or instantaneous power. Interest in the quantum
capacity for nonsteady state communication developed rather late.  We have
already mentioned Bremermann's heuristic formula\q{9-11} [see (84) below] which
purports to bound the capacity in terms of the {\it energy\/} available to the
signal.  Bremermann's arguments, and Bekenstein's much later one,\q{12} which
gave a similar formula, were based on specific models.  Before getting into
all that it is useful, following Ref.15, to use general arguments to write down
a bound on communication via a single channel when steady state does not
hold.\par
 
\vglue 12pt
\line{\it 3.1. General Form of Bound on Burst Communication \hfil}
\vglue 5pt

Guided by the results reviewed in Sec.2, we take the view that the only specific
signal parameters are duration $\tau$ and energy $E$. The rest, \eg\
polarization of electromagnetic signals, wave vector direction, \etc , are
descriptive of the channel.  Thus different polarizations, quanta  species,
\etc are to be associated with separate channels: unpolarized light, even if
monochromatic and perfectly collimated, is regarded as propagating through two
channels, say left and right circularly polarized. And an hypothetical
communication system involving monochromatic collimated beams of neutrinos
will  entail one channel for each neutrino species (flavor). This precaution is
useful in removing energy degeneracies in the subsequent treatment.\par

How is the maximum  information $\maxi$ a signal may bear related to $E$ and
$\tau$?  Since information is dimensionless, $\maxi$ must be a function
of dimensionless combinations of $E$, $\tau$, channel parameters and
fundamental  constants. We exclude channels which transmit massive
quanta, \eg  electrons, because rest mass is energy in a form not useful for
communication, so  that the strictest limits on capacity and the \cost\ are
expected for massless signal carriers. Hence Compton lengths do not enter into
the argument. Also in order to maximize the information flux we focus on
broadband channels, and exclude any frequency cutoff and its  associated
length. If we also exclude the gravitational constant from the argument on the grounds that
gravity can only bring about small effects (see Sec.6.2. for a deeper argument) there is a single
dimensionless combination of the parameters that can enter: $\xi=E\tau/\hbar$. It follows that
$$
\maxi=\Im (E\tau/\hbar), \eqno(47) 
$$ 
where $\Im(\xi)$ is some nonnegative valued function characteristic of the 
channel.  We call it the {\it characteristic information function\/} or CIF.\par

The reader may find it surprising that the ratio $c_s/c$, where $c_s$ is the 
propagation speed of signals, \eg the speed of sound, was not considered in our 
argument. Obviously the ratio, if different from unity, is a property of the 
channel, not of individual signals. Therefore, it is regarded as determining 
the {\it form\/} of the one--argument function $\Im(\xi)$. It will soon become
clear that in many cases $c_s/c$ does not appear at all in the CIF.  In fact
Pendry's argument reviewed in Sec.2.4. makes it clear that signal speed
becomes irrelevant in the limit of long signal duration or steady state.\par 

Let us check formula (47). Consider steady state  communication. Because of the
statistically stationary character of the signal,  it should be possible to infer
the peak communication rate by considering  only a finite section of the signal
bearing information  $\maxi$ and energy $E$. It should matter little how long a
stretch in $\tau$ is  used so long as it is not short. This can only be true if
$\imax\equiv\maxi\tau\recip1$ is fully determined by  the power $P\equiv E\tau\recip1$.
This is consistent with  $\maxi=\Im(\xi)$ only if $\Im(\xi)= B\surd \xi$ where
$B$ is a constant; only then does $\tau$ cancel out. It follows that 
$\imax= B(P/\hbar)^{1/2}$ which is precisely the Pendry formula (16). The 
argument is, however, too general to say anything about the value of 
$B$ which depends sensitively on the channel's parameters.\par 

The dividing line between steady state communication, and communication by 
means of very long signals is not sharp. This suggests that long signals 
must also obey a Pendry type formula, albeit approximately. Indeed, long ago 
Marko\q{6} proposed that $\maxi\propto(E\tau/\hbar)^{1/2}$ for long duration 
signals. As we shall see in Sec.3.4., for $\xi=E\tau/\hbar\gta 100$, 
$\Im(\xi)\rarrow B\surd\xi$. For $E\tau/\hbar\lta 100$ signal end effects 
become significant, and $\Im(\xi)$ departs from the form $\surd\xi$.

\vglue 12pt
\line{\it 3.2. Signals With Specified Mean Energy \hfil}
\vglue 5pt

The energy $E$ which enters in Eq.(47) is subject to various interpretations. 
Is it the precise energy of the signal, the mean energy (mean with respect to
a probability distribution), or the maximum available energy?  In this
and the following subsections we consider the implications of specifying the signal by its mean
energy $\=E$.  The case when $E$ is the maximum available signal energy is
the subject of Sec.4.\par

In order for a signal to be able to carry information, there must be
various possible signal states.  Each state $a$ has its own well defined
energy $E_a$ and is assigned an {\it a priori\/} probability $p_a$ satisfying
$\sum_{a}p_a=1$.  The mean energy is defined by 
$$
\=E=\sum_{a}{p_a\,E_a}.         \eqno(48)
$$
What is the capacity for signals with specified $\=E$?  Evidently we are called
to maximize the Shannon entropy Eq.(1) subject to the normalization constraint,
Eq.(48), and any other constraints deriving from the nature of the problem.  If
there is noise, one must deal with it along the lines reviewed in Sec.2.6. 
For simplicity we focus here on  noiseless channels.  Are there any other
relevant constraints, for example, those imposed by the nature of the
reception?\par

Clearly the formal distribution ${p_a}$ is physically relevant if all states $a$ can be
{\it detected and distinguished\/} by the receiver.  If several states can be
confused, the distribution should assign them equal probabilities, and this
should be taken into account in the maximization process.   Here we shall be
concerned with the more profound question of whether the vacuum signal state, 
\eg the "no photon" signal state in an electromagnetic channel, can be used to
signal. Can the receiver distinguish between a situation where a signal has arrived with all the
relevant modes in the vacuum (or ground) state and one in which no signal was received?  Only if
the answer is affirmative is it appropriate to assign nonvanishing probability to the vacuum
signal state.  The question comes up because there are situations where even if undetectable,
the vacuum can still be used for signaling.\par

For example, in a man--made channel transmitting a train of signals at equally
spaced intervals, the absence of energy in a particular time interval (not
the first or last) implies that {\it that} signal is in the vacuum state. 
The embedding  of a signal in a series is not the only way to use the vacuum.
Let two friends A and B agree that if A passes his exam, he will phone B
between 2 and 3 p.m.  If B's phone fails to ring in that period, he
has acquired a bit of information (A has failed), by receiving the vacuum state
of the signal. If in a scattering experiment at an accelerator no relevant
events are detected, information is obtained (upper bound on the cross section)
by means of the vacuum state of the signal. What is common to these examples is
that the signal is anticipated by virtue of its being part  of a structure
(series), by prior agreement (phone, if you pass), or by causality
considerations (no scattering expected, unless accelerator has been is on). A
signal of this sort is aptly termed a {\it heralded signal}. For heralded
signals the vacuum state, even if not directly detectable, can be put to use
in signaling just as any other state.\par

Consider now a signal whose arrival time is unanticipated. The observation of
the recent supernova outburst in the Large Magellanic Cloud, or the beta decay
of a particular {\it radioactive\/} nucleus provide two examples of events that could
not have been foreseen by the observer. Here the vacuum state cannot be
inferred by elimination since the receiver does not know when to expect it, and
so cannot carry  out measurements, \eg counting photons with a photomultiplier or
measuring the momentum of a beta particle.  Hence such a signal, if detected, is
always received in a nonvacuum state: the signal {\it heralds itself}. We shall
call such signals {\it self--heralding}.  We see that steady  state
communication is based on heralding signals since a sudden absence of a signal
during a very long transmission can be used to convey information. It is also
clear that burst communication at unanticipated times must be based on
self--heralding signals.\par

\vglue 12pt
\line{\it 3.3. Generic Properties of the CIF \hfil}
\vglue 5pt 

When we consider \sh\ signals, the vacuum signal state must be excluded from the
list of signal states. Formally this means  $p_{{\rm vac}}\equiv \Pr(E_a=0)=0$.
Let us maximize the Shannon entropy (1) over the nonvacuum states subject to
normalization of probability and the condition (48). The result can be written
in  a form applicable to both types of signals:
$$
\pa=C\times\cases{e^{-\mu E_a}&$E_a\neq 0$;\cr 1-\zeta &$E_a=0$,\cr} \eqno(49)
$$
where $\zeta=0$ for heralded signals, and $\zeta=1$ for \sh\ ones. Although 
other values of $\zeta$ seem to have no physical relevance, all the 
calculations to follow are unified if we keep $\zeta$ general.\par

Let us define the ``partition function'':
$$
Z\equiv\sum_a e^{-\mu E_a}, \eqno(50)
$$
where $\mu$ is a parameter analogous to inverse temperature in statistical mechanics.  The
normalization constant is now given by
$$
C=(Z-\zeta)^{-1}.           \eqno(51)
$$
When $\zeta\not= 0$ this is a bit different from the statistical mechanics result. As in
statistical mechanics, the expression for the mean energy can here be cast into the form
$$
\=E=\partial\ln C/\partial\mu, \eqno(52)
$$
which determines $\mu$ in terms of the prescribed $\=E$.\par

The calculation of $\maxi$ from the Shannon entropy corresponding to
distribution (49) gives
$$
\maxi=\mu\=E \log_2 e -\log_2 C-C(1-\zeta)\log_2 (1-\zeta)\ \ \ {\rm bits}. \eqno(53)
$$
Formally the last term vanishes for both $\zeta=1$ and $\zeta=0$. Of 
course, this does not mean that \sh\ and heralded signals bear identical 
information because, for given $\=E$, the two will have different $\mu$'s 
[see(52)].  Eqs.(52)--(53) give, in parametric form, $\maxi(\=E)$, and thus determine the {\it
form\/} of the CIF.\par 

Several properties of the CIF follow immediately. For example, differentiating
(53) with respect to  $\=E$ and using (52) we get for $\zeta=0,1$ that
$\partial\maxi/ \partial \=E=\mu$. Since $\mu$ must be positive (otherwise $Z$
would diverge and $C$ vanish), we find that $\Im(\xi)$ is always an
increasing  function ($\tau$ is a fixed parameter in the present exercise).
This conclusion also holds formally for $0<\zeta<1$.\par 

A look at (50)--(51) shows that in the limit of small $\mu$ (large $\=E$ or
$\xi$), the partition function overwhelms $\zeta$. Thus at large argument the
CIF's for heralded and \sh\ signals must merge. It will become clear that they
go over into the CIF associated with the Pendry formula,
\hbox{$\Im(\xi)\propto\surd\xi$} (see Sec.3.4.).

Taking the second derivative of Eq.(53), and observing that necessarily
$\partial \=E/\partial\mu<0$ by the analogy between $\mu$ and inverse
temperature, we discover that the CIF is always a  convex function of its
argument.  Again, this conclusion is formally valid  for $0\leq\zeta\leq
1$. Note that the CIF for infinitely long signals, $\Im(\xi) \propto\surd\xi$,
has this property. An immediate consequence of convexity  is that a signal of
mean energy $N\=E$ and duration $N'\tau$ carries less information than
$NN'$ signals of energy $E$ and duration $\tau$.\par 

To say more about the CIF we now interpret states like $a$ as (pure) quantum field
states, and denote them by $\ket{a}, \ket{b},\thinspace\dots\ $ with 
\i{a priori} probabilities $\pa, p_b,\thinspace\dots$  This is the full statistical
description of our problem.   Note the difference between this situation and the usual scenario in
statistical mechnaics.  There one uses the full density operator, and for consistency the
von Neumann quantum formula for  entropy.\q{34} Here we only use the density operator's
diagonal terms, namely the probabilities $\{p_a, p_b,\thinspace\dots\}\ $.  The off--diagonal parts
describe correlations which are foreign to the business at hand.  Were we to include them in the
description, we would get contributions to the information of  signals  which could not be
ferreted out by a receiver whose job is to distinguish one  pure state from another.\par

To make things simpler let us assume the signaling field is a free field.
If it is subject to interactions (arguably it must be if communication
is to be possible), we assume that our choice of \i{propagating} normal
modes manages to eliminate any cross interaction  terms, \eg normal modes
in an elastic solid. The field hamiltonian thus corresponds to a
collection of noninteracting harmonic oscillators, one for each field mode. 
Depending on what it takes to do this, the quanta will be free particles,  \eg
photons, or quasiparticles, \eg phonons.

Consider a single mode $j$. To it corresponds a harmonic oscillator 
hamiltonian $H_j$ with a certain frequency $\omega_j$. One type of states of 
mode $j$ are the occupation number states $\ket{j\alpha}$ defined by 
$H_j\ket{j\alpha}=n_\alpha\hbar\omega_j\ket{j\alpha}$ where $n_\alpha$ is a 
nonnegative integer. Other choices like coherent and squeezed states\q{24} 
are not eigenstates of the mode hamiltonian. However, any such state 
$\ket{j\beta}$ does have a well defined mean energy $\ej$:
$$
\ej = \bra{j\beta}H_j\ket{j\beta}. \eqno(54)
$$
We can now build the signal states $\ket{a}$ by exploiting the independence of 
the $H_j$, namely,
$$
\ket{a}=\ket{i\alpha}\otimes\,\ket{j\beta}\otimes\,\ket{k\gamma}\,\cdots \eqno(55)
$$
where $i, j, k,\,\ldots$ label modes, $\alpha, \beta, \gamma\,\ldots$ 
label one--mode states, and $a, b,\,\ldots\ $ label signal (many--mode) 
states.

The probabilities $\pa$ of the signal states are assumed to be normalized to 
unity, but it is unnecessary for the signal states to form a complete 
set in the sense of quantum theory. However, completeness obviously favors
higher communication rates by making a maximum number of states available, and
will  be assumed henceforth. We start by defining the {\it mean\/} energy of the 
signal:
$$
\=E = \sum_a\pa\,(\energy).  \eqno(56)
$$
Two averages are involved here: a quantum expectation value over the 
one--mode states which yields $\energy$, and a statistical average over the 
{\it a priori} probabilities for the signal states, $\pa$. Clearly, only the 
latter are involved in the calculations of $\maxi$. Thus from  our point of view
the expression $\energy$, though formally a  quantum expectation value, can be
treated as a definite energy $E_a$.

Turn now to the partition function, $Z$. The sum over $a$ is equivalent to one 
over all combinations of $j$ and $\alpha$. Thus in a manner analogous to well 
known thermodynamic calculations, $Z$ can be written as $\prod_j\,Z_j$ where
$$
Z_j(\mu)\equiv \sum_\beta e^{-\mu\ej}
=\sum_\beta\,\exp\big(-\mu\bra{j\beta}H_j\ket{j\beta}\big). \eqno(57)
$$
Contrary to naive expectations, the sum in Eq.(57) is not invariant under a
unitary transformation of the  $\ket{j\beta}$ because the exponentiation
process precedes the trace.  This means that the channel capacity may vary
with the type of quantum states $\ket{j\beta}$ used.  In the next two sections we study
communication via occupation number states.  They are contrasted with coherent states in Sec.3.5. 
Sec.3.6. presents a theorem showing that occupation number states are indeed optimal ones (maximum
communication rate for given energy), but are not unique in this respect.

\vglue 12pt
\line{\it 3.4. Occupation Number States\hfil} 
\vglue 5pt

Occupation number states are relevant, for example, for an optical fiber communication channel with a
photoelectric tube equipped with photon counting electronics as a detector. Our full
attention will here be given to the propagation of signals and we shall ignore questions involved
in the reception.  These last have been treated by Yamamoto and Haus\q{24} and Saleh and
Teich.\q{26}\par 

If the states $\ket{j\beta}$ are chosen as occupation number states, 
$\bra{j\beta}H_j\ket{j\beta}=n_\beta\hbar\omega_j$. For a bosonic field 
$n_\beta$ can be any nonegative integer.  Thus for bosons $Z_j$ reduces to the
partition function of a harmonic oscillator at temperature $\mu\recip1$:
$$
Z_j=\sum_{n=0}^\infty e^{-\mu n\hbar\omega_j}
=\left[1-e^{-\mu\hbar\omega_j}\right]^{-1}. \eqno(58)
$$
The calculation for fermions is quite similar.\q{15}  To calculate $Z$ we first
sum $\ln Z_j$ over modes, and then exponentiate  the result.  In general the sum
has to be done numerically.  However, in the small $\mu$ limit we can perform the sum
analytically in the continuum approximation.\par
 
For small $\mu$ the exponent in Eq.(58) changes gradually with $\omega_j$ so
that we may replace the sum over $Z_j$ by an integral according to the usual rule
\hbox{$\sum_j{ }\rarrow \tau\int{d\omega/2\pi}$}.  The integral is a familiar
one from the statistical mechanics of a one--dimensional Bose gas, and the
final result is 
$$
Z(\mu)=\exp\left({\pi\tau\over 12\mu\hbar}\right). \eqno(59)
$$
Our brief derivation here glosses over the question of dispersion (signal speed
depending on frequency).  It can be shown\q{15} that all effects of dispersion
cancel out if the various modes are properly sequenced.\par

Clearly for small $\mu$ (more precisely small $\mu\hbar\tau^{-1}$), $Z\gg\zeta$ so
that $C\approx Z\recip1$ for \sh\ signals. For heralded  signals this is, of
course, an exact result. Eq.(52) now gives
$$
E=\pi\tau/12\hbar\mu^2.           \eqno(60)
$$
Calculating $\maxi$ from Eq.(53) and eliminating $\mu$ between the results
gives in the continuum limit
$$
\maxi\rarrow(\pi E\tau/3\hbar)^{1/2} \log_2 e\ \ \ {\rm bits}.  \eqno(61)
$$
Apart from the numerical constant this is just Marko's\q{6} expression for 
$\maxi$. It reduces to the Pendry formula (16) under the substitutions 
$E/\tau\rarrow P$ and $\maxi/\tau\rarrow\imax$.   Thus as anticipated, for large $\xi$\ \
$\Im(\xi)\rarrow {\rm const.}\surd\xi$ and the difference between heralded and  \sh\ signals
disappears. Comparing Eq.(51) and (59) we see that the differences between heralded and \sh\ cases
disappear when $\tau/\mu\hbar\gg 10$. From (60) we see that 
the merging should be apparent for $\xi\gta 10^2$, which can also 
be taken as the criterion for approach to the limit (61). Thus the long duration
signals for which the results of Sec.2.4. apply are those with $E\tau/\hbar\gta
10^2$.\par

For $\xi < 10^2$ the continuum approximation is inappropriate and we must
go into some detail regarding the form of the spectrum $\{\omega_j\}$.  The
burst signal as seen from a fixed point may be represented by some function
$F(t)$ which has compact support in time \ie it is nonvanishing only in the
interval $[0,\tau]$.  In fact, it is mathematically convenient to regard $F$ as
periodic with period $\tau$.  This ``periodic boundary condition'', well
known from quantum physics, captures the essence of the finiteness of the
duration while keeping the mathematics simple.  Resolve $F(t)$ into its
Fourier components involving the angular frequencies ${2\pi j\tau^{-1}}$ for all
positive integers $j$ (negative integers are superfluous -- recall that under
second quantization of a Bose field negative frequencies just duplicate the
modes). The $j = 0$ (dc) mode may be be ignored; it can be argued (see Sec.5.4.) that it
corresponds to a condensate of the field to which no entropy (information) can be ascribed.  So the
spectrum is \hbox{$\omega_j = {2\pi j\tau^{-1}}$} with \hbox{$j =1,2 \thinspace\dots$}, and with no
degeneracies.\par

Using (58) we now write the partition function (50) as 
$$
\ln Z = - \sum_{j=1}^\infty\ln (1- e^{-b j})             \eqno(62)
$$
and the mean energy (52) as
$$
\=E={2\pi\hbar\over
\tau}\bigg({Z\over Z-\zeta}\bigg)\sum_{j=1}^\infty{j\over e^{b j}-1},
\eqno(63)
$$
where $b\equiv {2\pi\mu\hbar\tau^{-1}}$.  The parameter $b$ is to be chosen
so that the desired value of $\=E\tau/\hbar$ is reproduced by Eq.(63).\par

The continuum approximation (long duration signal) is accurate in the limit
$b\rarrow 0$.  To deal with the case when $b$ is not small (brief
signal), we carry out in Appendix B the sums in Eqs.(62)--(63) by means of the
Euler-Maclaurin summation formula to obtain an approximation that transcends
the validity of the continuum approximation. The results are
$$
\=E \approx{2\pi\hbar\over \tau}\left({Z-\zeta\over Z}\right) \left[{\pi^2\over
6b^2} -{1\over2b} + {1\over 24}\right] \eqno(64)
$$ 
and
$$
\ln Z \approx  {\pi^2\over 6b} + {1\over 2}\ln b - {b\over 24} -
0.91894.\eqno(65) 
$$
We have checked numerically that Eqs.(64)--(65) are very accurate
representations of (62)--(63) for $b < 1$, and even at $b\sim 4$ their
accuracy is better than 1\%; however, the accuracy deteriorates rapidly
for larger $b$.  At any rate, expression (64) for $\=E$ is strictly
positive as it should be.\par

\vglue 10.0truecm
\baselineskip=10pt
\centerline{\eightrm Fig.1. The characteristic information function for occupation number
states in the\hfil}
\centerline{\eightrm periodic boundary condition approximation as calculated from
Eqs.(50)--(53).\hfil}
\baselineskip=13pt
\vskip 0.5truecm

The leading term in (64) and (65), which dominates for small $b$,
corresponds precisely to Eqs.(59)--(61).  It thus reproduces the results of the
continuum approximation, and gives back the Pendry formula (16). 
In the general case Eqs.(64)--(65) together with Eqs.(51)--(53) give the CIF in
parametric form.  How does it look when $b$ is not small? 
Considering $\tau$ as fixed, let us first look at the case of heralded signals
($\zeta = 0$).  This simplifies Eqs.(51), (53) and (63) considerably so that we
get $$
\maxi = \left({\pi^2\over 3b} + {1\over 2}\ln b - 1.41894\right)
\log_2e\ \ {\rm bits}.       \eqno(66)    
$$
Solving Eq.(64) for $b$ in terms of $\xi=E\tau/\hbar$ and susbtituting in (66) we
get the form of the CIF:
$$
\Im(\xi) = (R\log_2e -{1\over 2}\log_2 R - 1.18808)\ \ {\rm bits},    
\eqno(67) 
$$
with
$$
 R\equiv 1+\left(1-{\pi^2\over 36} + {\pi\xi\over 3}\right)^{1/2}.  
\eqno(68)
$$
Since Eqs.(64)--(65) are accurate for $b < 4$, we can use Eq.(67) for $\xi=\=E\tau/\hbar > 0.12$. 
As expected, $\Im\rarrow\surd\xi$ for large $\xi$ in which limit we recover Pendry's formula.

For \sh\ signals the factor $(Z-1)/Z$ in Eq.(64) cannot be ignored unless $Z$
is so large that the continuum approximation is acceptable.  Hence, to get a
full picture of the CIF in this case, as well as for heralded signals with
small $\xi$, it is best to go back to the numerical evaluation of Eqs.(62)--(63). In Fig.1 we plot
$\maxi$ vs. $\=E\tau/\hbar$ on a log--log scale. The dotted line is the CIF in the periodic
boundary condition approximation for \sh\ signals while the dashed line refers to heralded
signals.  The solid line is the limiting formula (61) which is seen to be a strict upper bound on
$\maxi$, and an excelent approximation to the CIF for signals with $\xi > 10^3$
(these correspond to $\maxi > 50$ bits). As $\xi$ decreases, $\maxi$ of
finite duration signals falls below the naive prediction of (61) by a factor
which reaches 2.5 for self-heralding signals with $\xi = 10$.  The corresponding
true $\maxi$ is $\approx 2$ bits. Thus signals carrying modest information must
be treated as finite duration signals, rather than by the steady state
communication capacity.\par

Fig.2 displays the energy cost per bit $\emin$ as a function of $\maxi$
in the self-heralded (dotted line) and heralded (dashed line) cases. The
solid line corresponds once again to the limiting formula (61).  Clearly for
finite duration signals, the energy cost per bit exceeds that implied by the
theory of steady--state communication. It may be seen that for self-heralding
signals there exists a lower bound on the \cost\ of $\emin\approx 4.39\hbar\tau^{-1}$ 
which is attained for $\maxi\approx 3.5$ bits. No such bound exists for {\it
heralded} signals: the energy cost per bit can be low for heralded signals with
only fractions of a bit. Such low information signals {\it are} meaningful. 
For example, if a question has three alternative answers with the first being
98\% probable, then 0.3 bits suffice to single out the answer [see Eq.(1)].

\vglue 10.0truecm
\baselineskip=10pt
\centerline{\eightrm Fig.2. The energy cost per bit calculated under the conditions of Fig.1.}
\baselineskip=13pt
\vskip0.5truecm

The periodic boundary condition assumes the modes in the signal have sharp
frequencies.  In truth if the signal has finite duration they should contain a
continuum of frequencies.  In Ref.15 Gabor's method of time-frequency
cells\q{35} has been used to justify the results obtained with the periodic
boundary condition.

\vglue 12pt
\line{\it 3.5. Coherent States\hfil} 
\vglue 5pt

Coherent states can also be used for communication.  In fact, in some sense they were the first to
be used: a radio trasmitter produces an approximation to a coherent state of the electromagnetic
field. In quantum optics the use of coherent and the closely related squeezed states in communication
has been the subject of much interest.\q{24-26}  Let us investigate the maximum information that 
may be coded with coherent states.\q{24}  To avoid certain technical problems we concentrate on
heralded signals ($\zeta=0$).

A coherent state\q{36} $\ket a$ is defined as the tensor product {\it a la} Eq.(55) of eigenstates
of the anhilation operators $\hat a_j$ of all the $N$ modes involved; thus 
$$
\hat a_j \ket a = \alpha_j\ket a; \qquad j=1, 2,\dots N. \eqno(69)
$$
As is well known, a coherent state can be expanded in occupation number states.\q{36}  In our
case 
$$
\ket a = \prod_j^N{e^{-{1\over 2} |\alpha_j|^2}\sum_{n_j}{ {\alpha_j^n\over
\sqrt{n_j!}}\ket{n_j}}}. \eqno(70)
$$
The energy $E_a$ associated with $\ket a$ is the mean value of the Hamiltonian in $\ket a$: 
$$
E_a = \hbar\sum_j^N{\omega_j|\alpha_j|^2}.\eqno(71)
$$
\par

To calculate the information we shall need the partition function defined in (50).  In view
of (57), it may be written as\q{37} 
$$
Z(\mu) = \prod_j^N{\int_{-\infty}^{\infty}{d\alpha_1 d\alpha_2\over\pi}
e^{-\mu\hbar\omega_j(\alpha_1^2+\alpha_2^2)}}, \eqno(72)
$$
where $\alpha_{j1}$ and $\alpha_{j2}$ are the real and imaginary parts of $\alpha_j$ and we have
adopted the customary measure in $\alpha_1-\alpha_2$ space.\q{36}  Doing the Gaussian integrals
gives  
$$
Z(\mu) = \prod_j^N{{1\over\mu\hbar\omega_j}}.\eqno(73)
$$
As before the lagrange multplier $\mu$ is fixed in terms of the specified mean energy by
$$
\=E = -{\partial\ln Z\over\partial\mu} = {N\over\mu}.\eqno(74)
$$
\par

Now we use (53) to calculate the maximum information that can be stored in the system with mean
excitation energy $\=E$ by coding with coherent states\q{37}: 
$$
\maxi = N\log_2 e + \sum_j^N{\log_2\left({\=E\over N\hbar\omega_j}\right)}\quad {\rm
bits}.\eqno(75)
$$
We note that for fixed $N$ this information becomes formally negative when $\=E$ is so small
that the energy per mode becomes much smaller than $\hbar\omega_j$ for a substantial fraction of
the modes.  We interpret this problem as due to the overcompleteness of the set of coherent
states.\q{36}  At any rate it does not seem to be a problem for larger $\=E$.\par

It is interesting to compare this result with the maximum information codable using an equal number
$N$ of occupation number states. The partition function is the product of the $Z_j$ given by (58). 
The mean energy,
$$
\=E=\sum_j^N{\hbar\omega_j\over e^{\mu\hbar\omega_j}-1},   \eqno(76)
$$
is to be viewed as determining $\mu$ in terms of $\=E$.  Substitution in (53) gives
$$ 
\maxi=\mu\=E\log_2 e-\sum_j^N{\log_2\left(1-e^{-\mu\hbar\omega_j}\right)}\quad\ {\rm bits}, 
\eqno(77)
$$
which in conjunction with (76) provides a parametric prescription for
$\maxi(\=E)$.    Evidently this $\maxi(\=E)$ is quite different from the one for coherent
states, Eq.(75).  When all $N$ modes have very similar frequencies (narrowband channel) it is easy
to solve (76) for $\mu$, and subtitution in (72) gives
$$
\maxi=\left[\sum_j^N{\log_2\left(1+{\=E\over N\hbar\omega_j}\right)} + {\=E\over
\hbar\omega_j}\log_2\left(1+ {N\hbar\omega_j\over E}\right)\right]\quad \ {\rm bits},  \eqno(78)
$$
which could actually have been guessed from (45).  It is a simple exercise to show that this
$\maxi$ exceeds that for coherent states for any $N$ and $\=E$.\par

\vglue 12pt
\line{\it 3.6. The Optimum States for Signaling\hfil} 
\vglue 5pt

The previous discussion raises the natural question of which set of states leads to the
maximum information transmission, other things being equal. It has often been stated that
occupation number states are the best.\q{24,26}. The argument in support of this is that the
capacity (45) may be obtained by maximizing Shannon's entropy subject to normalization of
probability and stipulated mean occupation number (or mean energy for a narrowband channel). 
However, in this maximization the probabilities are regarded as depending on occupation
number.\q{24}  Were the states characterized by some other quantum number, it is not certain that
the resulting distribution and maximal entropy would be the same.  The problem here is
essentially that stated at the end of Sec.3.3.  We now state and prove a
theorem\q{37} that clarifies the situation: occupation number states are indeed one of the sets of
states which optimize information transmission (or information storage for that matter).\par

Let us imagine the change in $\maxi$ [as defined by given by (50)--(53)] due to an arbitrary
variation of the set of states $\ket a$.  Since $\mu$ is representation dependent, its variation
must be included. Thus  
$$
\delta\maxi = \left[\delta(\mu \=E) + \delta\ln (Z-\zeta)\right]\log_2 e.\eqno(79) 
$$
However, according to (50) 
$$
\delta Z = -\sum_a(E_a\delta\mu +\mu\,\delta E_a) e^{-\mu E_a},   \eqno(80)
$$
where
$$
\delta E_\alpha = \bra{\delta a}H\ket a+\bra a H\ket{\delta a},      \eqno(81)
$$
whereas by (52)
$$
\=E=-{1\over Z-\zeta}{\partial Z\over\partial\mu}.  \eqno(82)
$$
We find after substitution of all these in (79) that $\delta\mu$ cancels out. Therefore, the
condition for $\maxi$ to be an extremum is
$$
\delta \sum_a e^{-\mu E_a}\bigr|_\mu = 0.  \eqno(83)
$$
\par

The problem of extremizing $\maxi$ with respect to the class of states is thus equivalent to
extremization of the partition function at fixed $\mu$.  This result is reminiscent of the
thermodynamic rule that an equilibrium state characterized by a maximum of entropy for given
mean energy amounts to a minimum of the Helmholtz free energy $F$ at given temperature. Since the
partition function is just $\exp(-F/kT)$, we see that a maximum of the partition function is
involved.  This analogy tells us that the extremum sought in (83) is really a maximum. We may
now prove the following theorem.\par

{\tenbf Theorem 2.} $\maxi$ is maximized when the set of signaling states $\{\ket a\}$ is chosen as
a complete set of eigenstates of the Hamiltonian $\hat H$.\q{37}\par

{\tenbf Proof.}  From the variation principle in quantum mechanics\q{36} $E_a\equiv\bra
a \hat H\ket a$ is extremal when $\ket a$ is an eigenstate of $\hat H$.  And if we insist that
$\bra a \hat H\ket a$ be extremal with respect to a complete set $\{\ket a\}$, then this is the
set of (orthonormal) eigenstates of $\hat H$. Therefore, choosing $\{\ket a\}$ as the set of
eigenstates of $\hat H$ makes the partition function in (83) extremal with respect to small
variations of the $\ket a$, thus satisfying (83). In fact, the partition function is
maximized by this procedure.  For according to the variational principle, the ground
eigenstate gives the lowest possible $E_a$, the ground state eigenvalue.  The next $E_a$ is the
minimum possible within the set of states orthogonal to the ground state, and so on.  It is
clear that this set of minimum $E_a$'s gives maximum $Z$.  Therefore, by using a complete set of
eigenstates of $\hat H$ as signaling states, we get maximum $\maxi$.\qed\par

We should mention that with the is choice of signaling states, the partition function Eq,(50) is
formally identical to the partition function from statistical mechanics, $Tr(\exp(-\hat H/kT))$
as is clear by using the energy representation.  Occupation number states of a free field are a
special case of eigenstates of $\hat H$.  Therefore, in the communication systems under
consideration, channel capacity is maximized by using occupation number states (and measuring
occupation number at the receiver). 

\vglue 12pt
\line{\tenbf 4. The Linear Bound on Communication\hfil }
\vglue 5pt

As the example in Sec.3.4 shows, the CIF of a channel contains quite a lot of detail about the
channel's communication capabilities; by the same token its computaion is quite an elaborate task.
Sometimes the CIF description of communication is an expensive luxury.  We might prefer a less
detailed statement about the capacity which is easier to come by. It is under circumstances like
these that the linear bound introduced by Bremermann\q{9-11} is important.  According to
Bremermann, one can set a universal bound on channel capacity depending only on the signal's
energy.  His argument is that a signal, when looked at in quantum terms, must contain at least one
quantum of some sort (in the language of Sec.3. it must be self--heralding). Thus for alloted
signal energy $E$, the angular frequencies that can appear are bounded from above by $E/\hbar$.
Bremermann interprets this as the bandwidth $\dw$ of the system relevant in the Shannon capacity
formula (5).  Regarding the signal--to--noise factor in the formula, Bremermann uses an obscure
argument whose crux is to equate Shannon's noise with the energy uncertainty $\delta
E\geq\hbar/\tau$ required by the \te\ for a signal of duration $\tau$.  His final result is\q{11}
$$
\imax\leq{E\over 2\pi\hbar}\log_2 (1+4\pi)\ \ {\rm bits\ s}^{-1}.\eqno(84) 
$$
This bound involves no details of the channel's construction.  Bremermann's argument has been
critized\q{8,14,15} for relying on the classical Shannon formula to get an ostensibly quantum
result, and for the obscurity surrounding the connection of noise power with the \te, itself a
principle that invites confusion.\q{38,39}\par

However, there are other roads to the linear bound. For example, an independent argument\q{12}
for a bound like (84) relies on causality considerations combined with the bound on the entropy $H$
that may be contained by a physical system of proper energy $E$ and circumscribing radius $R$,
namely,
$$
H \leq {2\pi ER\over\hbar c}.          \eqno(85)
$$
Originally inferred from black hole thermodynamics,\q{16, 19} this bound
has since been established by detailed numerical experiments\q{13} and analytic
arguments\q{13,20-23}.  According to Shannon's information theory, the peak 
entropy $H_{\rm max}$ for a system limits the maximum information $\maxi$ that
can be stored in it.  Now by transporting a system with information inscribed in
it, one has a form of communication.  Because the system  cannot travel faster
than light, it sweeps by a given point in time $\tau\geq R/c$ (issues
connected with the Lorentz-Fitzgerald contraction are carefully dealt with in
Ref.12.). Thus an appriate ``receiver'' can acquire from it information at a
rate not exceeding $\tau^{-1}H_{\rm max}\log_2 e$ (as usual, $\log_2 e$
converts from nits to bits).  Substituting from Eq.(85) we have
$$
\imax\leq {2\pi E\over\hbar}\,\log_2e\ \ {\rm bits\ s}^{-1}          \eqno(86) 
$$
which is quite similar to (84).  The drawback of this ``derivation''
is that it deals only with a very special sort of communication: information
transfer by bulk transport.\q{14}\par

\vglue 12pt
\line{\it 4.1. Heuristic Derivation\hfil }
\vglue 5pt

We now offer a {\it heuristic\/} argument\q{40} for a communication bound
of the form (86) which does not rely on the entropy bound Eq.(85).
Suppose the information we wish to transmit is inscribed in a bosonic carrying
field by populating its energy levels with quanta; each quantum
configuration represents a different symbol.  Let $\tau$ and $E$ be the
signal's duration and energy, respectively, $\epsilon$ the lowest {\it
non--zero} one--quantum energy level, and $\Delta\epsilon$ the smallest energy
separation between levels beneath $E$. Evidently the total number of occupied
levels is $N \leq {E\over \Delta \epsilon}$, while the total number of quanta
is $M \leq E/\epsilon$. The total number of configurations $\Omega$ must thus bounded from
above by the number of configurations of a system composed of $M = E/\epsilon$
identical bosons distributed among $N = {E\over \Delta \epsilon}$ levels. 
This last is given by a formula well known from Bose statistics.  Therefore,
$$
\Omega \leq {(N + M - 1) !\over M!\,(N-1)!}. \eqno(87)
$$
All these configurations are {\it a priori\/} equally likely so that the peak
entropy of the signal is bounded according to
$$
H_{\rm max}\leq\ln [(N + M - 1)!] -\ln M! - \ln[(N-1)!]. \eqno(88) 
$$
Assuming $N$ and $M$ are large, the logarithms may be approximated by Stirling's formula.
Substituting the bounds on $N$ and $M$, equating $H_{\rm max}$ with the peak information, and
converting to bits we get  $$
\maxi\leq {E\over \sqrt{\epsilon \Delta \epsilon}}
\Big[({\epsilon \over \Delta \epsilon})^{1/2}\ \log_2 (1 + {\Delta \epsilon
\over \epsilon}) + ({ \Delta \epsilon \over \epsilon})^{1/2}\ \log_2 (1 +
{\epsilon \over \Delta \epsilon})\Big]\ \ {\rm bits}.
\eqno(89)
$$\par

The function $f(x) = x^{-1/2}\log_2 (1+x)$ appearing here is already familiar from Sec.2.3. It
follows from its properties that the term in square brackets in Eq.(89) can be no larger than
$\approx 2.32$. Now in order to be able to decode the information, the receiver must be able to
distinguish between the various enrgy levels, which calls for energy measurement with precision
$\delta E\leq\Delta\epsilon$. According to the time--energy uncertainty
principle, the finiteness of the measurement interval $\tau$ imposes an
uncertainty $\delta E\geq h/\tau$.  Thus, $\Delta\epsilon\geq h/\tau$ in order
that the useful information approach $\maxi$.  Furthermore, if $R$ is the spatial
extent of the signal, we can use the momentum--position uncertainty relation
to set the bound ${\epsilon/c}\geq\hbar/R$.  In addition, on grounds of
causality the inequality $\tau\geq R/c$ must apply.  Therefore,
$\sqrt{\epsilon\Delta\epsilon}\geq\sqrt{2\pi}\hbar\tau^{-1}$.  It follows from
(89) that
$$
\imax\leq {0.925 E\over\hbar}\ \ {\rm bits\ s}^{-1},           \eqno(90)
$$
which is of the same form as (86).\par

This argument, appealing as it is, suffers from two drawbacks: it is only valid
for large N and M, where Stirling's approximation may be trusted, and it makes
use of the popular but nonrigorous version of the time-energy uncertainty
relation. We now turn to two exact derivations of the linear bound.\par

\vglue 12pt
\line{\it 4.2. The Bound for Signals with Prescribed Mean Energy \hfil}
\vglue 5pt

The discussion here will refer to \sh\ signals only.  We assume the
signal states $\ket a$ to be occupation number states specified by the
list of occupation numbers $\{ {\rm n}_1 ,{\rm n}_2 ,\thinspace\dots \}$ for the
various modes.  Each state $\ket a$ is ascribed a  particular {\it a priori}
probability $p_a$ subject to the normalization condition $\sum_a p_a =1$. The
information that may be carried by the signal is limited by Shannon's
entropy (1). The energy of the state  $\ket a$ is evidently $\sum_j n_j
\epsilon_j$ with $\epsilon_j = \hbar \omega_j$. The signal's mean energy is
$$
\=E=\sum_{a}\ p_{a}\sum_j n_j \epsilon_j.                      \eqno(91)
$$\par

We maximize  $H/\=E$ subject to (91) by means of the variational principle
$$
\delta \left(-{\sum_a p_{a} \ln p_a \over \sum_a p_a
\sum_j n_j\epsilon_j} - \lambda \sum_{a} p_{a}\right) = 0,      \eqno(92)
$$
where $\lambda$ is a Lagrange multiplier. Notice that, in contrast with similar calculations in
statistical mechanics, one is not here enforcing an energy constraint. Calling $(H/E)_{\rm max}
\equiv \mu$, variation of $p_{\alpha}$ gives
$$
(\ln p_{a}+ \mu \sum_j n_j\epsilon_j + \lambda\=E) \delta  p_{a} = 0.  
\eqno(93)
$$
Thus for nonvacuum states the probability distribution reads,
$$
p_{a} = C e^{-\mu\sum_j n_j\epsilon_j}, \eqno(94)
$$
where $C$ is a normalization constant into which we have absorbed the
Lagrange multiplier $\lambda$.  Of course, when all $\nj=0$, the probability is
taken to vanish (\sh\ signal).\par 

When we compute the Shannon information from (53) with the maximizing
distribution (94) we get
$$
\maxi = \mu \=E - \ln C.          \eqno(95)
$$
However, $\mu$ is defined as the maximal $H/\=E$, and since $\=E$ in (93)
refers to the mean energy of the maximal $H/\=E$ situation, it follows that
$C=1$ for consistency.   Thus despite the similarity between our distribution and
Boltzmann's, the ``inverse temperature'' $\mu$ here is not a free parameter, but
is fixed by the normalization of probability. Now according to (49) and (51) for
\sh\ signals, the constraint $C=1$ implies that $Z=2$.  Writing $Z=\prod_j Z_j$
and using the form (58) we find that $\mu$ is a root of the equation
$$
-\sum_j\ \ln \left(1 - e^{-\mu \epsilon_j}\right) = \ln 2.         \eqno(96)
$$
Because each term in the sum increases with $\mu$, it follows that the root is
unique.\par

Let us now use the periodic boundary condition approximation according to which
$\epsilon_j = 2\pi j\hbar \tau^{-1}$ with $j=1,2,\thinspace\dots\quad$ Numerical
summation gives that  $\mu\pi\hbar\tau^{-1} \approx 0.4931$. Since $\mu$ is the
maximum of $H/E$, it follows after conversion to bits that
$$
\imax\leq 0.2279\ E/\hbar\ \ {\rm bits\ s}^{-1},           \eqno(97)
$$
which is of the same form as the heuristic bound (90), but tighter.\par

The equality in (97) would corresponds to a line of slope unity in Fig.1. just tangent to the
\sh\ curve at its knee.  It may be seen that for moderate signal information ($<10$ bits), the
linear bound gives a considerably better idea of the CIF than does the Pendry formula.  The
opposite is true for large information.  The linear bound may be saturated only for signals which
carry about 3 bits.  The point that maximum communication rate is attainable only for a signal
carrying of the order of one bit was made early by Landauer and Woo.\q{41}

Had we worked with heralded signals we would have found that there is no
maximal $H/E$ for given $\tau$.  This is already clear from Fig.1.  Does this
weaken the generality of the claimed linear bound?  In this respect it is
important to notice that the violation shows up only for very low signal mean
energy.  This means that only the lower energy levels are populated, and
sparsely at that. Now in Bose statistics of one level, the ratio of mean energy
$E$ to energy standard deviation $\Delta E$ is $N^{1/2}$ where $N$ is the total
number of quanta. Thus when the violation of the linear bound appears, the
system has few quanta and the  energy spread is not small compared to the mean
energy itself.  Hence mean energy is far from representing the actual energy. 
The conclusion must be that the characterization of a signal by mean energy is
not appropriate in that regime where the linear bound seems to break down. 
We can deal with this problem as follows.\par

\vglue 12pt
\line{\it 4.3. The Bound for Signals with Specified Energy Budget \hfil}
\vglue 5pt

Instead of specifying the signal by its mean energy, a misleading concept for
low excitations, one can instead specify the energy ``budget'' or energy
``ceiling'' for signaling -- the maximum available energy per signal.  Shannon's
entropy Eq.(1) reduces in this case to Eq.(2) since all signal states with
energies below the maximum are equally likely.  The problem reduces to counting
the number of signal states as a function of the energy budget.  This is a
difficult problem in general, as has long been known from its analog in
microcanonical statistical mechanics.  This counting was carried out
numerically for a few examples relevant to communication by Gibbons,\q{42} and
later by one of us.\q{13}  Recently progress has been made towards the limited
goal of establishing analytically {\it bounds\/} in the number of quantum states
up to a given ceiling energy for three dimensional systems.\q{23}  Here we
review a one dimensional version of these results which applies to signals
of finite duration.\q{40}  In Sec.5. we shall use the more general result\q{23} to
discuss limitations on information storage.\par  

Let $\Omega(E)$ be the number of distinct quantum states of a system accessible with energy
not exceeding $E$.  Evidently $\Omega(E)$ depends on
the one--quantum energy spectrum $\{\epsilon_j\}$.  According to Eq.(2) the
maximum (microcanonical) information that may be coded in the system
corresponds to the entropy
$$
H_{\rm max} = \ln \Omega(E). \eqno(98)
$$
\par
 
Now focus attention on configurations with a fixed number of indistinguishable
quanta $m$. If the one--particle levels  are ordered  by energy, so that
$\epsilon_{j_1}\leq\epsilon_{j_2}$  when \hbox{$ j_1 < j_2$}  (degenerate levels
are to be ordered arbitrarily), an $m$--quanta configuration is specified by the
set of occupied one--quantum levels $\{\epsilon_{j_1}, \epsilon_{j_2},\dots\}$  (of course, some
of them may be repeated, corresponding to multiple occupation of a level). The
number $\Omega_m(E)$ of $m$--quanta states with total energy $\leq E$ can be
written as

$$
\Omega_m(E)\equiv \sum_{j_1\leq j_2\leq\thinspace\dots j_m} \Theta[E -
\epsilon_{j_1} -\epsilon_{j_2} \thinspace\dots - \epsilon_{j_m}]\qquad
m\geq 1, \eqno(99) 
$$
where $\Theta$ is Heavyside's function. The disposition of the limits on the
summation has the effect of avoiding double counting of states which differ
only by the exchange of (identical) quanta.  We shall assume a nondegenerate
vacuum so that $\Omega_0(E) = 1$ for $E\geq 0$.  The number of one--quantum
states with energy up to $E$,
$$
n(E) \equiv \Omega_1(E) = \sum_{a=0}^\infty\Theta(E - \epsilon_j),\eqno(100)
$$
will play a key role in further discussion in Sec.5.  We assume there is no
zero--mode \ie $\epsilon_j > 0$.  Thus $n(E) = 0$ for $E\leq 0$.\par

The problem of finding the number of accessible quantum states, $\Omega(E)$, can
evidently be reduced to that of counting all possible $m$--quanta states: 
$$
\Omega(E) = \sum_{m=0} ^{\infty} \Omega_m(E). \eqno(101)
$$
Explicit calculation of $\Omega(E)$ by this means is not, in general, practical.  However,
bounds can be set on it by the following procedure.  Relaxing the energy
ordering in Eq.(99), we define the useful auxiliary quantity $N_m(E)$ which
overcounts the number of $m$--indistinguishable quanta configurations \hbox{$[\Omega_m(E) <
N_m(E)]$},  
$$
N_m(E)\equiv \sum_{j_1, j_2, \thinspace\dots} \Theta[E -
\epsilon_{j_1}-\epsilon_{j_2} \thinspace\dots - \epsilon_{j_m}] \qquad m\geq
1,      \eqno(102)
$$
and
$$
N_0(E)\equiv \Theta(E). \eqno(103)
$$
In analogy with equation (101), it is natural to define 
$$
N(E) \equiv \sum_m N_m(E) \eqno(104)
$$
which overcounts the number of accessible quantum states:
$$
\Omega(E) < N(E) .\eqno(105)
$$\par

The advantage of this procedure is that $N(E)$  satisfies a very simple integral
equation (see Appendix C):  
$$
N(E) = \Theta(E) + \int_0 ^E N(E - E')({dn\over dE'}) dE'.  \eqno(106)
$$
Let us take the Laplace transform of this equation.   Denoting the Laplace
transform of a function $f(E)$ by $\s f(s)$, and making use of the convolution
theorem, and of the fact that $n(0+) = 0$, we obtain 
$$
\s N(s) = {1\over s[1 - s \s n(s)]} .                         \eqno(107)
$$
This equation plays a central role in our discusion.\par

As we saw in Sec.4.2., in the periodic boundary condition approximation
for a signal with period $\tau$, the one--quantum spectrum is \hbox{$\epsilon_j =
j\epsilon$}; \hbox{$j =1,2,\thinspace\dots$}, where \hbox{$\epsilon = 2\pi\hbar
\tau^{-1}$}. Thus, the Laplace transform of the one--quantum particle number
function is 
$$
\s n(s) = \int_0^\infty\ dE\,e^{-Es} \sum_{j = 1} ^\infty \Theta( E
-j\epsilon),                                                      \eqno(108) 
$$
which can be cast in the form
$$
\s n(s) = s^{-1} \sum_{j=1} ^\infty e^{-j\epsilon s} .\eqno(109)
$$
Performing the sum in (109), substituting in (107), and inverting the Laplace transform 
$\s N(s)$ we have 
$$
N(E) = {1\over 2 \pi i} \int_{\gamma - i\infty} ^{\gamma + i\infty}
{e^{s\epsilon} -1 \over s (e^{s\epsilon} - 2)} e^{Es} ds, \eqno(110)
$$
where $\gamma$ must be chosen to the right of the poles located in the $s$--plane at
$s=\epsilon^{-1}(\ln 2 + i\,2\pi k)$, $k=\{\dots, -2, -1,0, 1, 2,\dots\}$.  It should
be noted that there is no pole at $s=0$.\par

This integral has been evaluated by the contour method\q{40} and is reproduced in Appendix D; the
result is 
$$
N(E) = 2^{[[E/\epsilon]]}\,     \eqno(111)   
$$
where $[[x]]$ stands for the whole part of $x$. Thus whenever $0 \leq E/\epsilon \leq 1$ we get
that $ N(E) = 1$ and $H(E) = 0$. This confirms the feeling that as long as the first energy level
is not accessible, no information can be encoded at all. Recalling the definition of $\epsilon$,
the fact that $\maxi\leq\log_2 N(E)$ and the convention $\imax = \maxi/\tau$, we get
$$
\imax\leq {E\over 2\pi\hbar}\ \ {\rm bits\ s}^{-1}, \eqno(112) 
$$
which is consistent with Eqs.(86), (90) and (96). 

\vglue 12pt
\line{\it 4.4. Caveats on the Derivation of the Linear Bound \hfil}
\vglue 5pt

As it first appeared in Bremermann's work, the linear bound on capacity was
predicated on the supposition that a signal must consist of at least one
quantum.  Our discussion of \sh\ vs. heralded functions in Sec.3.2. makes
it clear that the vacuum state is a legal signal state under some
circumstances, so that Bremermann's supposition is not generic.  It would thus
{\it appear\/} that the linear bound cannot be valid under all
circumstances.  Indeed, in Sec.4.2. we found that for signals with specified
mean energy, the linear bound (whatever the exact numerical coefficient) can
be surpassed by heralded signals (see especially Fig.1).  It turns out, somewhat surprisingly,
(Sec.4.3) that for signals with specified energy ceiling, the bound is valid regardless of whether
the signal is heralded or not (the vacuum was included in the list of states).  Thus, the linear
bound turns out to be very general.\par

Three ingredients went into the proof of the linear bound in Sec.4.3.: the
periodic boundary condition approximation, the assumption that the zero
frequency mode cannot be used in signaling, and the characterization of signals
by occupation number.  Let us discuss them in turn.\par

By viewing the signal as periodic one obtains a simple form for the frequency
spectrum.  This sort of approach is quite common in physics.  Arguably, it would
have been more realistic to look at signals that turn on and off abruptly.  In
that case there are no sharp one--quantum energies; rather all levels are
broadened.  One way to proceed then is to use Gabor frequency--time cells\q{35}
to partition the phase space occupied by the signal.  To each such cell is
assigned a Gaussian modulated sinusoidal wave which takes over the role of the
pure sinusoidals in the Fourier representation of the periodic signal, and
embodies the idea that the energy levels must be broadened in inverse
proportion to the duration $\tau$.  If all cells are chosen to extend a time 
$\tau$, it is natural to choose the central frequencies of the Gaussian
wavepackets to correspond to the energies $\epsilon_j=2\pi j\hbar\tau^{-1}$,
precisely the frequencies figuring in the periodic boundary condition
approximation.\q{15}  The energy spread of a wavepacket is
then $\sim 2\pi\hbar\tau^{-1}$. With this choice it is easy to grasp the effect
of the periodic boundary condition approximation.\par

For energy ceiling $E$, a many--quanta state with \hbox{$\sum_j \epsilon_j > E$}
was excluded in the periodic boundary condition approximation.  However, if
the energy sum exceeds $E$ only by a quantity of order $2\pi\hbar\tau^{-1}$, the
state is allowed in the present description because it is possible for the true
energies of several of the quanta to be on the low side of the central energies
of their Gaussian packets.  Of course, the larger the excess of 
\hbox{$\sum_j \epsilon_j$} over $E$, the less probable the state.  For if the
state is a one--quantum state, the quantum's energy must lie on the outskirts
of the Gaussian packet to keep below $E$.  This situation has low probability. 
If we deal with a several--quanta state, the individual energies can lie closer
to the central energies, but there must be a trend toward the lower energy
side.  Thus, although the individual quanta are not at very improbable
energies, the product of several probabilities smaller than one will cause the
overall configuration to be unlikely. Thus in the exact treatment extra states
become available, but these states have low probability.\par

We must also note that some states which were permitted in the periodic boundary
condition approximation become, in the exact treatment, low probability states.
These are states with \hbox{$\sum_j \epsilon_j$} within a
quantity $\lta 2\pi\hbar\tau^{-1}$ on the low side of $E$.  This is because with
nonnegligible probability some of the quanta involved can lie on the high side
of their Gausssian peaks, and cause the true total energy to exceed $E$.  This
effect partly neutralizes the gain of states discussed above.  The conclusion
must be that the periodic boundary condition approximation is likely to only
somewhat underestimate the number of states.  We thus venture to conclude that
(112) is likely to be only a little below the true linear bound in the exact
treatment.\par

In our derivation of the linear bound in Sec.4.3. we excluded the
mode with $\epsilon_j = 0$.  If included it would have led to
an infinity of states for any energy.  This is because we can form arbitrarily
many states by having a varying number of quanta with zero energy; all these
are permitted being below the energy ceiling.  To understand why the zero
frequency mode must be excluded, one must distinguish between the situation
where the signal is periodic, and the one where it is sharply limited in time. 
In the first case the periodic boundary condition is exact; the zero frequency
mode in question sets the dc level of the signal.  This dc level cannot serve to
send information.  It is permanent, and does not turn on when the signal is
sent so that the signal's information is not coded in it.  At best the dc level
conveys some information about the channel, but not specific to the
signal.  The zero frequency mode thus has no role in signaling.  When
the signal is sharply bounded temporally, the spreading of frequencies
precludes the existence of a mode with exactly zero energy.  Even if interpreted as the center
of a Gaussian wavepacket, $\epsilon_j$ cannot vanish: that would entail negative
as well as positive frequencies.  Hence in the periodic boundary condition approximation, we
should exclude the $\epsilon_j = 0$ mode.\par

In our derivation the signal states were classified by occupation number.  As shown in
Sec.3.6., these maximize the capacity.  Therefore, the bound on capacity (112) must apply to
all kinds of signal states.     

\vglue 12pt
\line{\it 4.5. The Linear Bound and the Time--Energy Uncertainty Relation\hfil}
\vglue 5pt

What is the linear bound good for?  First of all, it serves as a convenient
rule of thumb for estimating peak performance possible for a communication
system based on one channel and signals of finite duration.  Indeed, very
often one may not want to delve into details of the system which would be
necessary to determine the full CIF.  At the risk of extreme generosity we can
then use the linear bound to estimate the capacity.  For example, if a budget
of 1 ev is available per signal, we estimate from (112) that $\imax <
3.8\times 10^{13}\, \ {\rm bits\ s}^{-1}$.  Certainly no optical channel has
been known to exceed this bound.    
 
Another application of the linear bound is a reinterpretation of the
\te. The canonical statement of the \te\ is\q{38}: {\it the product
of the dispersion in the energy of a system and the timescale over which the
expectation value of a system observable varies \/} is $\geq\hbar/2$. However,
the ``popular'' version of the \te\ has it that: {\it the energy (or the
dispersion in the energy) of a  system times the duration over which it is
measured\/} $\gta h$ (here $h$ is  the quantum of action $2\pi\hbar$).
The popular version is {\it not\/} a theorem, but it may be employed
pragmatically (at ones risk)  in a case by case basis.\q{39} Indeed, some
studies of quantum capacity\q{5,10,11} (and even one of our own arguments in
Sec.4.1.) make use of it. Sometimes it is  misleading. For example, it would
seem to forbid signals with finite $\tau$  and arbitrarily small $E$; yet
such (heralded) signals are possible and meaningful, as mentioned in
Sec.3.4.\par

With help of our result (112), we may reinterpret the \te\ as follows.  The
\cost\ in communication, defined as $E/\maxi$, has been mentioned often. 
Suppose we carry out a measurement, obtaining thereby some information, which
is then conveyed to the observer by a communication channel.  The {\it product
of the \cost\ of the signal and the time interval $\tau$ during which the
information is delivered is no smaller than the quantum of action $h$\/}. 
This follows from (112).  This statement is quite different from the canonical
statement of the \te\ quoted earlier.  We believe it is more useful in the
context of quantum measurement because it refers to energy of the signal, not to
energy dispersion of the sytem, and it talks of the time interval over which
the information is delivered, not merely of a timescale of the system.  \par

\vglue 12pt
\line{\it 4.6. The Linear Bound for Many Channels\hfil}
\vglue 5pt

Up to now we have dealt only with communication via a single channel. 
Communication systems are more often than not multichannel ones,\eg waveguide
with two polarizations, television , fiber optics link, optical nerve,
\thinspace\dots\quad  How does the capacity of a multichannel system compare with
that of a single one?  More specifically, how are formulae like the Pendry
bound (16), or the linear bound, to be adapted to the many--channel case?  Both 
Pendry\q{8} and Levitin\q{43} have rightly stressed the difficulty in
formulating a universal bound on $\dot I$ for information flow in three
dimensions (many spatial channels).  A similar point was made earlier by
Landauer and Woo\q{41}.  By contrast Bremermann\q{11} regarded his bound,
Eq.(84), as also valid in the multichannel case.  Rather than review the entire
issue of multichannel communication, let us clarify this controversy  by
analyzing the linear bound in the context of an array of parallel channels
which partake of the total mean energy $\=E$.\par

We assume the signal velocity $c_s$ is constant and identical for all channels.
This makes the task of synchronizing the arrival of signals via the  diverse
channels rather straightforward. Such synchronization is evidently a 
prerequisite for maximizing the communication rate (staggered signals stretch
the duration of the reception). Since the signal with the longest duration
sets the characteristic time $\tau $ out of which $\dot I$ is computed,
one can  always make the overall $\dot I$ larger by rearranging the information
among the various  channels so that the durations of the signal are similar in
all of them.  Therefore, we shall assume that the durations of the signals in the
various channels are the same. 

We shall study the linear bound only for {\it \sh\/} signals by the method
of Sec.4.2.  The difference now is that a state $\ket a$ is here defined as a
set of occupation  numbers for the various modes $j$ in all the channels. We
suppose there are $N$ channels which we label by Greek subscripts like $\nu$. 
We shall denote \hbox{$\prod_j (1-e^{-\mu\epsilon_j})^{-1}$} taken over channel
$\nu$ by the symbol $Y_\nu$.   First we shall consider what we shall call a
{\it  simple\/} communication system, one whose channel architecture is 
orderly enough that it suffices for the signal to exhibit one quantum in some
channel in order for its arrival to be unambiguously noted, \eg an optical
fiber link where the fibers are not twisted or otherwise
jumbled,\thinspace\dots\quad  The probabilities $\pa$ for all states
are still given by Eq.(94), except that the vacuum in all channels is assigned
vanishing probability.  Therefore, the normalization of probability gives
$$
C\prod_{\nu =1}^N Y_\nu - C=1, \eqno(113)
$$
where the last $C$ corrects for the inclusion of the vacuum in the product.
As in Sec.4.2. we find that necessarily $C=1$ in order for the situation to
correspond to maximum $H/I$.  From (113) follows a condition on $\mu$ just like
Eq.(96), but with the sum being also over channels. 

If all $N$ channels are of the same sort, all contribute the same factor
$Y_\nu$.  This reflects the fact that maximum information transmission obtains
when the signal energy is shared out equally among all channels.  This is like
endowing all channels with the same ``temperature'' $\mu^{-1}$ so that
all the $Y$'s are identical. In conclusion, the condition on $\mu$ is
$$
-N\,\sum_j\ \ln \left(1 - e^{-\mu \epsilon_j}\right) = \ln 2      \eqno(114)
$$
where now the mode sum is over one channel.  Since the mode sum (including the
minus sign) is monotonically decreasing with $\mu$, it follows that when \hbox{$N\gg
1$}, $2\pi\mu\hbar\tau^{-1}\gg 1$ (for $N=1$,  $2\pi\mu\hbar\tau^{-1}=0.986$
according to Sec.4.2).  In this case the mode sum is dominated by its first
term, which may be approximated by $\exp(-2\pi\mu\hbar\tau^{-1})$.  We thus
obtain a simple expression for $\mu$.  Recalling that $\mu=(H/I)_{\rm max}$ and
converting to bits we have $$
\imax\leq {E\over 2\pi\hbar}\,\log_2\left({N\over\ln 2}\right);
\qquad N\gg 1.      \eqno(115) 
$$
This bound should be compared with (97). The well known tendency for logarithmic
growth of the capacity with the number of channels is evident here.  The
difficulty in stating a universal capacity or bound for multichannel
communication\q{8,41,43} is thus clear.\par

In the paradigm just considered, which is relevant for many man--made
communication systems, the basic requirement is that only the state which is
vacuum in all channels is to be excluded.  Yet in many naturally  occurring
communication systems, whose channel ``architecture" is complex or disorderly,
this would be too weak a requirement. A case in point is provided by the bundle
of electromagnetic channels through which an astronomer acquires information
about a supernova explosion in a distant galaxy. The relevant channels are
a set of photon channels whose propagation directions all fall in the tiny solid
angle subtended in the sky by the galaxy.  Before the outburst became visible,
the astronomer did not know which of all the channels available to him are
operative. As he becomes aware of the explosion, the operative set  is fixed by
the presence of photons appearing at random in some of a small subset of all
channels that our astronomer was monitoring. A second example may be provided
by the optical nerve (a bundle of electrochemical channels) which conveys
information to the visual cortex in the brain from a vast number of optical
receptors in the eye's retina.  When under dim illumination an object becomes
visible in part of the eye's field of view, the firing of neurons in a few
randomly chosen fibers belonging to the subset of the nerve that monitors the
relevant part of the field delimits which group of channels is operative for the
particular sighting.\par  

Abstracting from our examples, we define a {\it spatially blurred 
communication system\/} as one in which the operative set of channels
is determined when a signal is received by the presence of at least one
quantum in each of at least a fraction $r$ of the channels, with the populated
channels being selected at random.  We call $r$ the {\it filling fraction\/} of
the communication system.  Determination of $r$ must depend on details of the
physics and required  reliability of the communication system. Here we shall
only be concerned with the dependence of $\imax$ and $\emin$ on the assumed
filling fraction.\par

To formulate the theory of blurred communication systems, we imagine a system
with a large number $N$ of channels.  Allowed signals must have at
least one quantum in each of $M$ channels chosen at random out of the $N$. The
value $M$ is chosen so that $M/N$ approximates $r$ as $N$ and  $M$ are made
large. This  realization of the system will be justified if the ultimate results
depend only on the ratio of $M/N$, and not on $N$ and $M$ separately. 
Recalling the development in Sec.4.2., we see that the normalization condition
for the $\pa$, which determines $\mu$, can be put in the form (recall $C=1$)
$$
\prod_{\mu =1}^M \ Y_\nu  - \sum_{\nu>\rho>\cdots }\ Y_\nu  Y_\rho \cdots =1,
\eqno(116) 
$$
where each term in the sum has $M-1$ distinct factors. The first term in 
Eq.(116) is, appart from the factor $C$, the total probability of all
conceivable states. From it is deducted the formal [according to
Eq.(94)] probability for states with at least $N-M+1$ channels empty of quanta,
states excluded by definition.\par

Again to get a simple expression suppose that all $N$ channels are similar so
that all the $Y$'s are equal. Then the equation collapses  to
$$
Y^N-{N \choose {M-1}}Y^{M-1}=1. \eqno(117)
$$
Since we are assuming large $N$ and $M$, the right hand side of (117) is
plainly  negligible ($Y > 1$).  Approximating the factorials in the combinatoric
symbol  with help of Stirling's formula, we can cast this equation in the 
form
$$
\ln Y\approx G(r)\equiv r(1-r)^{-1}|\ln r|+|\ln(1-r)|, \eqno(118)
$$
which replaces Eq.(113).\par

Since \hbox{$Y\equiv\prod_j (1-e^{-\mu\epsilon_j})^{-1}$}, we see that to
obtain $\mu$ we merely have to replace the $\ln 2$ term in (114) by $G(r)$. 
Thus $\mu$ becomes a definite single valued function of $r$.  It has to be
calculated numerically by performing the sum in (114).  Writing the final
result as
$$
\imax\leq \alpha(r){E\over 2\pi\hbar},                          \eqno(119)
$$
we find that for \hbox{$r = \{10^{-3}, 10^{-2}, 10^{-1}, 1/2\}$},
\hbox{$\alpha = \{4.85, 2.95, 1.41, 0.645\}$}.  The notable feature here is that
$\imax$, and likewise $\emin$, do not depend directly  on the number of
channels $N$, but only on the fraction of them which are required to contain
quanta in order to certify arrival of a signal.  This, of course, justifies our
way of implementing  the spatially blurred communication system.  It can also be
seen that for a blurred communication channel, the usual linear bound is correct
except for a weak dependence of the coefficient on the filling fraction.  In
this sense Bremermann's claim\q{11} that multichannel communication is subject
to the linear bound is on the mark. 

\vglue 12pt
\line{\tenbf 5. Limitations on Information Storage \hfil}
\vglue 5pt

The past two decades have witnessed a breakthrough in computer and data storage
technology; one advance has been the great reduction in size of information
storage devices.  According to this trend individual atoms or molecules
may one day become short term information--storage devices. Can this trend
continue indefinitly, or is there is some physical limitation on the size
devices of given information capacity may reach in the future?   It
seems plausible that as the size of information storage devices approaches
elementary particle proportions, the end must come to the miniaturization
process.\par

Now the maximum entropy for a system quantifies the maximum information
that may be coded using all its microscopic degrees of freedom.  Therefore,
regardless of details of precisely how and where the information is held, a
bound on entropy like (85) limits the maximum amount of information that may be
inscribed in and retrieved from a physical system in terms of its maximum linear
size and its energy. This bound is in harmony with the intuitive feeling that
the entropy of a physical system must be limited by the available volume in
phase space which, in turn, ought to depend on the system's dimensions and
proper energy.   But how sure are we of the correctness of the
bound?\par

\vglue 12pt
\line{\it 5.1. Drawbacks of the Canonical Ensemble Method \hfil}
\vglue 5pt

Originally inferred from considerations of black-hole physics,\q{19} the bound
was immediately subjected to scrutiny from the point of view of statistical
physics.  Early microcanonical numerical calculations of the specific entropy
of free quantum fields confined to cavities of various shapes were carried out
by Gibbons\q{42}.  These, and later more extensive ones,\q{13} have supported
the bound in every case.  In order to obtain more generic results, one of us
applied in detail the canonical approach of statistical physics to quantum
fields confined in various cavities.\q{19,44}  In the canonical approach
(system parametrized by temperature or mean energy) the  validity of the bound
hinges on the sign and the value of the vacuum (Casimir) energy. If this last is
positive and not very small on the scale of the typical mode frequencies, then
the bound  is obeyed with the maximum $H/E$ occurring at low excitation
energy.\q{19, 44}  However, field theoretic  calculations for various cavities
and fields frequently show that the vacuum energies are negative \q{45-47} so
that violations of the bound can occur at sufficiently low temperatures.  Even
if the vacuum energy vanishes exactly, or if one chooses to interpret the $E$
as the excitation energy above the vacuum, it is easy to see that violation of
the bound is possible at low temperatures.  According to the canonical ensemble
the ratio $H/E \approx  T^{-1}$ at low $T$, so that the bound is violated at
sufficiently small temperature.\q{47-49} although the energy range over
which the violation occurs is extremely narrow.\q{44}\par

The very significance of canonical results in this regime is put into
question by the observation that at low temperatures fluctuations are so large
that mean energy is not a good indicator of actual energy.  Recall that the ratio
of a system's energy fluctuation $\Delta E$ to its mean energy $\=E$ is $\Delta
E/\=E\approx N^{-1/2}$, where $\=N$ is the mean number of quanta.  At low
$\=N$ the energy fluctuations could  be larger than the mean energy
itself.  Put another way, at low excitations the customary equivalence between
canonical and microcanonical ensembles cannot be relied upon.  Now the
canonical ensemble owes its popularity more to the convenience it affords in
calculations (which are always much more complicated, if not hopeless, in
microcanonical ensemble), than to the conviction that it gives a more correct
entropy. Whereas the microcanonical ensemble method relies only on very general
assumptions like ergodicity, the canonical ensemble may be deduced from it
only on basis of additional hypothesis like the validity of saddle point
approximation, positivity of specific heat, \etc  Sometimes the canonical
ensemble fails entirely:  the hydrogen atom cannot be canonically described.
Therefore, the microcanonical approach appears to be the primary theoretical
framework.  Henceforth we conduct our discussion using microcanonical
methods.\par

\vglue 12pt
\line{\it 5.2. One Particle Information Storage -- Examples \hfil} 
\vglue 5pt

In the simplest instance information can be stored in a one--particle system
by virtue of the multiplicity of quantum states.  In an early investigation of
bound (85) in microcanonical ensemble, Qadir\q{20} considered a single free
quantum mechanical particle confined to a volume $V$ and having energy up to
$E$.  Using the uncertainty principle to determine how many states are
available, he took the logarithm of this quantity as the entropy of the
system.  He concluded that $H\sim\ln(ER)$ where $R\approx V^{1/3}$.  This is
consistent with bound (85).  Perhaps a more interesting problem concerns
a particle subject to some attractive force.  Then the energy on the right hand
side of the bound is decreased while the entropy is not necessarily affected,
so that one tests the bound under more severe conditions.  Additionally,
this problem brings us closer to more realistic information storage systems. 

If a molecule could be harnessed as an information storage device, the coding 
would have to exploit the multiplcity of available molecular levels. Because
these states usually differ in energy, it is relevant to ask what is the
maximum information which may be encoded for a given available energy. Suppose
we apply the bound (85) taking care to include in $E$ all the energies, \ie
rest energies as well as excitation energies. Of course, in real atoms and
molecules most of the energy is rest energy, and so (85) predicts, for typical
atomic (molecular) dimensions and masses, that the limit is some $10^6$ bits.
This certainly exceeds the logarithm of the number of atomic (molecular) states
below ionization in known atoms and molecules, so bound (85) is easily
satisfied.  The seemingly discrepant case of the hydrogen atom with its
infinity of levels is easily accounted for by remembering that the highly
excited (Rydberg) states correspond to dimensions large by atomic
standards.  But it is interesting to consider an hypothetical atomic systems
whose constituents' rest masses coould be adjusted at will. Would not reduction
of such masses eventually bring (85) into conflict with the actual value of
\hbox{$(H/E)_{{\rm max}}$ ?}\par

To elucidate this question we now consider, following Ref.19, the cumulative
number of states $N(E)$ up to energy $E$ for one--particle quantum--mechanical
systems described by Schr\"odinger's equation. Our examples are meant to
capture  the essential features of the electronic, rotational, and vibrational
degrees of freedom we meet in atoms and molecules. We want to see whether the
peak value of $(H/E)_{{\rm max}}$ is indeed bounded by $2\pi R/\hbar c$, as
predicted by (85). As expected, the inclusion in $E$ of the rest energy of the
particle, however small, is essential for the bound to be obeyed, so we choose
the zero of the energy scale accordingly.\par

\vglue 12pt
\line{\rm 5.2.1. Particle in a One Dimensional Potential Well \hfil} 
\vglue 5pt

Our first example concerns a particle of mass $m$ in a one--dimensional
potential well.  Superconducting quantum interference devices (SQUIDs) can be
modelled as particle--in--well systems.   Let us assume that the particle is
constrained to a range of radius $R$ on either side of an appropriately chosen
point, regardless of its energy $E$.  A  simple way to count the number of
states $N$ up to and including $E$ is to use the WKB formula\q{36} 
$$
\int_{x_{{\rm min}}}^{x_{{\rm max}}} \sqrt{2m(E - V(x)} dx = 2\pi\hbar (N+1/2)
,\eqno(120)
$$
where $V(x)$ is the potential and $x_{\min} ,x_{\max}$ are the roots
of $V(x) = E$. Evidently, only the whole part of N given by (106) is
meaningful. For the moment we ignore the inherent inaccuracy of the WKB formula
for low--lying states.\par

Evidently, the range of $x$ is less than $2R$.  Further, $E - V \leq e$ where $e$ is the energy
measured with respect of the bottom of the potential well. Thus,
$$
N \leq  {\sqrt{2me} R\over\pi\hbar}.   \eqno(121)
$$
It is also clear that $E = e + mc^2$. Defining the dimensionless quantities
$e_\ast = e/mc^2$ and $R_\ast = R m c/\hbar$ we have
$$
{\ln N(E)\over E} \leq {YR\over 2\hbar c};\qquad Y \equiv 
{\ln (2e_\ast R_\ast^2/\pi^2)\over R_\ast (1+e_\ast)}.          \eqno(122)
$$
It is clear that within the Schr\"odinger theory we can only consider the case
$e_\ast < 1$ (nonrelativistic particle). Let us now maximize $Y$ with respect to
$e_\ast$. The maximum occurs at the $e_\ast$ determined by
$$
2R_\ast^2e_\ast^2 = \pi^2e_\ast \exp ( 1 + 1/e_\ast), \eqno(123)
$$
and amounts to $(R e_\ast)^{-1}$. Because $e_\ast < 1$, the right hand
side of (123) is never smaller than $72.93$ and so $Y < 0.1656$.
Therefore, after transforming to bits,
$$
I < 0.119\,{E R\over \hbar c}\ \ {\rm bits} \eqno(124)
$$
for all $e$. Thus a particle confined to a potential well satisfies bound (85)
regardless of the choice of $m$.

\vglue 12pt
\line{\rm 5.2.2.  Rigid rotator \hfil} 
\vglue 5pt

Consider now a two-dimensional system, a rigid rotator with moment of inertia
$I$ and mass $m$ confined within a sphere of radius $R$.  This can serve to
model the rotational levels of a molecule ($m$ is the molecular mass), a
futuristic information recorder. The rotational energy levels are given by $e =
j(j+1)\hbar^2/2I$ with $j = 0,1,\thinspace\dots$ labeling angular momentum;
the levels are $2j+1$ degenerate. The total energy is $mc^2 + e$. Obiously,
$N(E)$ is just the sum of $2j+1$ from  $j=0$  to the largest $j$ for which
$mc^2 + e$ does not exceed $E$. Denoting this by $j_\ast$ we find $N(E) =
(j_\ast + 1)^2$. Now, we are interested in the  peak value of ln$N(E)/E$.  This
obviously occurs for an $E$ which is a rotational level (if $E$ is increased
slightly, the factor $E$ depresses the ratio while $N(E)$ does not grow unless
the next level has been reached.) Thus with the notation $I_\ast = I/ mR^2$ and
$R_\ast = Rmc/\hbar$ we may put  
$$
{\ln N(E)\over E} \leq {2XR\over \hbar c}; \qquad X = {\ln(j_\ast +1) 
\over R_\ast +j_\ast(j_\ast +1)/2I_\ast R_\ast }. \eqno(125)
$$

As a function of $j_\ast$, $X$ peaks at the $j_\ast$ determined by
$$
(2j_\ast+1)(j_\ast+1)\ln(j_\ast+1) - j_\ast(j_\ast +1) = 2I_\ast R_\ast^2   \eqno(126)
$$
and
$$
X_{\rm max} = 2I_\ast R_\ast (j_\ast +1)^{-1}(2j_\ast +1)^{-1}.  \eqno(127)
$$
Of course, if (112) does not give integral $j_\ast$, then $X_{\rm max}$ cannot
be quite reached, and (127) actually gives us an upper bound on $I_\ast R_\ast$
for specific $j_\ast$.  However, if $I_\ast$ and $R_\ast$ are so adjusted that
the peak can be reached and $j_\ast =0,1,2,3, \thinspace\dots$, then $I_\ast
R_\ast =0, 1.08, 5.24, 13.4,\thinspace\dots\quad$ with the increasing trend
continuing indefinitely. Because the radius of the gyration cannot exceed $R$,
$I_\ast < 1$ so we get upper bounds on $I_\ast R_\ast$ itself to substitute in
(113). In this way we find, after converting the result to bits, that
$$
I < 0.499\,{E R\over \hbar c}\ \ {\rm bits}. \eqno(128)
$$
This is in harmony with bound (85).

\vglue 12pt
\line{\rm 5.2.3.  Three--dimensional harmonic oscillator \hfil} 
\vglue 5pt

Consider next a three--dimensional isotropic harmonic oscillator of rest mass
$m$ and frequency $\omega$. This could model a defect in a crystal lattice
used as an information cache. Its energy levels are 
$$
e = (n_1 + n_2 + n_3 +3/2)\hbar\omega \eqno(129)
$$ 
where $n_i = 0,1,\thinspace\dots\quad$ Again, the total energy is $E = mc^2 +
e$.  $N(E)$ is evidently the number of ways in which the $n_i$ can be added in
such a way that the total energy does not exceed $E$. Again, the peak  $\ln
N(E)/E$ is reached when $E$ exactly corresponds to some energy  level. Let
$F(n)$ be the number--theoretic function giving the number of ways in which
three labeled non--negative integeres may be added to give the integer $n$.
Then  
$$
{\ln N(E)\over E} \leq K {R\over\hbar c}\eqno(130)
$$
with
$$
K \equiv {\ln F(n) \over R_\ast[1 +(n + 3/2) y]}, \eqno(131)
$$
where $y \equiv\hbar/mc^2$ and $R_\ast$ is defined as before.\par

The effective radius of the oscillator can be taken as the
oscillation amplitude given by the virial theorem, $R^2 = e/m\omega^2$, or
equivalently
$$
R_\ast = (n + 3/2)^{1/2} y^{-1/2}.             \eqno(132)
$$
Since $F(n) < (n+1)^3$, our problem reduces to finding the maximum of
$$
K = {3  y^{1/2}\ln (n+1) \over (n +3/2)^{1/2}[1 +(n + 3/2) y]} \eqno(133)
$$
with respect to $y$ and $n$.  As a function of $y$, $K$ peaks at $y =
(n+3/2)^{-1}$, i.e., where the oscillator's energy just equals the rest energy.
Although this point is already outside the nonrelativistic domain, it should be
clear that the formal peak value so obtained bounds the $K$ realizable by the
nonrelativistic oscillator.  And with $y$ optimized, the result peaks for
$n = 2$.  Transforming the result to bits we have 
$$
I < 0.369\,{E R\over \hbar c}\ \ {\rm bits}. \eqno(134) 
$$
Again  this is in harmony with the bound (85) for all $m$ and $\omega$.\par

Actually the range of applicability of our example transcends the
harmonicity assumption. Any spherically symmetric potential well resembles a
harmonic potential near the bottom. Since the peak of $\ln N(E)/E$ is reached
at low excitation, it is likely that some anharmonicity of the potential does
not change (134) much.\par

What is the moral of our examples?  It is that, when the rest energy
is included in the energy $E$, the number of states accessible to a
quantum--mechanical system of size $R$ with energy limited to $E$ is less than
$\exp(2\pi E R/\hbar c)$. The inclusion of the rest mass in $E$ is essential. 
Without it any bound like (85) can be surpassed by adjusting parameters of the
system, \ie by making the moment of inertia of a molecule large. However,
the rest energy can be made small by "molecular engineering" without
upsetting our result. Since our examples can be tailored to electronic,
vibrational, and rotational levels, we have just shown that the information
that could be coded in an atom or molecule is indeed bounded by (85). For real
atoms and molecules the maximum must fall considerably below (85). In fact, if
we consider only electronic levels for which the electronic mass is the
relevant one, then short of ionization bound (124) limits the information to a
few tens of bits.\par

\vglue 12pt
\line{\it 5.3. Many Quanta Systems: Numerical Experiments \hfil} 
\vglue 5pt

Turn now to many particle systems.  Kahn and Qadir\q{21} investigated the
number of states in noninteracting quantum mechanical many--particle
systems.  They counted available states by the semiclassical (continuum)
approximation, and found support for bound (85), but expressed the opinion that
the bound can only be strictly correct in that approximation.  Now as is well
known, the semiclassical approximation is poor for low lying states.  It turns
out (see  below) that this is precisely the regime in which the ratio $H/E$
peaks.  This makes it clear that the semiclassical approximation is not
particularly well suited to analyze the bound.  As will become clear in this
section and in Sec.5.4., the bound is an exact result in the full quantum
treatment.\par

Following Ref.13, we consider here not quantum mechanical many--particle
systems, but rather relativistic quantum free field systems.  There are several
reasons.  First, the extension to relativistic fields does not incurr extra
computational challenges in the absence of interactions.  Second, quantum
fields provide a realization of black--body radiation which, being a high
entropy system, is a prime challenger of bound (85).\q{19} The
third reason concerns the computational process. In considering ways to
optimize computers, a useful reference would be a computing machine, itself
composed of elementary quanta, in which information is coded in the occupation
numbers of the various modes, and in which the elementary operations consist of
shifting quanta from one mode to another. It is difficult to believe that any
forseeable computer composed of macroscopic components could be more
energetically efficient, or faster at storing, retrieving, or processing
information. Thus, it is interesting to assess the information capacity of an
assembly of quanta or, equivalently, the maximal entropy for given available
energy.\par

We thus consider a collection of quanta of some field confined inside a cavity
of some shape.  The stationary one--particle modes of the system will have a
discrete spectrum $\{\epsilon_j\}$ and to each mode $j$ there will correspond a
degeneracy $g_j$.  In view of indistinguishability of quanta, a many--quanta state is
specified fully  by the occupation numbers $\{n_j\}$ of the various modes.  If there are no
interactions, the energy of the state is $\sum\,n_j\epsilon_j$.\par

Let $\Omega(E)$ represent the number of distinct quantum states accessible to the system with
energy (measured from the vacuum) not exceeding $E$.  We assume the vacuum in
nondegenerate so that $\Omega(0)=1$.  If the quanta are (indistinguishable) bosons, the number of
ways to realize a set of occupation numbers $\{n_j\}$ is\q{34}
$$
D\{n_j\} = \prod_i {(n_i+g_i-1)!\over n_i!\,(n_i-g_i)!}.    \eqno(135)
$$
For fermions the exclusion principle eliminates a number of possibilities so
that $D\{n_j\}$ is smaller.  As we raise the energy a jump of $\Omega(E)$
equal to $D\{n_j\}$ occurs as $E$ coincides with some $\sum\,n_j\epsilon_j$.  Then
$\Omega(E)$ stays constant until the next such coincidence.  Therefore,
$\Omega(E)$ looks like a ladder function.  Ordinarily entropy is defined as
the logarithm of the function $D\{n_j\}$.  Since this is a very discontinous
``comb'' function, we prefer to take $S=\ln\Omega(E)$, as done by Gibbons in
his early numerical experiments.\q{42} This agrees with Eq.(2).\par

Evidently for a many--modes field, $\Omega(E)$ is a very complicated
combinatoric function.  It is hopeless to try to calculate it
analytically.  Here we describe numerical experiments carried out in Ref.13
which extended Gibbons' early ones.  The procedure adopted for bosons was the
following:\par

$\bullet$ List the energies $\epsilon_j$ of the modes and their degeneracies $g_j$. 
Only massless fields were considered; massive quanta ``waste'' energy in the
rest mass which could have been used to reach more states, and their
$\Omega(E)$ can only be smaller.\par

$\bullet$ Populate the modes according to a pattern which guarantees inclusion
of all many--quanta states up to some energy ceiling.  We describe here the
pattern used for Bose quanta; fermions are further constrained by the
exclusion principle and thus their $\Omega(E)$ must be lower. 

$\bullet$ Bin the many--quanta states found in narrow energy bins, and deduce an
approximation to $\Omega(E)$.\par

The populating strategy was the following. First a single quantum was
succesively promoted through the modes with $\epsilon_j$ increasing until it exceeded
the ceiling.  The number of states appearing at each stage of the promotion was
computed from Eq.(135) and these numbers were binned by energy.  Then the
first quantum was returned to the lowest lying mode, and a second quantum was
added to that mode.  Next the first quantum was promoted mode by mode until
the energy ceiling was reached.  At this point the second quantum wss promoted
by one mode, and the first was returned to that same mode.  Then the first
quantum was promoted again in the previous pattern.  Numbers of states were
computed from (135) at each promotion and binned by energy.  When promotion
of the second quantum with return of the first to that same mode already led
to energy in excess of the ceiling, a third quantum was added at the lowest
lying mode and the first two quanta were returned to it.  The pattern of
promotions, first quantum first, was repeated.  At each stage a new quantum
would be added.  When addition of a new quantum and return of the previous
quanta to the lowest lying mode caused the energy to exceed the ceiling, the
process was stopped.  This populating pattern assures that all quantum
configurations allowed by the principle of indistinguishability are
counted.  A count of the number of states accumulated in all bins beneath
energy $E$ gave an approximation for $\Omega(E)$.  This function was thus
reconstructed up to the chosen ceiling.\par

\vglue 10.0truecm
\baselineskip=10pt
\centerline{\eightrm Fig.3. Specific entropy vs. energy for a scalar field in a 
rectangular box of dimensions}
\centerline{\eightrm 1x0.95x0.9 with Neumann boundary conditions. Numerical values
assume $\hbar=c=1$.}
\baselineskip=13pt
\vskip 0.3truecm

The spectra for fields in various cavities were described in Ref.19, and are summarized in Ref.13. 
In a spherical cavity of radius $R$ the generic form of the spectrum is
$$
\epsilon_j = {\hbar cj_{n\ell}\over R}.                      \eqno(136)
$$
For the scalar field with Dirichlet boundary conditions $j_{n\ell}$ means the $n$-th
zero of the spherical Bessel function of order $\ell$;  $\ell=0,1,2,\thinspace\dots\ \ $ and the
modes are \hbox{$(2\ell +1)$--fold} degenerate.  For Neumann boundary condition $j_{n\ell}$ is to be
interpreted as the $n$-th zero of the {\it derivative\/} of the spherical Bessel function of order
$\ell$ (degeneracies and range of $\ell$ are as for the Dirichlet case).  The neutrino field in a
sphere can also be analyzed under a special boundary condition\q{19} giving the spectrum (136)
based on the zeroes of the spherical Bessel functions, each with degeneracies $2(2\ell +1)$;
again $\ell=0,1,2,\thinspace\dots\ \ $ For the electromagnetic field in a highly conducting
cavity, the tangential electric field must vanish on the boundary.  The spectrum is then of the
form (136) with $\ell=1,2,3\thinspace\dots\quad$ except that for each value of $\ell$ there
are eigenvalues corresponding to zeroes of  both the spherical Bessel function and of its
derivative.  Each of these is $2\ell+1$ degenerate. \par

In a rectangular cavity with sides $A$, $B$ and $C$ the mode energies of both
fields are 
$$
e=\pi\hbar\,\left({i^2\over A^2} + {j^2\over B^2} + {k^2\over C^2}\right)^{1/2}.           
\eqno(137)
$$
For the scalar field with Dirichlet boundary conditions $i, j, k = 1, 2,
3,\thinspace\dots\quad$ and the modes are  nondegenerate  (excepting accidental
degeneracy arising from conmensurate $A$, $B$ or $C$).  For the
electromagnetic field with vanishing tangent electric field on the boundary,
$g_{ijk}=2$ for $i, j, k = 1,2,3,\thinspace\dots\quad$ and $g_{ijk}=1$ when one
of $i$, $j$ or $k$ vanishes.  Modes with two or three vanishing quantum numbers
are excluded.  (Higher degeneracies are possible only when some of the sides are
conmensurate).

In the numerical experiments the energy ceiling was set at some seven times the
value of the lowest mode energy.  This range included some 50--200 modes (not counting
degeneracies) in most cases.  The number of quantum states so included was of order $10^4$. Some
200 energy bins were used which provided sufficient resolution.  An example of 
the detailed behavior of $\ln\Omega(E)$ is seen in Fig.3.\par

Generically $\ln\Omega(E)$ starts at zero for $E=0$, and rises with $E$ in an
oscillatory fashion but faster, on average, than linearly.  Then the rise rate
moderates and $\ln\Omega(E)$ tends asymptotically to a $E^{3/4}$ behavior. 
This last is easily understood.  For large energy there are many possible
states so that the thermodynamic limit sets in: the collection of quanta
behaves like black body radiation.  Since for black body radiation the energy
$E\propto T^4$, but the entropy $H\propto T^3$, we have $\ln\Omega(E) =
H\propto E^{3/4}$.   At any rate, it is clear that the specific entropy
$\ln\Omega(E)/E$ must always have an absolute peak at some not too large
$E$.  This is certainly consistent with bound (85).  Values of the peak $H/E$
were obtained numerically in a number of examples by scanning the numbers
stored in the bins, and some are displayed in Table 1.  It may be seen that
they always comply with bound (85).\par
\vskip 0.2truecm
$$\vbox{\eightrm\halign{#\hfil\qquad &#\hfil\qquad &#\hfil\qquad &#\hfil\qquad &#\hfil\qquad &#\cr
\multispan6\hfil Table 1. Peak specific entropy: numerical result, analytic estimate
and bound (85).$^*$\hfil\cr 
\multispan6\hrulefill\cr 
Field & Cavity & Boundary & $(H/E)_{max}$ & $\zeta(4)^{1/4}$ & $2\pi R$ \cr
\multispan6\hrulefill\cr
scalar & unit sphere & Dirichlet & 0.454 & 0.476 & 6.283\cr
'' &     '' & Neumann & 0.701 & 0.721& 6.283\cr
electromagnetic &     '' & conducting & 0.716 & 0.749& 6.283 \cr
neutrino &     '' & see Ref.19 & 0.547 & 0.566& 6.283\cr
scalar & 1x1x1 & Dirichlet & 0.269 & 0.280 & 5.441 \cr 
    '' & 1x0.95x0.9 & Dirichlet & 0.249 & 0.266 & 5.174\cr
    '' &     '' & Neumann & 0.428 &0.457 & 5.174\cr
electromagnetic &     '' & conducting & 0.365 & 0.380 & 5.174 \cr
scalar & 1x0.66x0.2 & Dirichlet & 0.122 &0.131 & 3.828 \cr
\multispan6\hrulefill\cr
\multispan6 $^*$ Largest dimension of rectangular cavities taken as unit of lenght; here we take
$\hbar=c=1$.\hfil\cr
}}$$

\vglue 12pt
\line{\it 5.4. Many Quanta Systems: Analytic Results \hfil} 
\vglue 5pt

Numerical examples dealing with specific cavity geometries cannot prove the
bound.  Therefore, we discuss two analytic results that go a long way to establishing its
validity.

\vglue 12pt
\line{\rm 5.4.1. Analytical Estimate of Peak Specific Entropy \hfil} 
\vglue 5pt

An approximate proof of bound (85) in microcanonical ensemble can be based on an analytic argument
developed in Ref.13.  This shows that the peak value of $H/E$ can be approximated by
$[\zeta(4)]^{1/4}$, where $\zeta(\kappa)$ is an analogue of the Riemann zeta function constructed
from the mode spectrum of the field:   
$$
\zeta(\kappa) = \sum_j\,\epsilon_j^{-\kappa}.         \eqno(138) 
$$
In three dimensional space the number of eigenvalues below $E$ grows as $E^3$;
therefore the sum representing the zeta function converges only for $\kappa >
3$.  If all we desire is an upper bound on $H(E)$ of a given system, the rule
$(H/E)_{\rm max}\approx[\zeta(4)]^{1/4}$ is a great work saver.  It usually overestimates
$(H/E)_{\rm max}$ by only a few percent [see Table 1 and also Ref.13]. Therefore, in most cases the
information codable in a quantum system whose interactions can be neglected is analytically
approximated by
$$
(H/E)_{\rm max}\approx[\zeta(4)]^{1/4}.   \eqno(139) 
$$
The analytic estimate (139) and bound (85) are displayed for a few examples in Table 1.\par

When only a rough estimate of $\zeta(4)$ is required, it is usually sufficient to cut off the sum in
(138) after a few terms.  This is because the terms in $\zeta(4)$ drop off
rapidly.  Also the function is raised to power 1/4 so errors in it are
diluted.  In this approximation it is easy to see why a bound of form (85) must
hold.  On dimensional grounds the first eigenvalue $\epsilon_1$ has to be of order
$\hbar cR^{-1}$ for a massless field, where $R$ is the radius of the circumscribing sphere.  Thus
$[\zeta(4)]^{1/4}$ must be a few times $R/\hbar c$.  Then to within a numerical factor (139)
reduces to (85).

\vglue 12pt
\line{\rm 5.4.2. Microcanonical Proof of Bound on Specific Entropy \hfil} 
\vglue 5pt 

We now present an exact proof of bound (85), a much simplified and more
rigorous version of that given in Ref.13.  Our system consists of a free
field (scalar, electromagnetic, $\thinspace\dots\, $) confined within a cavity of
arbitrary shape by appropriate boundary condition.  The mathematical framework
we shall use is that developed in Sec.4.3 for pulsed communication.  Recall
that instead of trying to calculate the cummulative number of quantum states
$\Omega(E)$ up to a given energy budget $E$, we switched attention to the
auxiliary quantity $N(E)$ which overestimates $\Omega(E)$.  $N(E)$ was shown to
be the solution of the integral equation (106) in which the number of modes with
$\epsilon_j=\hbar\omega_j\leq E$, $n(E)$, enters in the kernel. In our
one dimensional problem in Sec.4.3 we knew the spectrum $\{\hbar\omega_j\}$;
here each cavity will have a different one--quantum spectrum.\par

For concreteness think of a scalar field $\psi$.  We are then interested in solving the
eigenvalue problem
$$ 
-\nabla^2 \psi_j = {\omega_j}^2\psi_j   \eqno(140)
$$
subject to the general boundary condition
$$
\left[{\partial\psi\over\partial n} + \alpha \psi\right]_{\partial\Sigma}=0, 
 \eqno(141)
$$ 
where $\Sigma$ is the domain of the cavity and $\partial\Sigma$ its boundary;
$\partial/\partial n$ represents the normal derivative at $\partial\Sigma$.
For $\alpha=0$ (141) corresponds to Neumann boundary conditions; for
$\alpha\rarrow\infty$ it reduces to Dirichlet boundary conditions.  For
generic $\alpha$ we have Robin boundary conditions.\par

Since it is hopeless to try to deal with the problem on a cavity--by--cavity
basis, we ask what generic properties of the spectrum can be inferred by
solving for the spectrum corresponding to a fictitious spherical cavity that
completely encloses the true system.  This last is a much simpler
problem, and there turns out to be a simple relation between the special result
for it and the generic result.\par 

We first rephrase the eigenvalue problem as a problem in the
calculus of variations which provides a powerful method for comparing
eigenvalues for related problems.  Define the {\it functional\/}\q{50} 
$$ 
\omega^2 \left[\chi\right] =  {\int_{\Sigma} \nabla{\chi} \cdot \nabla{\chi} +
\alpha \int_{\partial \Sigma} \chi^2 \over \int_{\Sigma}\chi^2 }. \eqno(142)
$$
The extremal values obtained when $\omega^2\left[\chi\right]$ is varied with
respect to $\chi$ are known to reproduce the eigenvalue spectrum of problem
(140).  If $\alpha=0$ and the variation is performed with $\chi=0$ on
$\partial\Sigma$, one gets the Dirichlet spectrum.  If instead free variation of
$\chi$ on $\partial\Sigma$ is allowed, one gets the Neumann spectrum.\q{50} 
Taking $\alpha\not= 0$ and allowing free variation of $\chi$ at $\partial\Sigma$
gives the Robin spectrum. The minimum of $\omega^2\left[\chi\right]$ is the
lowest eigenvalue $\omega_1$ of the corresponding boundary value problem, and
to it corresponds the ground state eigenfunction $\chi_1$.  This
eigenfunction may thus be approximated by substituting a very flexible trial function in the
functional. Higher eigenvalues $\omega_j$ and eigenfunctions $\chi_j$ are obtained by varying the
above functional with a trial function $G$ orthogonal to all those  obtained
previously, \ie $\int_\Sigma G\,\chi_k = 0$ for all $k < j$.\par

There is an alternative way to characterize the sequence of eigenvalues
variationally: the minimax principle.\q{50}  The $i$--th trial function for
minimizing $\omega^2[\chi]$ is chosen as satisfying the appropriate
boundary conditions and orthogonal to any set of $i-1$ independent functions $G$
with the same boundary conditions.   Then the minimum obtained is maximized over all choices of the
$G$'s and there result $\omega_i^2$ (the $i$--th eigenvalue in magnitude) and the corresponding
eigenfunction $\chi_i$.  According to the minimax principle if $\=\omega_i$ is the $i$-th
eigenvalue of a variational problem in which the trial functions belong to a
certain class of admissible functions $\{\=G\}$ \eg all continuous and
differentiable functions satisfying one of conditions (141) on the cavity
boundary, and $\omega_i$ is the $i$-th eigenvalue for the {\it same\/} problem
with respect to a second class of functions $\{G\}$ which are subject to
additional constraints, then $\=\omega_i\leq\omega_i$.  In other words, adding
constraints to the trial functions can only lift all the spectrum.\par

In our case the class of trial functions for the real cavity will be required to
vanish not only at the spherical cavity surface, but also in the region between
the real and the reference spherical cavities.  Because of the additional
constraint,
$$
\=\omega_i\leq\omega_i,                  \eqno(143)
$$
where $\=\omega_i$ now denotes the eigenfrequency for the sphere corresponding
to $\omega_i$ of the real cavity.\par

Now recall the zeta function defined by (138).  We evidently have the
inequality\q{42}
$$
\zeta(\kappa){\omega_j}^\kappa
=\sum_{i=1}^\infty\left({\omega_j\over\omega_i}\right)^\kappa > 
\sum_{i=1}^j\left({\omega_j\over\omega_i}\right)^\kappa > j       \eqno(144)
$$
since in the right hand side there are precisely $j$ terms no smaller than
unity.  Let us now define a ``reference" spectrum,
$$
\omega^\ast_j\equiv\left[{j\over\=\zeta(\kappa)}\right]^{1/\kappa},  \eqno(145)
$$
where $\=\zeta(\kappa)$ is the zeta function for the spherical cavity. 
According to (143) $\=\zeta(\kappa)\ > \zeta(\kappa)$ so that by (144) the
reference ``spectrum'' must satisfy 
$$
{\omega^\ast}_j\leq\omega_j.                        \eqno(146)
$$
Now since the reference spectrum is everywhere lower than the true
spectrum, the corresponding cummulative number of quantum states 
$\Omega^\ast(E)$ for the former must satisfy 
$$
\Omega(E) < \Omega^\ast(E)                       \eqno(147)
$$
(a given energy $E$ can be split in more ways if all the mode energies are
lower).\par

In entire analogy with $N_m(E)$ of Eq.(102), it is now possible to define
${N^\ast}_m(E)$ by simply replacing the true mode spectrum by the reference one.
If we now sum over number of quanta as in Eq.(104), we get $N^{\ast}(E)$ which
must evidently bound $\Omega^{\ast}(E)$ from above. In view of (147) we thus have
$$
H_{\rm max}(E)=\ln\Omega(E) < \ln N^\ast(E).                \eqno(148)
$$\par

The $N^\ast(E)$ is the solution of the integral equation (106) with the
reference mode number density $dn^\ast(E)/dE$ as kernel.  Definition (145) is
equivalent to
$$
n^\ast(E)\equiv\=\zeta(\kappa)E^\kappa.                  \eqno(149)
$$
We proceed to solve the integral equation (106) for $N^\ast(E)$ by Laplace
transforms.  Since the Laplace transform of $n^\ast(E)$ is
$$
\s n^\ast(s) = \Gamma(\kappa+1)\,\=\zeta(\kappa) s^{-(\kappa +1)},\eqno(150)
$$
we have in analogy with Eq.(107)
$$
\s N^\ast(s) = {s^{\kappa-1}\over s^\kappa-\kappa!\,\=\zeta(\kappa)}.    
\eqno(151) 
$$
\par
$N^\ast(E)$ is to be obtained by inverting the Laplace transform $\s N^\ast(s)$, namely,
by evaluating the integral of $\s N^\ast(s)\exp(sE)/2\pi i$ along a contour parallel to the
imaginary axis to the right of the $\kappa$ poles of $\s N^\ast(s)$. These poles are distributed
uniformly around a circle in the complex $s$--plane with radius
$(\kappa!\,\=\zeta(\kappa))^{1/\kappa}$. Their phases are those of the $\kappa$
distinct $\kappa$--th roots of unity
$\sigma_1,\sigma_2,\thinspace\dots,\sigma_\kappa$. It is convenient to
translate the contour to large negative $s$ while indenting it to avoid the
poles as illustrated in Fig.4 for the concrete case $\kappa=4$. In this way
only the residues contribute to the inverse transform; the contribution of the 
vertical part of the contour vanishes in the limit of large negative real part
of $s$. In view of this,
$$
N^\ast (E) =\kappa^{-1}\sum_{n=1}^\kappa\exp\left\{\sigma_n[\kappa!\,\=\zeta
(k)]^{1/\kappa} E \right\}.                    \eqno(152)
$$

\vglue 10.0truecm
\baselineskip=10pt
\centerline{\eightrm Fig.4. Contour for evaluating the inverse Laplace transform in Eq.(151).}
\baselineskip=13pt
\par
\vskip 0.5cm

Since the zeta function is only defined for $\kappa > 3$, let us choose
$\kappa=4$ [this leads to the tightest bound on $\Omega(E)$] and set  $x\equiv
(4!\,\,\=\zeta (4))^{1/4}$ . Then  by exploiting various trigonometric and
transcendental identities\q{51} we have
$$
N^\ast(E) = {\exp(xE) + \exp(-xE) + \exp(ixE) + \
exp(-ixE)\over 4}
$$
$$
= {\cosh(xE) + \cos(xE)\over 2} \leq {\cosh(xE) +1 \over 2} 
$$
$$
= \cosh^2(xE/2)\leq\exp(xE) =
\exp\left\{[4!\,\=\zeta(4)]^{1/4}E\right\}.\eqno(153)
$$
According to (148) the specific entropy should satisfy,
$$
{H(E)\over E} < [4!\,\=\zeta (4)]^{1/4} .\eqno(154)
$$
We see that the rigorous bound here obtained for $H/E$ is just a factor of 2.2 above
our analytic estimate, Eq.(139).  According to the argument following Eq.(139), the bound
is seen to be of the same form as (85), and amounts to a proof of it.\par

Neumann boundary conditions for the scalar field raise an interesting question.  The lowest
eigenvalue is $\omega_1=0$ with corresponding homogeneous eigenfunction.  As a result the formal
zeta function is always infinite, and bound (154) is uninteresting.  We have argued
elsewhere\q{13,23, 52} that ``zero modes'' like this one have to be excluded from consideration when
calculating the entropy because they correspond to a field condensate analogous to the superfluid
condensate, not to modes that can be populated by a definite number of quanta.  It follows that
for consistency, zero modes must be excluded from the zeta function, a procedure that was followed
in constructing the fifth column of Table 1.  Unruh\q{53} has argued that a zero mode furnishes the
opportunity to build an infinity of quantum states with the same energy, and thus violates bound
(85).  In fact the states dispayed by Unruh can be understood, in analogy with the phenomenon of
symmetry breaking, to belong to different systems.\q{52}  Thus they do not contribute to the
entropy of one system.  It must be realized that were the infinity of states truly possible,
systems with a zero mode would be endowed with infinite entropy.  There is simply no evidence
from closely akin systems, \eg superfluids, that this is the case.\par  

Is bound (85) respected by systems other than the ones considered in Table 1. ?  What about
systems of fermions ?  We know that, because of the exclusion principle, a fermion with a single
helicity will have smaller $(H/E)_{\rm max}$ than a boson with a single helicity having the same
spectrum.  Of course the Dirac equation has, in general,  a different spectrum than, say, the scalar
equation. The results quoted in Table 1. for the neutrino field (the boundary condition is
described in Ref.19) show that the fermion character does not, in this example, endanger the
bound..  What about massive quanta?  Because they put a goodly fraction of the energy in rest mass,
rather than in ``phase space'', massive fields should give lower $(H/E)_{\rm max}$ than massless
ones, provided the rest masses are included in $E$ as in Sec.5.2.  So far the discussion
has referred to one field.  If we put in a  cavity a sufficiently large number of distinct field
species, the bound can be violated.\q{48,54}  This is because the zeta function for a mixture of
$N$ fields with identical spectra is $N$ times the individual field's zeta function.  Thus
the estimate (139) rises as $N^{1/4}$, and must eventually surpass (85) (this usually happens
only for hundreds of field species).  However, the point has been made\q{13,44} that
since the number of fermion generations is limited (almost certainly only 3),
the number of elementary particle species is below $10^2$.  This is not enough
to allow bound (85) to be surpassed.  Indeed, the argument has been reversed to
set bounds on the number of particle generations starting from the bound on
specific entropy (85).\q{55,56}

\vglue 12pt
\line{\it 5.5. The Effects of Interactions\hfil }
\vglue 5pt

Although the specific entropy bound is now well established for free fields,
its status in the presence of interactions is not so clear.  Interactions are important in many
information storage and communication technologies, \eg  SQUIDs and nonlinear optical media.  It is
clear that interactions raise new challenges for bound (85).  First, even weak interactions can lead
to the formation of bound states, thus creating ``new species''.  If enough of these form, the bound
might be violated by the mechanism mentioned in Sec.5.4.  Second, nonlinear interactions are bound to
complicate the energy spectrum and this might lead to the overthrow of the bound.  Finally, in the
presence of interactions the additivity of energies of quanta which was crucial for the work in Secs.
5.2-5.4 falls through.  Despite these hurdles quite a bit has been learned about the validity of the
bound in the presence of interactions, and this is summarized in the following subsections.

\vglue 12pt
\line{\rm 5.5.1. The Hadron System\hfil }
\vglue 5pt

The blatant example in nature of manifold species arising from the binding
of a few elementary components is the hadron spectrum with its myriad
resonances.  Therefore, an analysis was made of the number of quantum states
in a hadron gas as a function of energy budget.\q{13}  It being hopeless to
deal explicitly with the strong interactions, the bootstrap philosophy was
adopted: the influence of the strong interaction is approximately accounted
for by considering simultaneously the full complement of hadronic species. 
The hadronic spectrum is well described by Hagedorn's semiempirical density
of levels which includes spin and isospin multiplicity as well as
hadron--antihadron duplicity.  If hadron energy $e$ is measured in MeV,
Hagedorn's formula for the number of levels in the interval $de$ is\q{57,58}
$$
\mu(e)\approx 26300(2.5\times10^4 + e^2)^{-5/4}\exp(e/160)\,de.         \eqno(155)
$$

A realization of the hadron spectrum was constructed which agreed with
(155).  This gave the list of energy levels $\{\epsilon_j\}$ of the system.   The
levels were then populated according to the methods described in Sec.5.3. 
Kinetic energy was ignored (it is nonegligible unless the hadron gas is
relativistic).  The algorithms described in Sec.5.3. led to $(H/E)_{\rm
max}=0.007\, MeV^{-1}$.  The hadron gas must evidently be confined to a space
with radius no smaller than $1\times 10^{-13}$ cm.  Therefore, $2\pi R/\hbar c
> 0.032\, MeV^{-1}$.  We see that the hadron gas obeys the specific entropy
bound (85).  This example shows that even if many species can form because of
interaction, they tend to be sufficiently separated in energy for the bound to be
upheld.

\vglue 12pt
\line{\rm 5.5.2. Solitons as Information Storage Devices\hfil }
\vglue 5pt

Does nonlinearity by way of its effect on the spectrum allow violations of the bound?  One
of the simplest covariant models of self interacting field is the quartic self-interacting one; it
obeys the equation
$$
\qed\,\Phi - m^2 \Phi + \lambda\Phi^3 = 0, \eqno(156)
$$
where $\qed$ is the d'Alembertian and $m^2$ and $\lambda$ are real positive constants, the last
measuring the strength of the interaction. This equation has static self--confining field
configurations.  In one space dimension these are solitons.  For example, in empty space Eq.(156)
has, apart from the two ``vacuum'' stationary solutions \hbox{$\Phi = \pm m\lambda^{-1/2}$}, the
time independent soliton solution
$$
\Phi_0(x) = {m\over\sqrt \lambda} {\rm tanh}{mx\over \sqrt 2},
\eqno(157)
$$
which interpolates between the two vacuua.  The soliton is a nonperturbative creature; it cannot be
obtained by perturbation theory based on the vacuua.  It has finite energy
$\sqrt{8/9}m^3\lambda^{-1}$. \q{59} Since only a fraction $3\times 10^{-4}$ of the total energy lies
in  $x > \sqrt 8 m^{-1}$, we define the soliton radius as $R_s= \sqrt 8 m^{-1}$.\par

Note that the Lorentz invariance of Eq.(156) allows one to immediately obtain a traveling soliton
solution by simple transformation.  Thus the question of information storage is closely bound up with
that of communication.  Solitons are of more than academic interest here.  When one burns a
fingertip, the information is conveyed by a solitary wave of axon potentials traveling from finger
to brain along the nervous fibers.\par

Our soliton offers a model for a self--contained information storage system.  The possibility of
storing information in the soliton arises because excitations of it are possible, thus providing a
variety of states for information coding.  Consider a small perturbation about the soliton
configuration: $\Phi=\Phi_0(x)+\eta(x,t)$.  It satisfies the linearized equation
$$   
\qed\,\eta - m^2 \eta + 3\lambda\Phi_0^2\eta = 0. \eqno(158)
$$
Look for eigenmodes of these equation of the form $\eta_j(x,t)=\Xi_j(x)\,e^{-i\omega_j t}$.  The mode
functions satisfy
$$
\omega_j^2 \Xi_j = c^2\left[-\nabla^2 + m^2 - 3\lambda\Phi_0^2\right]\Xi_j.  \eqno(159)
$$
As usual they form a complete set and are orthogonal.\par

In the canonical quantization approach, the field operator corresponding to $\eta$ may be expanded
as 
$$
\hat \eta(x) = \sum_j \left[a_j \Xi_j(x) + a_j^\dagger \Xi_j^*(x)\right], \eqno(160)
$$
where $a_j$ and $a_j^\dagger$ satisfy the canonical commutation relations for the harmonic
oscillator.  The field hamiltonian is
$$
H = {1\over 2}\int \left[\dot\Phi^2+({\bf\nabla}\Phi)^2-m^2\Phi^2+{1\over 2}\lambda\Phi^4\right]
d^3 x.   \eqno(161)
$$
Separating the terms quadratic in $\eta$, substituting (160), and normal ordering, one gets
$$
\hat H = \sum_j\hbar\omega_j a^{\dagger}_j a_j. \eqno(162)
$$
Accordingly, we interpret  $a^{\dagger}_j$ and  $a_j$ as creation and annihilation operators for
quasiparticles ``riding'' on the soliton.  To this approximation the energies of these quanta
are additive.\par

With $\Phi_0$ of Eq.(157), (159) is Schr\"odinger's equation for a particle moving in the
potential  $V = -3\,{\rm sech}^2(mx/\surd 2)$, a standard problem in quantum mechanics.\q{38} In
our specific problem there are two bound states and a continuuum.\q{59,60} The eigenvalues and
eigenfunctions are summarized in Table 2: \vskip 0.2 truecm 
$$
\vbox{\eightrm\halign{#\qquad\hfil & #\hfil\cr
\multispan2 \hfil Table 2. Eigenvalues and eigenfunctions of soliton perturbations.$^*$\hfil\cr 
\multispan2\hrulefill\cr
eigenvalue & eigenfunction\cr
\multispan2\hrulefill\cr
$\omega_0^2 = 0$ & $\Xi_0(z) ={{\rm sech}^2 z}$;\cr
$\omega_1^2 ={3m^2\over 2}$ & $\Xi_1(z) = {\sinh z\ {\rm sech}^2z}$;\cr
$\omega_k^2 =m^2({k^2\over 2}+2)$ & $\Xi_k(z) = 3 e^{ikz}(\tanh^2 z -{1\over 3} - {k^2\over 3} -
ik\tanh z)$\cr
\multispan2\hrulefill\cr
\multispan2 $^* z\equiv mx/\surd 2$ and $-\infty < k < \infty$ is a continuous index.\hfil\cr
}}
$$
\vskip 0.2truecm

The zero mode is not useful for storing information; it does not correspond to a soliton
excitation but is rather related to translational invariance\q{59} as clear from
the fact that $\Xi_0(x) \propto [\Phi_0(x+ \delta x) -\Phi_0(x)]/\delta x$. 
Thus, it makes no sense to talk of quanta occupying this level.  While the $\Xi_1$ mode is a bound
state confined to the soliton's extent, the continuum levels $\Xi_k$ are not confined within
the soliton radius $R_s$ but spread to infinity.  Accordingly, if we wish to use the soliton to
store information without help of a confining box, we can only use its first excitation $\Xi_1$.\par

This being clear, the number of possible information--holding configurations based on the soliton
equals the number of quanta that might populate the first excited level. To this number we must add
unity to account for the  (background) soliton configuration itself.  Unlike the situation for
self--heralding signals where the ground state is not counted, here the background's existence can be
detected, \eg it carries energy. Therefore, one should include this contribution to the total number
of states. Thus, the number of possible configurations within an energy budget $E$ above the
(unexcited) soliton energy is $N(E) = 1 + [[E/\omega_1]]$ where, again, $[[x]]$ stands for the
integral part of $x$. The information that may be stored is thus 
$$
\maxi = \ln (1 + [[{E\over\omega_1}]])\log_2 e\ \ \ {\rm bits}.  \eqno(163)
$$ 
Since $\ln (1 + [[x]]) \leq x$ we find, in light of the definition of $R_s$ and the value of
$\omega_1$, that
$$
\maxi < 0.416 {ER_s\over \hbar c}  \eqno(164)
$$
which is consistent with bound (85).\par

To exploit the continuum states for information storage, one must confine them (after all we are
discussing storage in a finite space).  One way to do this would be to put the soliton in a box. 
However, it would then be impossible to meet the boundary condition $\Phi_0=0$ at both ends.  To get
over this hurdle we might consider instead a soliton--antisoliton pair.  (An antisoliton is the
solution of the field equation differing from (157) only in sign.)  We would put the soliton at one
end of the box with its node $\Phi_0=0$ on the box wall, and we similarly locate the antisoliton at
the other end with its node at the farther wall.  Since the pair are well separated, they constitute a
good approximation to a static exact solution of the equation.  We would then go on to study
excitations of the soliton--antisoliton system within the box.  However, the stationary
configuration envisaged is far from being a generic stationary configuration of the field theory in
a box.  What is the complete set of such configurations, and what do their excitations look like? 

\vglue 12pt
\line{\rm 5.5.3. Scalar Field with Quartic Self--Interaction in a Box\hfil }
\vglue 5pt

Bekenstein and Guendelman\q{22} investigated this issue by looking at the massless {\it charged\/}
scalar field with self--interaction confined to a box. When the interaction is quartic the equation
is (156) again, except that instead of $\Phi^3$ we must write $\Phi^2\Phi^*$.  Classically this
field exhibits a continuum of configurations within a finite interval of energy, so that it should
violate bound (85).  Quantum mechanically the situation is different: the requirement that
configurations have integral charge discretizes the spectrum of stationary states.  Each stationary
state and the excitations of it form a particular charge sector.  Within each sector permitted
energies are sufficiently separated to give the bound a fighting chance.\par

The energy spectrum can be solved analytically in the case of a one--dimensional box.  Call its size
$L$.  In one dimension the quartic coupling constant $\lambda$ is dimensional: $L_\ast\equiv
(\hbar c\lambda)^{-1/2}$ is a scale of length.  Bekenstein and Guendelman
computed the lower spectrum of stationary state energy levels explicitly for the
range of dimensionless box sizes $ 0.05 < L/L_\ast < 20$, and also obtained
asymptotic formulae for box size outside this range.  Although the spectrum is
complicated, the levels are well spaced, and the quantity $\ln\Omega(E)/E$
behaves qualitatively as in Fig.3. The values of $(H/E)_{\rm max}$ are below
the specific entropy bound, $2\pi(L/2)/\hbar c$, by a sizeable factor regardless
of the value of $L/L_\ast$.  Taking into account excitations within each charge
sector does not increase the entropy much, and the bound (85) continues to be
respected.  This example shows that nonlinear interactions do not necessarily
violate the bound on specific entropy even when they introduce extraneous scales into the problem
and change the nature of the energy spectrum.\par

More complicated situations involving interacting fields in a box may be handled by path integral
techniques.\q{61}  There is an indication that bound (85) will be respected for a large class
of interactions. 

\vglue 12pt
\line{\rm 5.5.4. The Gravitational Interaction \hfil }
\vglue 5pt

Does the gravitational interaction help to transcend the bound? 
After all, gravitation is highly nonlinear and introduces a special scale
of length, the gravitational radius.  First we should mention that from the
beginning\q{19} it was clear that a nonrotating neutral black hole, the most
bound of systems, just saturates bound (85).  Wald, Sorkin and Jiu\q{62}
considered the entropy of a self gravitating sphere of black body radiation in
equilibrium.  They concluded that this would respect bound (85) provided the
solution of the hydrostatic equilibrium equations is nonsingular.  Singular
solutions were considered by Zurek and Page\q{63} who did not find any whose
entropy exceeded that allowed by the bound.  Admittedly, none of the results
mentioned is very generic.  However, it is interesting that systems in which
highly nonlinear interactions play a major role do not show much predisposition
for violating bound (85).

\vglue 12pt
\line{\it 5.6. Information Storage in One Dimension \hfil }
\vglue 5pt

One dimensional information storage systems are quite important.  In a magnetic tape information
is basically stored in one dimension.  The DNA molecule is a more striking example in which
the sequencing of four types of molecules codes the genetic information.  A more general system of
this sort is one in which $N$ ``molecules'' picked from $n$ species are arranged in a chain.  There
are $n^N$ such sequences so that $H_{\rm max}=N\ln n$.  If $m$ is the typical molecular mass, and
$\varsigma$ is the typical molecular radius, the system's energy is $E=Nm c^2$ and its half--length
(if not curled up) is $L/2=N\varsigma$.  Of course, $\varsigma>\hbar/m c$ (a molecule must be
larger than its Compton length).  Therefore $2\pi E(L/2)/\hbar c > 2\pi N^2$.  This far exceeds
$H_{\rm max}$ unless the number of molecule species is exponentially large.  Therefore, the system
in question satisfies the information bound.\par

It is also possible to store information in the oscillations of the molecules about their
equilibria (it is unclear whether this option is exploited in biological systems).  Here
interactions are quite important.  To make headway in the analysis, we assume that it is possible to
define normal modes for the oscillations, so that the excitations are free phonons.  Phonons are
characterized by momentum $p$ which may be of either sign; there are only longitudinal phonons in one
dimension.  Assuming a fixed sound speed $c_s$, the corresponding energy is $\eofp=c_s |p|$.  We
calculate the maximum entropy by the approach of Sec.2.4. Eq.(11) gives the thermal entropy $s(p)$ of
a mode at temperature $T$; this corresponds to maximum entropy for given mean energy.  Integrating
$s(p)$ with measure $Ldp/2\pi\hbar$ from $p=-\infty$ to $p=+\infty$, integrating by parts, and
rescaling we get
$$
H_{\rm max}={2LkT\over\pi\hbar c_s}\int^X_0{{xdx\over e^x-1}},  \eqno(165)
$$
where $X\equiv p_{\rm max}/kT$.  There is a peak momentum because the continuum description of the medium
through which the phonons propagate breaks down at the ``lattice constant'' $2\varsigma$.  Hence
$X\approx {\hbar c_s\over 2\varsigma kT}$. A similar calculation gives 
$$
E={L(kT)^2\over\pi\hbar c_s}\int^X_0{{xdx\over e^x-1}}\eqno(166)
$$
for the thermal energy
\par

In order for the continuum approximation to apply at all, we need $kT\gg\hbar c_s/L$ so that at
least the lowest lying phonon levels be highly populated.  If in addition $kT\ll\hbar
c_s/\varsigma$ (both conditions can hold provided $N\gg 1$), we have $X\gg 1$ so that
the upper limit in the integral can be extended to infinity; the integral then equals $\pi^2/6$
(see Appendix B).  Eliminating $kT$ between (165) and (166) leads to 
$$
H_{\rm max}=\left(2\pi E L/3\hbar c_s\right)^{1/2}; \qquad \hbar c_s/L\ll E\ll\hbar
c_s L/\varsigma^2.                \eqno(167)  
$$
This is reminiscent of Pendry's formula (16); however, the dependence on the signaling speed does
not disappear here.  In the limit of large temperature $X\ll 1$ so that we can replace the integral by
$X$.  We then get 
$$
H_{\rm max}\approx L/\pi\varsigma; \qquad E\gg\hbar c_s L/\varsigma^2,   \eqno(168) 
$$
so that the information approches asymptotically a number a little smaller than the number of
molecules in the chain.  Evidently the thermal entropy never quite competes wuith the ground state
entropy.  Thus our earlier argument indicates that the information bound is always obeyed.\par

It will be noticed that inclusion of the rest energy of the substrate structure in the energy entering
into the bound is crucial to the latter's correctness.  The bound does not necessarily work if $E$ is
taken as the excitation energy alone.  It is thus interesting to study a case where there is no
mass in the substrate.  An example is information coded in the states of a scalar field confined
to a one dimensional cavity of length $L$.  A calculation analogous to the above gives
$$
H_{\rm max}=\left(2\pi E L/3\hbar c\right)^{1/2}; \qquad E\gg \hbar c/L.          
\eqno(169)
$$
This is the real information--storage analog of the Pendry formula (16).  Does it obey the information
bound?  Yes.  The constraint on $E$ guarantees that the argument of the square root is
large.  Thus $H_{\rm max}< 2\pi EL/3\hbar c$ so that bound (85) is obeyed.  Of course, nothing has
been proven about the range $E\lta\hbar c/L$.  This case must be studied numerically by the
microcanonical approach, as in Sec.5.3.  The result is that for the full range of energies 
$$
(H/E)_{\rm max}=0.216\,L/\hbar c,     \eqno(170)
$$
which is consistent with bound (85).

\vglue 12pt
\line{\tenbf 6. The Spacetime View of Information \hfil}
\vglue 5pt

Thus far, as customary in the field, we have treated information storage and
communication as separate issues.  But clearly they are not.  A situation can be
described purely in terms of information storage only in the rest frame of the
storing device.  In another Lorentz frame information flows with the motion of
the device, and the communication facet surfaces.  In fact, the Lorentz
invariance of the laws of physics must mean that information storage and
communication are inextricably linked, and proper understanding of one of them
suffices for understanding of the other.  Thus far no unified treatment of this
sort exists.   But there are some insights into how information is intertwined
with the concept of spacetime.  We describe these here.

\vglue 12pt
\line{\tenit 6.1. Influence of Uniform Motion on Communication \hfil}
\vglue 5pt

One of our basic results in communication is that the information in a burst
signal is bounded by Eq.(47).  However, we never made it clear in what Lorentz
frame one is to calculate $E$ and $\tau$.  Normally transmitter and receiver
are at rest in the same frame, and the question is not important.  However, the
transmitter can be in a spacecraft rapidly moving with respect to the
earthbound receiver.  Although the relative velocities in this example are
nonrelativistic, it does illustrate that the question of Lorentz frame is not a
trivial one.\par

However, it is easy to show that, under wide circumstances,  Eq.(47) is a Lorentz
invariant statement.  For example, consider a ``medium'' such as a fluid or
dielectric solid in which signals propagate with fixed speed $c_s$ and no
dispersion. The carrier quanta could be phonons propagating in the fluid, or
``dressed'' photons propagating in a dielectric channel, \etc  We assume there 
are no currents (flows) in the medium so that all of it is at rest in a given 
Lorentz frame A. Consider another Lorentz frame B moving to the right relative 
to A with speed $V$. Withouth loss of generality we may assume that their 
origins coincide at time $t_A=0$.\par

Let a right--moving signal's front pass the  origin of A at that same time. We assume $V<c_s$;
the opposite case can be  studied with appropriate changes. At some time $t_A=t_1$ the signal's
rear end  will pass the origin of $A$, at which time the origin of B has reached  position
$x_A=Vt_1$. At some later time $t_A=t_2$ the signal's rear has  caught up with
the origin of B which is then at $x_A=Vt_2$. Calculating  entirely in A we find
$(c_s-V)t_2=c_s t_1$ so that  
$$
t_2/t_1=(1-V/c_s)^{-1}. \eqno(171)    
$$
Evidently, the duration of the signal in A is just $\tau_A=t_1$. Because of 
time dilation, the duration in B is just $\tau_B=t_2\gamma\recip1$ where 
$\gamma\equiv(1-V^2/c^2)^{-1/2}$ is the Lorentz factor between the frames.
Then by virtue of (142) we have
$$
\tau_B=\tau_A(1-V/c_s)^{-1}\gamma^{-1}. \eqno(172)
$$

Let us now look at the energy. If in A the energy and momentum of a quantum are 
$\epsilon$ and $p$, respectively, then by virtue of the constancy of the
propagation  velocity, $\epsilon=c_s p$. In the absence of interactions the total
energy $E_A$ and momentum of the signal must  stand in the same ratio 
Therefore, by the Lorentz  transformation of energy and momentum, the signal
energy in frame B is 
$$
E_B=\gamma E_A(1-V/c_s). \eqno(173)
$$
We now see from (172)--(173) that $E_A\tau_A=E_B\tau_B$ which shows that the 
quantity $\xi\equiv E\tau/\hbar$ is the same in the propagation medium's frame
and in some other frame, eg that of the receiver in motion with
respect to the medium.  It is possible to demonstrate the invariance when B is
the {\it transmitter\/}'s frame by having frame B move to the left, and the
signal to the right, with  respect to A. Of course, the information $I$ is
itself a Lorentz invariant.  The end result is that the formula
$\maxi=\Im(E\tau/\hbar)$ is Lorentz invariant. In particular, it has the same
form in the frames of the medium (if other than vacuum), the transmitter, and
the receiver.

When the signal moves precisely with the speed of light, \eg photons in empty 
space, $\dots\,$, the above argument may be rephrased by taking A as the 
transmitter's frame, while B is some other frame, like the receiver's. The 
calculations go through formally as before, and demonstrate the Lorentz 
invariance of $\maxi=\Im(E\tau/\hbar)$ in this case also.

\vglue 12pt
\line{\tenit 6.2. Influence of Gravitation on Communication \hfil}
\vglue 5pt

Up to now we have implicitly assumed that the signal propagates in flat 
spacetime (no gravitational field). Consider now its propagation through 
an external stationary gravitational field, \eg signaling from the surface of a planet to an
orbiting spacecraft, or its propagation in the expanding universe (time dependent but
spatially homogeneous gravitational field).  In either case we assume the transmitter and
receiver to be at rest (in the cosmological example this means at rest in the frame of the
microwave background). Redshift effects will make the $E$ and  $\tau$ at reception differ
from those at transmission. However, $E\tau$ will be the same. To verify this focus on a
single Fourier component of the wavepacket representing a particular signal state (in the
cosmological case we refer to spatial Fourier component). Evidently, the variation  of the
phase from front to the rear of the packet must be conserved in transit. At a fixed point
in the  transmitter's frame A, the overall phase  variation is just $\omega_A\tau_A$ where
$\omega_A$ is the angular frequency or time derivative of the phase in frame A. Analogously
at a fixed point in the receiver's frame B, the change of phase amounts to
$\omega_B\tau_B$.  Now for a  single quantum $\epsilon=\hbar\omega$. Therefore, if field
self--interaction may be neglected ($E$ is the sum of $\epsilon$'s), if the signal transit
is adiabatic  (no quantum transitions between various states), and if dispersion is absent
(signal does not spread), then $E\tau$ will be conserved in transit. The same adiabaticity
assumption guarantees that information is not lost. Thus formula (47) is equally valid as
applied to transmitter or receiver  (or in any motionless frame in between).

By combining this result with our previous one on Lorentz invariance interpreted
locally,  we conclude that $\maxi=\Im(E\tau/\hbar)$ must be valid in all Lorentz frames,
and in the presence of external stationary or time dependent but spatially homogeneous
gravitational fields.

The previous argument neglected the self--gravitation of the signal: the
gravitational field was taken as external.  Although in everyday signals
self--gravitation is indeed negligible, the issue of self--gravitating
signals is of great intrinsic interest.  Once self--gravitation is present, 
Newton's constant $G$ enters into the discussion alongside $\hbar$ and $c$. 
Reviewing the argument developed in Sec.3.1., and excluding again lengths
derived from Compton lengths or frequency cutoffs, we conclude that there are
now {\it two\/} independent dimensionless combinations of $E$, $\tau$ and
natural constants.  These can be taken as $\xi=E\tau/\hbar$ and $\varpi\equiv
GEc^{-5}\tau^{-1}$.  Thus the CIF must be a function of $\xi$ and $\varpi$:
$\maxi\equiv\Im(\xi,\varpi)$.\par

The parameter $\varpi$ is of the order of the ratio of the gravitational binding energy of the 
signal to the signal energy $E$ and, therefore, a good measure of self--gravitation.  This assumes
that the size of the signal is $c\tau$; if the propagation speed $c_s$ is smaller than
$c$, $\varpi$ is a lower bound on the specific gravitational binding
energy.  Another interpretation: $\varpi$ is the ratio of the Schwarzschild
radius of the signal $GE/c^4$ to its size $c\tau$ (if $c_s < c $,
$\varpi$ is only a lower bound on the ratio).  From these comments it is
clear that $\varpi$ has a maximum value of order unity.  When $\varpi\ll
c_s/c$, self--gravitation is negligible, and the CIF reduces to a function of
$\xi$ only, as in Sec.3.

When $\varpi$ is maximal, the signal collapses into a black hole (signal
has shrunk down to its Schwarzschild radius), and it can longer convey any information.  Thus
$\Im\rarrow 0$ in that limit. It thus seems likely that the main effect of nonnegligible $\varpi$
is to reduce the CIF below its value for $\varpi=0$.  As of yet no calculation of the
dependence on $\varpi$ has been made.  

\vglue 12pt
\line{\tenit 6.3. Acceleration as a Communication Jammer \hfil}
\vglue 5pt
  
In Sec.6.1. we saw that communication between transmitter and receiver in motion
with respect to each other can be described with the same CIF as for
transmitter and receiver at rest.  What if the motion is accelerated? 
We might be tempted to argue that momentarily the transmitter and receiver are
related by a particular Lorentz transformation, so that the description via the
Lorentz invariant CIF can be employed.  However, this line of reasoning leaves
out a crucial point of principle.  As discovered by Unruh,\q{64} a receiver moving with
uniform acceleration $a$ (this is a statement independent of Lorentz frame) is subject to
quantum noise having all the properties of thermal radiation with temperature $T_U=\hbar
a/2\pi ck$.  Any communication with that receiver is thus affected by thermal
noise intrinsically connected with its motion.\par

Let us, for simplicity,  consider communication in the limit of very long
duration signals.  The relevant formalism is Lebedev and Levitin's for a
broadband noisy channel (see  Sec.2.5).  We recall that the argument is
concerned mainly with the receiver.  The channel capacity Eq.(25) is governed by
one parameter, the power $P$ received.  Now power is a Lorentz invariant\q{65}. 
Therefore, for a steady state transmitter, the $P$ received is also constant
although the receiver is constantly changing its speed.  Let us make the
substitution $kT_1\rarrow\hbar a/2\pi c$ in Eq.(25).  We get
$$
\imax ={a\over 12c}\left\{\left[1 + {48\pi c^2P\over \hbar a^2
}\right]^{1/2}-1\right\}\log_2 e\ \ {\rm bits\ s}^{-1}. \eqno(174)
$$
\par

For large $P$ (174) goes over to the Pendry formula (16). For low $P$ we have
$$
\imax\approx \left(2\pi cP/\hbar a\right)\log_2 e\ \ {\rm bits\ s}^{-1}.        \eqno(175)
$$
The transition occurs at a characteristic power $P_c=10^{-2}\hbar a^2 c^{-2}$.  Although for
everyday accelerations this is a tiny power, for the acceleration typical of electrons in atoms
($10^{25} {\rm cm\, s}^{-2}$), $P_c\sim 10^{12} {\rm ev\, s}^{-1}$ which is
quite large.  It thus maybe that the transfer of information to and from elementary particles
involved in natural processes is governed primarily by the limiting form (175).

\vglue 12pt
\line{\tenbf Acknowledgments \hfil}
\vglue 5pt

JDB thanks John A. Wheeler for suggestions.  MS thanks Jamil Daboul for comments, and the Physics
Department, Ben--Gurion University for hospitality.  Partial support by a grant from the Wolf
Foundation for the Advancement of Science and Art administered by the Israel National Academy of
Sciences is acknowledged. 

\vglue 12pt
\line{\tenbf References \hfil} 
\vglue 5pt
\medskip \ninerm \baselineskip=11pt

\r{1}\obscure{H. Nyquist}{Bell Syst. Tech. J., April}{}{324}{24}; \obscure{H. Nyquist}{Trans.
A.I.E.E}{47}{617}{28}; \obscure{R. V. L. Hartley}{Bell Syst. Tech. J., July}{}{535}{28}.
\r{2}\book{C. Shannon and W. Weaver}{The Mathematical Theory of 
Communication}{Univ. of Illinois Press}{49}.
\r{3}\obscure{T. E.Stern}{IEEE Trans. Inf. Theory}{IT-6}{435}{60}.
\r{4}\obscure{J. P.Gordon}{Proc.IRE}{50}{1898}{62}.
\r{5}\inbooked{J. P. Gordon}{Advances in Quantum
Electronics}{J. R. Singer}{Columbia University Press}{61}.
\r{6}\obscure{H. Marko}{Kybernetik}{2}{274}{65}.
\r{7}\obscure{D. S. Lebedev and L. B. Levitin}{Dokl. Akad. Nauk SSSR}{149}
{1299}{63}\enskip [Sov. Phys. Dokl. {\bf 8} (1963) 377].
\r{8}\jpa{J. B. Pendry}{16}{2161}{83}. 
\r{9}\inbookeds{H. J. Bremermann}{Self--Organizing Systems}{M. C. Yovitz, T. C.
Jacobi and G. D. Goldstein}{Spartan Books}{62}.
\r{10}\inbookeds{H. J. Bremermann}{Proc. Fifth Berkeley Symp. on Mathematical
Statistics and Probability}{L. M. LeCam and J. Neyman}{Univ. of California
Press}{67}.
\r{11}\ijtp{H. J. Bremermann}{21}{203}{82}.
\r{12}\prl{J. D. Bekenstein}{46}{623}{81}.
\r{13}\prd{J. D. Bekenstein}{30}{1669}{84}.
\r{14}\ijtp{R. Landauer}{21}{283}{82}.
\r{15}\pra{J. D. Bekenstein}{37}{3437}{88}.
\r{16}\prd{J. D. Bekenstein}{7}{2333}{73}.
\r{17}\prd{J. D. Bekenstein}{9}{3292}{74}.
\r{18}\cmp{S. W. Hawking}{43}{199}{75}.
\r{19}\prd{J. D. Bekenstein}{23}{287}{81}.
\r{20}\pla{A. Qadir}{95}{285}{83}.
\r{21}\ncl{I. Kahn and A. Qadir}{41}{493}{84}.
\r{22}\prd{J. D. Bekenstein and E. I. Guendelman}{35}{716}{87}.
\r{23}\prd{M. Schiffer and J.D. Bekenstein}{39}{1109}{89}.
\r{24}\rmp{Y. Yamamoto and H. A. Haus}{58}{1001}{86}.
\r{25}\inbooked{H. Takahashi}{Advances in Communication Systems}{A. V.
Balakrishnan}{Academic Press}{65}.
\r{26}\prl{B. Saleh and M. Teich}{58}{2656}{87}.
\r{27}\obscure{R. Landauer}{Phys. Scripta}{35}{88}{87}; \obscure{R. Landauer}{Ann. N. Y. Acad.
Sci.}{426}{161}{84}; \fop{R. Landauer}{16}{551}{86}.
\r{28}\obscure{L. Szilard}{Z.Phys.}{53}{840}{29}.
\r{29}\book{L. Brillouin}{Science and Information Theory}{Academic Press}{65}.
\r{30}\pr{E. Jaynes}{106}{620}{57}; \pr{E. Jaynes}{108}{171}{57}.
\r{31}\book{A. Katz}{Principles of Statistical Mechanics: The Information
Theory Approach}{Freeman}{67}.
\r{32}\book{G. Chaitin}{Algorithmic Information Theory}{Cambridge University
Press}{87}.
\r{33}\book{L. D. Landau and E. M. Lifshitz}{Statistical Physics, {\rm Part II}}
{Pergamon}{80}, p.326.
\r{34}\book{L. D. Landau and E. M. Lifshitz}{Statistical Physics, {\rm Part I, 3rd ed.}}
{Pergamon}{80}. 
\r{35}\obscure{D. Gabor}{Phil. Mag.}{41}{7, 1161}{50}.
\r{36}\book{E. Merzbacher}{Quantum Mechanics}{Wiley}{70}.
\r{37}M. Schiffer, ``Shannon's Information is not Entropy'', submitted to Phys. Lett.{\rm A},
1990. 
\r{38}\book{L. D. Landau and E. M. Lifshitz}{Quantum Mechanics, {\rm 3rd. ed.}}
{Pergamon}{77}.
\r{39}\inbookeds{E. P. Wigner}{Aspects of Quantum Theory}{A. Salam and E. P. 
Wigner}{Cambridge University Press}{72}.
\r{40}M. Schiffer, ``The Quantum Limit for Information Transmission'', submitted to Phys. Lett.{\rm
A}, 1990.
\r{41}\inbooked{R. Landauer and J. W. F. Woo}{Synergetics}{H. Haken}{Teubner}{73}.
\r{42}G. Gibbons (1980), unpublished.
\r{43}\ijtp{L. B. Levitin}{21}{299}{82}.
\r{44}\prd{J. D. Bekenstein}{27}{2262}{83}.
\r{45}\zp{ W. Lukosz} {262}{327}{73}.
\r{46}\prd{S.D. Unwin}{26}{944}{1982}.
\r{47}\anp{J. Ambjorn and S. Wolfram}{147}{1}{83}.
\r{48}\prd{D. Page}{26}{947}{82}.
\r{49}\prl{D. Deutsch}{48}{286}{82}.         
\r{50}\book{J. W. Dettman}{Mathematical Methods in Physics and
Engineering}{McGraw-Hill}{62}.
\r{51}\book{M. Abramowitz and I. Stegun}{Handook of Mathematical Functions}{Dover}{65}.
\r{52}\prd{M. Schiffer and J.D. Bekenstein}{42}{}{90}, in press.
\r{53}\prd{W. G. Unruh}{42}{}{90}, in press.
\r{54}\prd{W. G. Unruh and R. M. Wald}{25}{942}{82}.
\r{55}K. H. Mariwalla (1981), unpublished.
\r{56}\grg{J. D. Bekenstein}{14}{355}{82}.
\r{57}\nc{R. Hagedorn}{56A}{1027}{68}; \aa{R. Hagedorn}{5}{184}{70}.
\r{58}\prl{K. Huang and S. Weinberg}{25}{895}{70}.
\r{59}\obscure{R. Rajaraman}{Physics Reports}{21}{239}{75}.
\r{60}\prd{R. F. Dashen, B.Hasslacher and A. Neveau}{10}{4130}{74}.
\r{61}M. Schiffer, Dissertation, Ben Gurion University (1988).
\r{62}\grg{R. Sorkin, R. M. Wald and Z. Z. Jiu}{13}{1127}{81}.
\r{63}\prd{W. H. Zurek and D. N. Page}{29}{628}{84}.
\r{64}\prd{W. G. Unruh}{14}{870}{76}. 
\r{65}\book{J. D. Jackson}{Classical Electrodynamics}{Wiley}{62}.
\r{66}\book{I. S. Gradshteyn and I. M. Ryzhik}{Table of Integrals, Series and Products}{Academic
Press}{80}.

\vglue 12pt
\line{\tenbf Appendix A \hfil}
\vglue 5pt
\tenrm
\baselineskip=13pt

Here we prove Theorem 1.  First we write the composition law
$$
(1-e^{-\beta})e^{-\beta m}=(1-e^{-\alpha})\sum_{n=0}^m{e^{-\alpha(m-n)}Q(n)}.
\eqno(A.1)
$$
In (A.1) we now shift $m\rarrow m+1$, multiply the equation by $e^\beta$ and
substract from the original equation.  Separating out the term of index $m+1$,
and replacing the remaining sum over $n$ by means of Eq.(A.1), we are able to
solve for
$$
Q(m+1)={1-e^{-\beta}\over 1-e^{-\alpha}}(1-e^{\beta-\alpha})\,e^{-\beta(m+1)}.  
\eqno(A.2)
$$
This agrees with Eq.(46) for $m\neq 0$.  To finish the proof, we consider the $m=0$
case of Eq.(A.1).  By virtue of $n$ being fixed as 0, we immediately get the $m=0$ case of
Eq.(46)\qed.\par

\vglue 12pt
\line{\tenbf Appendix B \hfil}
\vglue 5pt
\tenrm
\baselineskip=13pt

Here we prove Eqs. (64)--(65).  The Euler Maclaurin summation formula with the
residue term left out is
$$
\sum_{n=1}^{N-1} f(n) = \int_1^{N} f(x) dx + {f(1) + f(N)\over 2} +
{1\over 2!}B_2f^{(1)}(x)|_{1}^{N} + {1\over
4!}B_4f^{(3)}(x)|_{1}^{N} + \thinspace\dots         \eqno(B.1)
$$
where $B_n$ are the Bernoulli numbers: $B_0=1$, $B_1=-{1\over 2}$, $B_2={1\over
6}$ and $B_p = 0$ for $p=3,5,7,\thinspace\dots\quad$   Suppose $f(x)$ is such
that it and all its derivatives vanish for large arguments.  Let us take
$N\rightarrow \infty$.  In such a situation Eq.(B.1) may  be cast in the form
$$
\sum_1 ^{\infty} f(n) = \int_0^{\infty} f(x) dx -
\int_0 ^1 f(x)dx - \sum_{p=1}^{\infty} {1\over p!} B_p f^{(p-1)}(1).  \eqno(B.2) 
$$
\par

We shall apply this summation formula to the function $f(x)=x\,(e^{\beta
x}-1)^{-1}$ which satisfies the mentioned conditions.  We first perform the
integral
$$
\int_0^\infty f(x) dx = \beta^{-2} \int_0^\infty {x\, dx\over e^x -1} =
{\pi^2\over 6\beta^2}. \eqno (B.3)
$$
Keeping in mind that $f(x)$ is the generating function of the Bernoulli
numbers, namely, 
$$
f(x)={x\over(e^{\beta x}-1)} = \sum_{k=0}^\infty {1\over k!}\,B_k x^k
\beta^{k-1},  \eqno(B.4) 
$$
we obtain for the second integral
$$
\int_0^1 f(x) dx = \int_0^1 {x\, dx\over(e^{\beta x}-1)}
=\beta^{-1}\sum_{k=0}^\infty {1\over k + 1 !} B_k \beta^k. \eqno(B.5)
$$
By virtue of (B.4), we may express
$$
\sum_{p=1}^\infty {1\over p!} B_p f^{(p-1)}(1) = \beta^{-1} \sum_{p=1}^\infty
\sum_{k=p-1}^\infty {1\over k + 1!} B_k \beta^k {k+1\choose p} B_p.   \eqno(B.6)
$$
Next with due care of the limits, we interchange the order of summation:
$$
\sum_{p=1}^\infty {1\over p!} B_p f^{(p-1)}(1) =\beta^{-1} \sum_{k=0}^\infty
\sum_{p=1}^{k+1} {1\over (k+1)!} B_k \beta^k {k+1\choose p} B_p  
$$
$$
= \beta^{-1} \left\{ B_0 B_1 + \sum_{k=1}^\infty\, {1\over (k+1)!}B_k \beta^k 
\sum_{p=1}^{k+1} {k+1\choose p} B_p\right\}
$$
$$
= \beta^{-1} \left\{ B_0 B_1 + \sum_{k=1}^\infty {1\over (k+1)!} B_k \beta^k
\left[ \sum_{p=0}^{k+1}   {k+1\choose p} B_p - B_0 \right] \right\}.  \eqno(B.7)
$$

Recalling that $B_0=1$ and $B_1=-{1\over 2}$ and using the identity\q{51}
between Bernoulli's numbers
$$
\sum_{p=0}^{k+1}   {k+1\choose p} B_p  = B_{k+1}, \eqno(B.8)
$$
we write
$$
\sum_{p=1}^\infty {1\over p!} B_p f^{(p-1)}(1) = \beta^{-1} \left\{ -{1\over
2} +\sum_{k=1}^\infty {1\over (k+1)!}B_k\beta^k(B_{k+1} - 1)\right\}.\eqno(B.9)
$$
Since $B_k=0$ for $k=3,5,\thinspace\dots$, $B_kB_{k+1}=0$ for $k\geq
2$.  Inserting (B.3), (B.5) and (B.9) into (B.2) we obtain finally
$$
\sum_{j=1}^\infty {j\over e^{\beta j} - 1} \approx {\pi^2\over 6\beta^2} -{1\over
2\beta} + {1\over 24}.\eqno(B.10)
$$
Since the derivative with respect to $\beta$ of $\ln Z$ in Eq.(62) is the
negative of the sum in (B.10), $\ln Z$ may be obtained by integration of the
former expression with respect to $\beta$. The integration constant was obtained
by performing the sum (62) numerically for, say, $\beta=1$.  We find 
$$
\ln Z \approx  {\pi^2\over 6\beta} + {1\over 2}\ln \beta -{\beta\over 24} -
0.91894.\eqno(B.11) $$

\vglue 12pt
\line{\tenbf Appendix C \hfil}
\vglue 5pt
\tenrm
\baselineskip=13pt

Here we establish the integral equation (106).  First, since ${dn\over dE}$ is a sum of delta
functions [see Eq.(100)], we have for $m \geq 1$ the identity  
$$
N_m(E) = \int_0 ^E dE_1 \int_0 ^E dE_2 \thinspace\dots \int_0 ^E dE_{m-1}  
$$
$$
({dn\over dE_2}) \thinspace\dots ({dn\over dE_{m-1}})\ n(E - \sum_{k=1}^{m-1}
E_k).      \eqno(C.1)
$$
This equation may be put in an  equivalent form by recalling that 
the function $n(E - \sum_{k=1} ^{m-1} E_k)$ vanishes for a negative
argument, \ie the inequality $E_p \leq E - \sum_{k=1} ^{p-1} E_k $ is always
satisfied.  Thus
$$
N_m(E) = \int_0 ^E dE_1 \int_0 ^{E-E_1} dE_2 \thinspace\dots \int_0 ^{E-E_1-E_2
\thinspace\dots - E_{m-2}} dE_{m-1}    
$$
$$
({dn\over dE_1}) ({dn\over dE_2}) \thinspace\dots ({dn\over dE_{m-1}})\ n( E -
\sum_{k=1} ^{m-1} E_k)    \eqno(C.2)   
$$
for $m\geq 1$ together with Eq.(103).  Putting these pieces together, $N(E)$ is
expressable as
$$
N(E) =\Theta(E) + \sum_{m=1}^\infty \int_0 ^E dE_1 \int_0 ^{E -E_1} dE_2
\thinspace\dots \int_0 ^{E-E_1-E_2\thinspace\dots -E_{m-2}} dE_{m-1}  
$$
$$ ({dn\over dE_1}) ({dn\over dE_2}) \dots
({dn\over dE_{m-1}})\ n( E - \sum_{k=1} ^{m-1} E_k). \eqno(C.3)
$$
 This messy expression is nothing but the iteration of the integral equation
$$
N(E) = \Theta(E) + \int_0 ^E N(E - E')({dn\over dE'}) dE'    \eqno(C.4)
$$
starting from $N(E)=n(E)$.  Eq.(C.4) is identical to Eq.(106).

\vglue 12pt
\line{\tenbf Appendix D \hfil}
\vglue 5pt
\tenrm
\baselineskip=13pt

Here we use the method of Ref.40 to evaluate the integral
$$
N(E) = {1\over 2\pi i} \int_{\gamma -i\infty}^{\gamma +i\infty}{e^{\epsilon
s}-1\over s(e^{\epsilon s} -2)} e^{sE}ds.    \eqno(D.1)
$$
Define the variables $\sigma \equiv \epsilon s -\ln 2$ and $ a \equiv E/\epsilon$. The above
expression then reads
$$
N(E) = 2^{a}\, I(a), \eqno(D.2)
$$
where
$$
I(a) \equiv {1\over 2\pi i} \int_{\gamma\epsilon -i\infty}^{\gamma\epsilon+i\infty}{(2e^\sigma
-1)\, e^{a\sigma}\over 2(\sigma +\ln2) (e^\sigma -1)}\, d\sigma.        \eqno(D.3)
$$
\par

\vglue 10.0truecm
\baselineskip=10pt
\centerline{\eightrm Fig.5. Contour for evaluating the contour integral in Eq.(D.1).}
\baselineskip=13pt
\vskip 0.3truecm

Now push the contour leftwards to minus infinity while indenting it so as not to overrun any
of the infinity of poles  $\sigma = i 2\pi k$ with $k$ integral, as shown in Fig.5. (note that
there is no pole at $\sigma=-\ln 2$).  By Cauchy's theorem the integral is 
$$
I(a) = \sum_{k = -\infty}^\infty {e^{i 2\pi a k}\over 2(i 2\pi k +\ln2)}. \eqno(D.4) 
$$
At this point we expand the exponential in sines and cosines and rationalize the complex
denominator.  Four series result of which two vanish by symmetry.  We are left with
$$ I(a) = {1\over 2\ln2} + {\eta^2\over \ln2}\sum_{k = 1}^\infty {\cos(2\pi a k)\over k^2 + \eta^2}
+{1\over 2\pi}\sum_{k = 1}^\infty {k \sin(2\pi a k)\over k^2 + \eta^2}, \eqno(D.5) 
$$
where $\eta \equiv \ln2/2\pi$.  Notice that $I$ depends only on the fractional part of its
argument because addition of any integer to $a$ leaves the above series unchanged.\par

Now recall the identities\q{66}
$$\sum_{k=1}^\infty {\cos(kx)\over k^2 + \eta^2}= {\pi\over 2\eta}{\cosh \eta(\pi-x)\over \sinh
\eta\pi} -{1\over2\eta^2};\qquad 0 \leq x \leq 2\pi.   \eqno(D.6) 
$$
and
$$
\sum_{k=1}^\infty {k \sin(kx)\over k^2 + \eta^2}={\pi\over 2}{\sinh\eta(\pi - x)\over
\sinh\eta\pi}; \qquad  0 \leq x \leq 2\pi.\eqno(D.7) 
$$
Expanding the hyperbolic functions, using the explicit value of $\eta$, and setting $x=2\pi a$,
we can use these to reduce (D.5) to the form
$$
I(a) = 2^{-[a]},            \eqno(D.8)
$$
where $[a]$ stands for the fractional part of $a$.  The whole part of $a$ drops out for the
reason mentioned above.  Going back to (D.2) we see that $N(E)$ is 2 to the whole part
of $a$. 

\par\vfill\eject\end